\pdfoutput=1
\documentclass[a4paper, UTF8, 11pt]{article}
\usepackage[title]{appendix}
\usepackage{float}
\usepackage{geometry}
\usepackage{amsmath}
\usepackage{mathtools}
\usepackage{amssymb}
\usepackage{amsfonts}
\usepackage{amsthm}
\usepackage{mathtools}
\usepackage{extarrows}
\usepackage{setspace}
\usepackage{indentfirst}
\usepackage[bottom]{footmisc}
\usepackage{titlesec}
\usepackage{natbib}
\usepackage{graphicx}
\usepackage{mathrsfs}
\usepackage{diagbox}
\usepackage{caption}
\usepackage{color}
\usepackage{xcolor}
\usepackage{enumitem}
\usepackage{bbm}
\usepackage{fancyhdr}
\usepackage{lastpage}
\usepackage{multirow}
\usepackage[colorlinks=true, urlcolor=magenta, citecolor=purple, linkcolor=blue, backref=none]{hyperref}
\usepackage[nodayofweek]{datetime}

\titleformat*{\section}{\large\bf}
\titleformat*{\subsection}{\normalsize\bf}
\setcitestyle{authoryear}
\allowdisplaybreaks[4]
\makeatletter
\renewcommand\p@enumii{}
\makeatother

\newtagform{test}{[}{]}
\usetagform{test}

\newcommand{\E}{\mathbb{E}}
\newcommand{\V}{\mathrm{Var}}

\newcommand{\p}{\mathbb{P}}
\newcommand{\ddd}{\;\mathrm{d}}

\newcommand{\mr}{\mathbb{R}}

\newcommand{\convd}{\xrightarrow{\mathrm{d}}}
\newcommand{\convp}{\xrightarrow{\mathbb{P}}}
\newcommand{\convas}{\xrightarrow{\mathrm{a.s.}}}
\newcommand{\convds}{\xrightarrow{\mathrm{d}^*}}

\newcommand{\defeq}{\xlongequal{\mathrm{def}}}

\newcommand{\lp}{\left(}
\newcommand{\lbk}{\left[}
\newcommand{\lbr}{\left\{}
\newcommand{\rp}{\right)}
\newcommand{\rbk}{\right]}
\newcommand{\rbr}{\right\}}
\newcommand{\lv}{\left\vert}
\newcommand{\rv}{\right\vert}

\newcommand{\indicator}{\mathbbm{1}}
\newcommand{\T}{\mathsf{T}}

\newtheoremstyle{thrysty}{5pt}{10pt}{\normalfont}{\parindent}{\bfseries}{:}{0.5em}{}
\theoremstyle{thrysty}
\newtheorem{thry}{Theorem}[section]

\newtheorem{lemma}[thry]{Lemma}

\newtheorem{assum}{Assumption}[section]

\newtheorem{algo}{Algorithm}[section]
\newtheorem{remark}{Remark}[section]
\newenvironment{pf}[1]{\textbf{Proof{#1}:}}{\qed}

\newcommand{\bN}{\mathbb{N}}
\newcommand{\bM}{\mathbb{M}}
\newcommand{\bW}{\mathbb{W}}
\newcommand{\bS}{\mathbb{S}}

\newcommand{\bZ}{\mathbb{Z}}
\newcommand{\bG}{\mathbb{G}}
\newcommand{\mI}{\mathcal{I}}

\newcommand{\rhat}{\widehat{r}}

\newcommand{\betahattall}{\widehat{\beta}_\mathrm{tall}}
\newcommand{\Chat}{\widehat{C}}
\newcommand{\chat}{\widehat{c}}
\newcommand{\ehat}{\widehat{e}}
\newcommand{\ehatbar}{\overline{\widehat{e}}}
\newcommand{\sighat}{\widehat{\sigma}}

\newcommand{\oy}{\mathscr{Y}}
\newcommand{\Ytall}{Y_\mathrm{tall}}
\newcommand{\Ywide}{Y_\mathrm{wide}}

\newcommand{\Hhatmiss}{\widehat{H}_\mathrm{miss}}
\newcommand{\Ftall}{{F}_\mathrm{tall}}
\newcommand{\Fwide}{{F}_\mathrm{wide}}
\newcommand{\Ltall}{{\Lambda}_\mathrm{tall}}
\newcommand{\Lwide}{{\Lambda}_\mathrm{wide}}
\newcommand{\Htall}{H_\mathrm{tall}}
\newcommand{\Hwide}{H_\mathrm{wide}}
\newcommand{\ewidei}{e_{\mathrm{wide},i}}
\newcommand{\Fhattall}{\widehat{F}_\mathrm{tall}}
\newcommand{\Fhatwide}{\widehat{F}_\mathrm{wide}}
\newcommand{\Lhattall}{\widehat{\Lambda}_\mathrm{tall}}
\newcommand{\Lhatwide}{\widehat{\Lambda}_\mathrm{wide}}

\newcommand{\Ftid}{\widetilde{F}}
\newcommand{\Ftidtall}{\widetilde{F}_\mathrm{tall}}
\newcommand{\Ftidwide}{\widetilde{F}_\mathrm{wide}}
\newcommand{\Ltid}{\widetilde{\Lambda}}
\newcommand{\Ltidtall}{\widetilde{\Lambda}_\mathrm{tall}}
\newcommand{\Ltidwide}{\widetilde{\Lambda}_\mathrm{wide}}
\newcommand{\ftid}{\widetilde{f}}
\newcommand{\ltid}{\widetilde{\lambda}}
\newcommand{\Dtall}{D_\mathrm{tall}}
\newcommand{\Dwide}{D_\mathrm{wide}}

\newcommand{\Vhattall}{\widehat{V}_\mathrm{tall}}
\newcommand{\Vhatwide}{\widehat{V}_\mathrm{wide}}
\newcommand{\bVhat}{\widehat{\mathbb{V}}}

\newcommand{\fhattallt}{\widehat{f}_{\mathrm{tall},t}}
\newcommand{\fhattalls}{\widehat{f}_{\mathrm{tall},s}}
\newcommand{\fhatwidet}{\widehat{f}_{\mathrm{wide},t}}
\newcommand{\fhatwides}{\widehat{f}_{\mathrm{wide},s}}
\newcommand{\fhattallsk}{\widehat{f}_{\mathrm{tall},s-k}}
\newcommand{\lhatwidei}{\widehat{\lambda}_{\mathrm{wide},i}}
\newcommand{\lhatwidej}{\widehat{\lambda}_{\mathrm{wide},j}}
\newcommand{\lhattalli}{\widehat{\lambda}_{\mathrm{tall},i}}

\newcommand{\Phihat}{\widehat{\Phi}}
\newcommand{\Gamhat}{\widehat{\Gamma}}
\newcommand{\tr}{\mathrm{tr}}
\newcommand{\Ops}{O_{\p^*}}
\newcommand{\ops}{o_{\p^*}}

\title{
\textbf{Confidence Intervals of Treatment Effects in Panel Data	Models with Interactive
	Fixed Effects}\footnote{The authors thank all seminar participants for their comments. We also thank
Xinyi Wang for assistance in translating our original MATLAB codes into Python and R codes.}
}    
\author{
	Xingyu Li\\National School of Development\\Peking University\\x.y@pku.edu.cn
	\and 
	Yan Shen\\National School of Development\\Peking University\\yshen@nsd.pku.edu.cn
	\and
	Qiankun Zhou\\Department of Economics\\Louisiana State University\\qzhou@lsu.edu
}
\date{\today}

\begin{document}
	
\setlength\abovedisplayskip{2pt plus 0pt minus 2pt}
\setlength\belowdisplayskip{2pt plus 0pt minus 2pt}

\maketitle

\thispagestyle{fancy}

\begin{abstract}
We consider the construction of confidence intervals for treatment effects
estimated using panel models with interactive fixed effects. We first use
the factor-based matrix completion technique proposed by %
\citet{bai2021matrix} to estimate the treatment effects, and then use
bootstrap method to construct confidence intervals of the treatment effects
for treated units at each post-treatment period. Our construction of
confidence intervals requires neither specific distributional assumptions on
the error terms nor large number of post-treatment periods. We also
establish the validity of the proposed bootstrap procedure that these confidence
intervals have asymptotically correct coverage probabilities. Simulation
studies show that these confidence intervals have satisfactory finite sample
performances, and empirical applications using classical datasets yield
treatment effect estimates of similar magnitudes and reliable confidence
intervals.


\end{abstract}

\pagebreak

\section{Introduction}

\label{sec:Introduction}

Treatment effects of certain policy interventions on economic entities are
often of major interest in economic and econometric studies. In panel data
models, estimation of and inference on treatment effects can be formulated
as missing data problems. For instance, consider a sample of $N$ units with
a policy intervention on units $N_{0}+1,\ldots ,N$ at time $T_{0}$, so that
the pre-treatment periods are $1,\ldots ,T_{0}$, the post-treatment periods
are $T_{0}+1,\ldots ,T$, and the control units are $1,\ldots ,N_{0}$. Let $%
y_{i,t}$ be the potential outcome in the absence of policy intervention for
unit $i$ at time $t$, and $y^{+}_{i,t}=y_{i,t}+\Delta _{i,t}$ be the
potential outcome under policy intervention for unit $i$ at time $t$.
We are interest in the treatment effect $\Delta _{i,t}$ for $(i,t)\in
\mathcal{I}_{1}=\{N_{0}+1,\ldots ,N\}\times \left\{ T_{0}+1,\ldots
,T\right\} $, and it is easy to see that calculating $\Delta _{i,t}=%
y^{+}_{i,t}-y_{i,t}$ requires knowing $y_{i,t}$, which is actually
unobservable (missing) for $(i,t)\in \mathcal{I}_{1}$.

Many existing methods of estimating $\Delta _{i,t}$ make efforts to
construct ``counterfactuals" for $y_{i,t}$, $(i,t)\in
\mathcal{I}_{1}$ (\textit{e.g.}, \citealp{abadie2010synthetic}; \citealp{hsiao2012panel};
\citealp{bai2021matrix}; \citealp{athey2021matrix}; among others). 
The underlying models in counterfactual
frameworks can be simplified as $y_{i,t}=\xi _{i,t}+e_{i,t}$, where $\xi
_{i,t}$ is the systematic part and $e_{i,t}$ is the idiosyncratic error.
Then the predicted values $\left\{ \widehat{\xi }_{i,t}:(i,t)\in \mathcal{I}%
_{1}\right\} $ serve the role as the counterfactuals for $\left\{
y_{i,t}:(i,t)\in \mathcal{I}_{1}\right\} $, and the treatment effects are
estimated by $\widehat{\Delta }_{i,t}=y^{+}_{i,t}-\widehat{\xi }_{i,t}$
for every $(i,t)\in \mathcal{I}_{1}$. 
According to model specifications, assumptions, and estimation strategies,
we can roughly classify the existing methods into the following 4 categories.

The first category is the synthetic control method proposed by
\citet{abadie2003economic} and \citet{abadie2010synthetic}.
It approximates $y_{N,t}$ by a convex combination of $\left( y_{1,t},\ldots ,y_{N-1,t}\right)
$, which can be obtained from a constrained regression.
Generalisations of the synthetic control method can be found in, for example,
\citet{amjad2018robust}, \citet{abadie2021penalized}, \citet{arkhangelsky2021synthetic},
\citet{ben2021augmented}, \citet{ben2021synthetic}, \citet{kellogg2021combining},
and \citet{masini2021counterfactual}. One can see \citet{abadie2021using} for a comprehensive review.
The second category is the the panel
data approach proposed by \citet{hsiao2012panel}, which constructs the counterfactual
by solving a linear regression of $y_{N,t}$ on $\left( y_{1,t},\ldots ,y_{N-1,t}\right)$.
This approach is later extended by \citet{ouyang2015treatment},
\citet{li2017estimation}, and \citet{hsiao2022multiple}.
Methods in the third category use the factor model technique
(\textit{e.g.}, \citealp{bai2002determining}; \citealp{bai2003inferential}; \citealp{pesaran2006estimation};
\citealp{bai2009panel}; \citealp{moon2015linear}) to construct counterfactuals.
Examples include \citet{kim2014divorce}, \citet{xu2017generalized}, and \citet{bai2021matrix}.
The last category starts from the statistical learning literature on matrix completion
(\textit{e.g.}, \citealp{candes2009exact}; \citealp{candes2010matrix}; \citealp{mazumder2010spectral};
\citealp{gamarnik2016note}),
and a recent example is \citet{athey2021matrix}, which completes the potential outcome matrix
via nuclear norm regularisation.

In practice, a mere point estimator of the treatment effect $\Delta _{i,t}$
is not sufficient either theoretically or empirically, and a valid inference
is needed, which usually requires the (asymptotic) distribution of $\left( \widehat{%
\Delta }_{i,t}-\Delta _{i,t}\right) $. As $\widehat{\Delta }_{i,t}-\Delta
_{i,t}=\left( \xi _{i,t}-\widehat{\xi }_{i,t}\right) +e_{i,t}$, the
idiosyncratic error term $e_{i,t}$ will dominate the distribution of $\left(
\widehat{\Delta }_{i,t}-\Delta _{i,t}\right) $ as $ N_{0},T_{0}\rightarrow
\infty $ whenever the systematic part can be consistently approximated. 
Due to this issue, 
most studies make inferences on time series average treatment effects as the
number of post-treatment periods $\left( T-T_0 \right) \to\infty$
 (\textit{e.g.}, \citealp{hsiao2012panel}; \citealp{li2017estimation};
\citealp{li2020statistical}, and \citealp{chernozhukov2018t}), 
or make inferences on cross-sectional average treatment effects as the
number of treated units $\left( N-N_0 \right) \to\infty$
(\textit{e.g.}, \citealp{xu2017generalized}; \citealp{arkhangelsky2021synthetic};
\citealp{ben2021synthetic}; \citealp{bai2021matrix}),
or assume the
normality of error terms (\textit{e.g.}, \citealp{fujiki2015disentangling};
\citealp{bai2021matrix}).
However, these approaches can not simultaneously meet
the following needs in empirical studies.

\begin{enumerate}[label=(\arabic*), labelindent=\parindent, leftmargin=*, nosep]

\item Inferences do not rely on any specific distributional assumptions (%
\textit{e.g.}, normality) on the error terms.

\item Inference are built on small number of post-treatment period, either
because the observed post-treatment time span is short or in order to avoid
confounding effects from other interventions in longer time span.

\item Inferences are performed not only for the (cross-sectional or time
series) average treatment effect, but also for the treatment effect on every
treated unit at each post-treatment period.
\end{enumerate}

To the best of our knowledge, there are only few inferential studies
simultaneously satisfy the three conditions above. %
\citet{abadie2010synthetic} use a cross-sectional permutation strategy,
where they apply the synthetic control method to every potential control
unit to approximate the true distribution of treatment effect estimator
under the null hypothesis of $\Delta _{N,t}=0$. Obviously, the validity of
this permutation approach relies on the cross-sectional exchangeability of
the units. Furthermore, if one wants to construct the confidence intervals
of the treatment effects in this fashion, then the equality between
distributions of outcomes in the absence of and under intervention up to a
location shift is required.
\citet{chernozhukov2021exact} apply time series permutation test to the
inference on treatment effects. For a given null hypothesis $H_{0}:\Delta
_{N}=\delta _{N}$, they plug $\delta_{N}$ into the observed data matrix to
compute the full outcome matrix in the absence of intervention, and then
perform time series permutations of the residuals from the treated unit to
obtain an approximation of the distribution of the test statistic under $%
H_{0}$ and the critical value. Confidence intervals can be obtained by grid
searching on a set of different values of $\delta_{N}$, which may lead to
intensive computation. As the latest work, \citet{cattaneo2021prediction}
build prediction intervals for synthetic control methods on conditional
prediction intervals and non-asymptotic concentration. Since concentration
inequalities only guarantee lower bounds of coverage probabilities, the
proposed prediction intervals may suffer from conservativeness.

This paper aims to provide a theoretically non-conservative and
computationally simple inferential procedure for the
estimated treatment effects that simultaneously satisfy the above
conditions, without imposing the cross-sectional exchangeability assumption. 
To be more specific, we apply bootstrap method to the treatment effect
estimators proposed by \citet{bai2021matrix}.
Note that \citet{bai2021matrix}'s model keeps a simple factor structure
but is quite general in the sense that
it allows for multiple treated units and heterogeneous time of intervention.
Bootstrapping is one of most commonly used tools to approximate an unknown
distribution via resampling, and is shown to be asymptotically valid under
mild conditions. We construct bootstrap-based confidence intervals of
treatment effects in panel data models with interactive fixed effects.
Specifically, we first apply the factor-based matrix completion technique
proposed by \citet{bai2021matrix} to obtain point estimators of treatment
effects. Then we follow
\citet{gonccalves2017bootstrap} to approximate the distribution of $\widehat{%
\Delta }_{i,t}-\Delta _{i,t}$ by (block) wild bootstrap and bootstrap, and
use the quantiles of bootstrapped distribution to construct confidence
intervals of $\Delta _{i,t}$ {for every} $(i,t)\in \mathcal{I}_{1}$.

Our paper contributes to the literature in the following folds. First, we
establish the validity of confidence intervals by proving that they have
asymptotically correct coverage probabilities as $\left( N_{0},T_{0}\right)
\rightarrow \infty $, and our proposed confidence intervals do not require
any specific distributional assumption on $e_{i,t}$ and are robust to small
number of post-treatment periods. Thus, it is theoretically non-conservative
and can meet the three needs for inferential purpose of treatment effects estimation
in empirical studies.
Second, we extend the bootstrap procedure in 
\citet{gonccalves2017bootstrap} to a more general case.
Since \citet{gonccalves2017bootstrap} intend to
make inference on factor-augmented regression models \citep{bai2006confidence} where
the factors serves as intermediate variables,
they only require valid bootstrap approximations of factors.
In this paper, we intend to make inference on the factor model (with missing values) itself,
and the estimators involve products of estimated factors, their associated loadings,
and a rotation matrix.
This implies that we need valid bootstrap approximations of factors, loadings,
and the rotation matrix, which creates theoretical challenges. Nevertheless, we successfully
establish the asymptotic validity of our proposed bootstrap for such a
scenario.

The finite sample properties of proposed bootstrap construction of
confidence intervals for estimated treatment effects are investigated
through Monte Carlo simulation, using different data generating processes
(DGPs) for models with or without exogenous covariates, and for models with
heteroscedastic errors or serially correlated errors. The simulation
results show that the proposed bootstrap procedure works remarkably
well for constructing confidence intervals for estimated treatments. Namely,
the empirical coverage ratios are quite close to nominal values in all cases
we considered regardless of whether the number of unobserved factors are
priorly given or estimated from data.

Finally, we revisit
the benefits of political and economic integration of Hong Kong with Mainland China in
\citet{hsiao2012panel}, as well as the effectiveness of the California Tobacco Control
Program (CTCP) on per capita cigarettes consumption and personal health
expenditures in \citet{abadie2010synthetic}. We note that the treatment effects estimated using
our new method are of similar magnitudes to those in the literature, while
our bootstrap procedure provides reliable confidence intervals showing that
the impacts of CTCP were significant over time for both per capita cigarettes
consumption and personal health expenditures, and that the impacts of
political and economic integration of Hong Kong with Mainland China vanished
over time, although significant in the first few years.

The rest of this article is organised as follows. Section \ref{sec:The Model}
establishes the model and the assumptions. Section \ref{sec:Estimation and
Inference in a Model without Covariates} focuses on the estimation and
inference in a model without covariates, and Section \ref{sec:Estimation and
Inference in a Model with Covariates} deals with a model with covariates. We
conduct Monte Carlo simulations in Section \ref{sec:Simulation Studies}, and
apply our method to classical datasets in Section \ref{sec:Empirical
Application}. Section \ref{sec:Conclusion} concludes. Mathematical proofs of
main results are left to online appendices.

\textbf{Notations}: we introduce some notations that are frequently used
throughout this paper. $\mathbb{Z}_{+}=\{1,2,\ldots \}$ is the set of
positive integers. $I_r$ is the $r\times r$ identity matrix.
$A^{\mathsf{T}}$ denotes the transpose of matrix $A$. For
a vector $z$, let $\left\Vert z\right\Vert =\sqrt{z^{\mathsf{T}}z}$ be the
Euclidean norm of $z$. For a matrix $A$, let $\left\Vert A\right\Vert =\sqrt{%
\mathrm{tr}\left( A^{\mathsf{T}}A\right) }$ be the Frobenius norm of $A$. $%
\mathbb{N}\left( \mu ,\Sigma \right) $ stands for a normal distribution with
mean $\mu $ and variance $\Sigma $. And $\xrightarrow{\mathrm{a.s.}}$, $%
\xrightarrow{\mathbb{P}}$ and $\xrightarrow{\mathrm{d}}$ denote almost sure
convergence, convergence in probability and convergence in distribution,
respectively. We let $M$ denote a generic finite positive constant, whose
value does not depend on $N$ or $T,$ and may vary case by case.

\section{The Model}

\label{sec:The Model}

Suppose there are observations $\left( \mathscr{Y}_{i,t},x_{i,t}\right) $
for $i=1,\ldots ,N$ and $t=1,\ldots ,T$. Let the dummy variable $d_{i,t}$
indicate the $i$-th unit's treatment status at time $t$ with $d_{i,t}=1$ if
under the treatment and $d_{i,t}=0$ if not. The observed data takes the form
\begin{equation}
\mathscr{Y}_{i,t}=y_{i,t}+d_{i,t}\Delta _{i,t},
\end{equation}%
where $y_{i,t}$ is the latent outcome of unit $i$ at time $t$ in the absence
of treatment, and $\Delta_{i,t}$ is the treatment effect on unit $i$ at
time $t$. For the ease of exposition, we assume $d_{i,t}=1$ for all $(i,t)\in
\mathcal{I}_{1}$ and $d_{i,t}=0$ for all $(i,t)\in \mathcal{I}\backslash
\mathcal{I}_{1}$, where
\begin{equation*}
\mathcal{I}=\left\{ 1,\ldots ,N\right\} \times \left\{ 1,\ldots ,T\right\}
,\qquad \mathcal{I}_{1}=\left\{ N_{0}+1,\ldots ,N\right\} \times \left\{
T_{0}+1,\ldots ,T\right\}
\end{equation*}%
with $1\leq N_{0}<N$ and $1\leq T_{0}<T$. That is, we assume the last $%
N-N_{0}$ units are intervened by the treatment at time $T_{0}$. 
Therefore, the observed matrix of outcomes is
\begin{equation*}
	\begin{bmatrix}
		y_{1,1} & \cdots & y_{N_{0},1} & y_{N_{0}+1,1} & \cdots & y_{N,1} \\
		\vdots & \ddots & \vdots & \vdots & \ddots & \vdots \\
		y_{1,T_{0}} & \cdots & y_{N_{0},T_{0}} & y_{N_{0}+1,T_{0}} & \cdots &
		y_{N,T_{0}} \\
		y_{1,T_{0}+1} & \cdots & y_{N_{0},T_{0}+1} 
		& y_{N_{0}+1,T_{0}+1}+\Delta_{N_{0}+1,T_{0}+1} &
		\cdots & y_{N,T_{0}+1}+\Delta_{N,T_{0}+1} \\
		\vdots & \ddots & \vdots & \vdots & \ddots & \vdots \\
		y_{1,T} & \cdots & y_{N_{0},T} 
		& y_{N_{0}+1,T}+\Delta_{N_{0}+1,T} & \cdots 
		& y_{N,T}+\Delta_{N,T}
	\end{bmatrix}.
\end{equation*}
Note that
the above specification can be generated to allow for heterogeneous
intervention time periods by letting $T_{0}$ be the time of the earliest
intervention. This is consistent with the definition of $\mathcal{I}_{1}$ in %
\citet{bai2021matrix} that $\mathcal{I}_{1}$ corresponds to the
\textquotedblleft largest possible" missing block of the matrix.

We are interested in measuring the treatment effects of the policy
intervention for the treated units after time $T_{0}$, which are $\Delta
_{i,t}$ for $(i,t)\in \mathcal{I}_{1}$. Note that $\Delta _{i,t}$ measures
the difference of the outcomes with and without the intervention of the
treatment in the post-treatment periods. Unfortunately, the outcomes with
and without the intervention of treatment cannot be simultaneously observed
in reality for the same unit at a given time. This is because, once the policy intervention
is in effect, then the researchers can only observe $\mathscr{Y}%
_{i,t}=y_{i,t}+\Delta _{i,t}$, not $y_{i,t}$. Thus, in order to estimate $%
\Delta _{i,t}$, we need to generate the counterfactual of $y_{i,t}$ for $%
(i,t)\in \mathcal{I}_{1}$, denoted as $\widehat{y}_{i,t}$, and thus the
treatment effect can be estimated as
\begin{equation}
\widehat{\Delta }_{i,t}=\mathscr{Y}_{i,t}-\widehat{y}_{i,t},\qquad (i,t)\in
\mathcal{I}_{1}.
\end{equation}

Our goal in this paper is to estimate $\Delta _{i,t}$ for $(i,t)\in \mathcal{%
I}_{1}$ when neither treated units $N_{1}=N-N_{0}$ nor the post treated
periods $T_{1}=T-T_{0}$ is large, where the former can be formulated as
average treatment effects across units (see \citealp{hsiao2022multiple} for
the aggregation of multiple treatment effects), and the latter can be formulated
as the average treatment effects across post-treatment periods
(see \citealp{fujiki2015disentangling}
and \citealp{li2017estimation}).

Following the literature of treatment effects estimation using panels (\textit{e.g.},
\citealp{abadie2010synthetic}; \citealp{hsiao2012panel}; and \citealp{bai2021matrix}),
we assume that $y_{i,t}$ is generated by the following panel data model with
interactive fixed effects:
\begin{equation}
y_{i,t}=x_{i,t}^{\mathsf{T}}\beta +c_{i,t}+e_{i,t},\qquad c_{i,t}=f_{t}^{%
\mathsf{T}}\lambda _{i},  \label{eq:model with covariates}
\end{equation}%
where $x_{i,t}$ is a $p$-dimensional vector of observed covariates of unit $%
i $ at time $t$, $e_{i,t}$ is the idiosyncratic error term of unit $i$ at
time $t$, $f_{t}$ is an $r$-dimensional time-variant unobserved factor at
time $t$, and $\lambda _{i}$ is an $r$-dimensional individual specific factor
loading of unit $i$, where $r$ is the number of unobserved factors and is
usually unknown to researchers.
\footnote{
Even if $r$ is unknown in practice, it can be consistently estimated using
the method described in Section 4 and 5 of \citet{bai2002determining} or the
method proposed by \citet{alessi2010improved}, and thus we can treat $r$ as
known in the theoretical analyses below. This can be formally justified as
follows. Let $\theta $ be a parameter in the model of interest and $\widehat{%
\theta }$ be the estimator of $\theta $ based on estimated number of factors
$\widehat{r}$. Note that $r$ takes values in $\mathbb{Z}_{+}$, and then the
consistency of its estimator $\widehat{r}$ implies that $\mathbb{P}\left(
\widehat{r}=r\right) \rightarrow 1$ as $\left( N,T\right) \rightarrow \infty
$. Therefore,
\begin{align*}
\mathbb{P}\left( \widehat{\theta }\leq x\right)  =\mathbb{P}\left( \widehat{%
\theta }\leq x,\widehat{r}=r\right) +\mathbb{P}\left( \widehat{\theta }\leq
x,\widehat{r}\neq r\right) 
 =\mathbb{P}\left( \left. \widehat{\theta }\leq x\right\vert \widehat{r}
=r\right) \mathbb{P}\left( \widehat{r}=r\right) +o(1) 
 =\mathbb{P}\left( \left. \widehat{\theta }\leq x\right\vert \widehat{r}
=r\right) +o(1).
\end{align*}%
}

Let $y_{i}=\left( y_{i,1},\ldots ,y_{i,T}\right) ^{\mathsf{T}}$, $%
X_{i}=\left( x_{i,1},\ldots ,x_{i,T}\right) ^{\mathsf{T}}$, and $%
e_{i}=\left( e_{i,1},\ldots ,e_{i,T}\right) ^{\mathsf{T}}$ for every $i\in
\{1,\ldots ,N\}$. Define $Y=\left( y_{1},\ldots ,y_{N}\right) $, $F=\left(
f_{1},\ldots ,f_{T}\right) ^{\mathsf{T}}$, and $\Lambda =\left( \lambda
_{1},\ldots ,\lambda _{N}\right) ^{\mathsf{T}}$. We make the assumptions
below.

\begin{assum}
\label{ass:factors and loadings} The factors and loadings satisfy the
following conditions.

\begin{enumerate}[label=(\arabic*), labelindent=\parindent, leftmargin=*, nosep]

\item $\mathbb{E} \left( \left\Vert f_t \right\Vert^8 \right)\le M$ for all $%
t\in \{ 1,\ldots, T \}$.

\item $\left\Vert \lambda_{i} \right\Vert \le M$ for all $i\in\{ 1, \ldots,
N \}$.

\item There exists an $r\times r$ positive definite matrix $\Sigma_{F}$, so
that
\begin{align*}
\frac{1}{T}\sum_{t=1}^T f_t f_t^\mathsf{T} \xrightarrow{\mathbb{P}}
\Sigma_{F}, \qquad \frac{1}{T_0}\sum_{t=1}^{T_0} f_t f_t^\mathsf{T} %
\xrightarrow{\mathbb{P}} \Sigma_{F}, \qquad \frac{1}{T-T_0}%
\sum_{t=T_0+1}^{T} f_t f_t^\mathsf{T} \xrightarrow{\mathbb{P}} \Sigma_{F}
\end{align*}
as $T$, $T_0\to\infty$.

\item There exists an $r\times r$ positive definite matrix $\Sigma_{\Lambda}$%
, so that
\begin{align*}
\frac{1}{N}\sum_{i=1}^N \lambda_{i} \lambda_{i}^\mathsf{T} \to
\Sigma_{\Lambda}, \qquad \frac{1}{N_0}\sum_{i=1}^{N_0} \lambda_{i}
\lambda_{i}^\mathsf{T} \to \Sigma_{\Lambda}, \qquad \frac{1}{N-N_0}%
\sum_{i=N_0+1}^N \lambda_{i} \lambda_{i}^\mathsf{T} \to \Sigma_{\Lambda}
\end{align*}
as $N$, $N_0\to\infty$.

\item The eigenvalues of $\Sigma_{F}\Sigma_{\Lambda}$ are distinct.
\end{enumerate}
\end{assum}

\begin{assum}
\label{ass:distributions of idiosyncratic errors} The distributions of the
idiosyncratic errors have the following properties.

\begin{enumerate}[label=(\arabic*), labelindent=\parindent, leftmargin=*, nosep]

\item The idiosyncratic errors have no cross-sectional dependence, \textit{%
i.e.}, the sequences $\left\{ e_{1,t} \right\}_{t=1}^T$, $\left\{ e_{2,t}
\right\}_{t=1}^T$, $\ldots$, $\left\{ e_{N,t} \right\}_{t=1}^T$ are mutually
independent.

\item For all $i\in \{1, \ldots, N \} $, the process $\left\{ e_{i,t}
\right\}_{t=1}^\infty$ is strictly stationary and ergodic.

\item For all $(i,t)\in \mathcal{I}$, the cumulative distribution function
of $e_{i,t}$ is everywhere continuous.

\item The error terms $\lbr e_{i,t}:(i,t)\in\mI \rbr$ are independent of the treatment status
$\lbr d_{i,t}:(i,t)\in \mI \rbr$.
\end{enumerate}
\end{assum}

\begin{assum}
\label{ass:moments of idiosyncratic errors} The moment conditions below hold
for the full sample indexed by $\mathcal{I}$, the balanced subsample indexed
by $\{1,\ldots, N_0\}\times \{ 1, \ldots, T_0\}$, the control subsample
indexed by $\{ 1, \ldots, N_0\}\times \{1, \ldots, T \}$, the pre-treatment
subsample indexed by $\{1, \ldots, N \}\times \{ 1, \ldots, T_0\}$, and the
missing subsample indexed by $\mathcal{I}_1$.

\begin{enumerate}[label=(\arabic*), labelindent=\parindent, leftmargin=*, nosep]

\item $\mathbb{E}\left( e_{i,t}\right) =0$ and $\mathbb{E}\left(
e_{i,t}^{8}\right) \leq M$ for all $(i,t)\in \mathcal{I}$.

\item For all $t\in\{1,\ldots, T \}$,
\begin{align*}
\sum_{s=1}^T \left\vert \mathbb{E} \left( \frac{1}{N}\sum_{i=1}^N e_{i,t}
e_{i,s} \right) \right\vert \le M.
\end{align*}

\item
\begin{align*}
\frac{1}{NT}\sum_{i=1}^N \sum_{t=1}^T \sum_{s=1}^T \left\vert \mathbb{E}
\left( e_{i,t} e_{i,s} \right) \right\vert \le M.
\end{align*}

\item For all $(s,t)\in \{1,\ldots, T\}^2$,
\begin{align*}
\mathbb{E} \left( \left\vert \frac{1}{\sqrt{N}}\sum_{i=1}^N \left[
e_{i,s}e_{i,t} -\mathbb{E}\left( e_{i,s}e_{i,t} \right) \right]
\right\vert^4 \right) \le M.
\end{align*}

\item
\begin{align*}
\mathbb{E}\left( \frac{1}{N}\sum_{i=1}^N \left\Vert \frac{1}{\sqrt{T}}%
\sum_{t=1}^T f_t e_{i,t} \right\Vert^2 \right) \le M.
\end{align*}

\item For all $t\in \{ 1,\ldots, T\}$,
\begin{align*}
\mathbb{E}\left( \left\Vert \frac{1}{\sqrt{NT}}\sum_{s=1}^T\sum_{i=1}^N f_s %
\left[ e_{i,s}e_{i,t}-\mathbb{E}\left( e_{i,s}e_{i,t} \right) \right]
\right\Vert^2 \right) \le M.
\end{align*}

\item
\begin{align*}
\mathbb{E}\left( \left\Vert \frac{1}{\sqrt{NT}}\sum_{t=1}^T\sum_{i=1}^N f_t
\lambda_i^\mathsf{T} e_{i,t} \right\Vert^2 \right) \le M.
\end{align*}
\end{enumerate}
\end{assum}

\begin{assum}
\label{ass:central limit theorems} The factors, loadings and errors satisfy
central limit theorems.

\begin{enumerate}[label=(\arabic*), labelindent=\parindent, leftmargin=*, nosep]

\item For all $t\in\{1,\ldots, T\}$, there exists an $r\times r$ positive
definite matrix $\Gamma_{t}$, so that
\begin{align*}
\frac{1}{\sqrt{N}}\sum_{i=1}^N \lambda_{i} e_{i,t} \xrightarrow{\mathrm{d}}
\mathbb{N} \left( 0, \Gamma_{t} \right), \qquad \frac{1}{\sqrt{N_0}}%
\sum_{i=1}^{N_0} \lambda_{i} e_{i,t} \xrightarrow{\mathrm{d}} \mathbb{N}
\left( 0, \Gamma_{t} \right)
\end{align*}
as $N, N_0 \to\infty$.

\item For all $i\in \{1,\ldots, N\}$, there exists an $r\times r$ positive
definite matrix $\Phi_{i}$, so that
\begin{align*}
\frac{1}{\sqrt{T}}\sum_{t=1}^T f_t e_{i,t} \xrightarrow{\mathrm{d}} \mathbb{N%
} \left( 0, \Phi_{i} \right), \qquad \frac{1}{\sqrt{T_0}}\sum_{t=1}^{T_0}
f_t e_{i,t} \xrightarrow{\mathrm{d}} \mathbb{N} \left( 0, \Phi_{i} \right)
\end{align*}
as $T, T_0 \to\infty$.
\end{enumerate}
\end{assum}

\begin{assum}
\label{ass:order conditions} The quantities $N$, $T$, $N_0$, $T_0$ satisfy
the conditions below.

\begin{enumerate}[label=(\arabic*), labelindent=\parindent, leftmargin=*, nosep]

\item $TN_0> r\left( T+N_0 \right)$ and $T_0 N > r \left( T_0+N \right)$.

\item $N$, $T$, $N_0$, $T_0$ are of the same order, \textit{i.e.},
\begin{align*}
\lim_{N,N_0\to\infty} \frac{N_0}{N}=c_1\in (0,1], \qquad
\lim_{T,T_0\to\infty} \frac{T_0}{T}=c_2\in (0,1], \qquad \lim_{N,T\to\infty}
\frac{N}{T}=c_3\in (0,\infty).
\end{align*}
\end{enumerate}
\end{assum}

\begin{assum}
\label{ass:covariates x} The conditions below hold for the full sample
indexed by $\mathcal{I}$, the control subsample indexed by $\{ 1, \ldots,
N_0\}\times \{1, \ldots, T \}$, and the pre-treatment subsample indexed by $%
\{ 1, \ldots, N\}\times \{1, \ldots, T_0 \}$.

\begin{enumerate}[label=(\arabic*), labelindent=\parindent, leftmargin=*, nosep]

\item $\mathbb{E}\left( \left\Vert x_{i,t} \right\Vert^8 \right) \le M$ for
every $(i,t)\in\mathcal{I}$.

\item $\inf \left\{ \mathcal{D}(\mathcal{F}) : \mathcal{F}\in \mathscr{F}
\right\}>0$, where
\begin{align*}
\mathcal{D}(\mathcal{F})&=\left[ \frac{1}{NT}\sum_{i=1}^N X_i^\mathsf{T}
\left( I_T-\frac{\mathcal{F}\mathcal{F}^\mathsf{T}}{T} \right) X_i \right] -%
\left[ \frac{1}{TN^2} \sum_{i=1}^N \sum_{k=1}^N X_i^\mathsf{T} \left( I_T-%
\frac{\mathcal{F}\mathcal{F}^\mathsf{T}}{T} \right) X_k a_{i,k} \right], \\
a_{i,k}&=\lambda_{i}^\mathsf{T} \left( \frac{\Lambda^\mathsf{T} \Lambda}{N}
\right)^{-1} \lambda_{k}, \\
\mathscr{F}&=\left\{ \mathcal{F}\in \mathbb{R}^{T\times r}: \left. \mathcal{F%
}^\mathsf{T} \mathcal{F} \right/ T =I_r \right\}.
\end{align*}
\end{enumerate}
\end{assum}

\begin{assum}
\label{ass:additional conditions of idiosyncratic errors} The conditions
below hold for the full sample indexed by $\mathcal{I}$, the control
subsample indexed by $\{ 1, \ldots, N_0\}\times \{1, \ldots, T \}$, and the
pre-treatment subsample indexed by $\{ 1, \ldots, N\}\times \{1, \ldots, T_0
\}$.

\begin{enumerate}[label=(\arabic*), labelindent=\parindent, leftmargin=*, nosep]

\item
\begin{align*}
\frac{1}{T}\sum_{t=1}^T \sum_{s=1}^T \max_{1\le i \le N} \left\vert \mathbb{E%
} \left( e_{i,t} e_{i,s} \right) \right\vert \le M.
\end{align*}

\item $\left\Vert \mathbb{E}\left( e_{i}e_{i}^\mathsf{T} \right)
\right\Vert_S \le M$ for all $i\in\{1,\ldots, N\}$, where $\left\Vert \cdot
\right\Vert_S$ is the spectral norm of a matrix.

\item
\begin{align*}
\frac{1}{T^2 N}\sum_{t=1}^T \sum_{s=1}^T \sum_{u=1}^T \sum_{v=1}^T
\sum_{i=1}^N \left\vert \mathrm{Cov} \left( e_{i,t}e_{i,s}, e_{i,u}e_{i,v}
\right) \right\vert \le M.
\end{align*}

\item
\begin{align*}
\frac{1}{TN^2}\sum_{t=1}^T \sum_{s=1}^T \sum_{i=1}^N \sum_{j=1}^N \left\vert
\mathrm{Cov} \left( e_{i,t}e_{j,t}, e_{i,s}e_{j,s} \right) \right\vert \le M.
\end{align*}

\item There exists a $p\times p$ positive definite matrix $\Omega$, so that
\begin{align*}
\frac{1}{\sqrt{NT}}\sum_{i=1}^N \left[ \left( I_T-\frac{FF^\mathsf{T}}{T}
\right) X_i -\frac{1}{N}\sum_{k=1}^N a_{i,k} \left( I_T-\frac{FF^\mathsf{T}}{%
T} \right) X_k \right]^\mathsf{T} e_i \xrightarrow{\mathrm{d}} \mathbb{N}
(0,\Omega).
\end{align*}
\end{enumerate}
\end{assum}

\begin{assum}
\label{ass:error is independent of factor and x} $\left\{ e_{i,t}: (i,t)\in%
\mathcal{I} \right\}$ is independent of $\left\{ f_t \right\}_{t=1}^T$ and $%
\left\{ x_{i,t}: (i,t)\in\mathcal{I} \right\}$.
\end{assum}

The moment and convergence conditions in Assumptions \ref{ass:factors and loadings},
\ref{ass:moments of idiosyncratic errors}, \ref{ass:central limit theorems},
\ref{ass:covariates x}, and \ref{ass:additional conditions of idiosyncratic errors} 
are used to establish probability bounds and asymptotic distributions.
They are quite standard in the literature for panel models
with interactive fixed effects, see, for example, \citet{bai2003inferential}%
, \citet{bai2009panel} and \citet{bai2021matrix}, among others.
The distribution and order conditions in Assumptions \ref{ass:distributions of idiosyncratic errors},
\ref{ass:order conditions},
and \ref{ass:error is independent of factor and x} mainly work for the validity of
our bootstrap procedure.

We make further remarks on some of the assumptions.
Assumption \ref{ass:distributions of idiosyncratic errors}(4) requires the strict
exogeneity of treatment status, and one can find similar conditions in Assumption
5 of \citet{hsiao2012panel} and Assumption 2 of \citet{xu2017generalized}.
Assumption \ref{ass:factors and loadings}(3) states
that the factors in the pre-treatment subsample and post-treatment subsample have the same
asymptotic second sample moment matrices as those in the full sample.
Assumption \ref{ass:factors and loadings}(4) states
that the factor loadings in the control subsample and treated subsample have the same
asymptotic second sample moment matrices as those in the full sample.
When either $T-T_0$ or $N-N_0$ is finite, Assumption \ref{ass:factors and loadings}(3) or 
\ref{ass:factors and loadings}(4) can be replaced with equality as
in the identifying restriction PC1 on Page 19 of \citet{bai2013principal} to fulfil the condition that
either the factors or factor loadings possess the same properties in the full sample as well as 
in the post-treatment periods or for the treated units.

\section{Estimation and Inference in a Model without Covariates}

\label{sec:Estimation and Inference in a Model without Covariates}

To highlight the essence of our approach for estimation and inference of
treatment effects, in this section we focus on a special case of Model
\eqref{eq:model with covariates}, where there is no covariate (\textit{i.e.}, $%
\beta=0$) and the model is reduced to a pure approximate factor model:
\begin{align}
y_{i,t}=c_{i,t} + e_{i,t}, \qquad c_{i,t}=f_t^\mathsf{T} \lambda_i.
\label{eq:model without covariates}
\end{align}
We will extend our suggested approach to the full model \eqref{eq:model with covariates} 
in Section \ref{sec:Estimation and Inference in a Model with
Covariates}.

\subsection{Estimation of Treatment Effects}

Following \citet{bai2021matrix}, we consider the factor-based approach to
estimate $y_{i,t}$ for $(i,t)\in\mathcal{I}_1$. Let $Y$, $F$, and $\Lambda$
follow their definitions in Section \ref{sec:The Model}. Since $y_{i,t}$ is
unobserved for all $(i,t)\in \mathcal{I}_1$, the south-east block of $Y$ is
missing in reality. Formally, we define two sub-matrices of $Y$:
\begin{align}
Y_\mathrm{tall}=%
\begin{bmatrix}
y_{1,1} & y_{2,1} & \cdots & y_{N_0,1} \\
y_{1,2} & y_{2,2} & \cdots & y_{N_0,2} \\
\vdots & \vdots & \ddots & \vdots \\
y_{1,T} & y_{2,T} & \cdots & y_{N_0,T}%
\end{bmatrix}%
, \qquad Y_\mathrm{wide}=%
\begin{bmatrix}
y_{1,1} & y_{2,1} & \cdots & y_{N,1} \\
y_{1,2} & y_{2,2} & \cdots & y_{N,2} \\
\vdots & \vdots & \ddots & \vdots \\
y_{1,T_0} & y_{2,T_0} & \cdots & y_{N,T_0}%
\end{bmatrix}%
.  \label{eq:submatrices}
\end{align}
That is, $\Ytall$ corresponds to the control subsample, and
$\Ywide$ corresponds to the pre-treatment subsample.
Let $\left( {F}_\mathrm{tall}, {\Lambda}_\mathrm{tall} \right)$ and $\left( {%
F}_\mathrm{wide}, {\Lambda}_\mathrm{wide} \right)$ be the factor and loading
matrices associated with $Y_\mathrm{tall}$ and $Y_\mathrm{wide}$,
respectively. It is easy to see that ${F}_\mathrm{tall}=F$, ${\Lambda}_%
\mathrm{wide}=\Lambda$, ${F}_\mathrm{wide}$ is a sub-matrix formed by the
first $T_0$ rows of $F$, and ${\Lambda}_\mathrm{tall}$ is a sub-matrix
formed by the first $N_0$ rows of $\Lambda$.

In lieu of \citet{bai2021matrix}, we can use the algorithm below to estimate
the treatment effects $\Delta_{i,t}$ for all $(i,t)\in\mathcal{I}_1 $.

\begin{algo}
\label{algo:estimation} Treatment effect estimation via factor-based matrix
completion (without covariates).

\begin{enumerate}[labelindent=\parindent, leftmargin=*, nosep]

\item Perform a singular value decomposition of $\dfrac{Y_\mathrm{tall}}{%
\sqrt{T N_0}}$, and let $D_\mathrm{tall}$ be an $r\times r$ diagonal matrix
with the $r$ largest singular values of $\dfrac{Y_\mathrm{tall}}{\sqrt{T N_0}%
}$ on the diagonal in descending order. Then let $P_\mathrm{tall}$ and $Q_%
\mathrm{tall}$ be $T\times r$ and $N_0\times r$ matrices containing the left
and right singular vectors of $\dfrac{Y_\mathrm{tall}}{\sqrt{T N_0}}$
respectively, corresponding to $D_\mathrm{tall}$. Compute
\begin{align}
&\widehat{F}_\mathrm{tall}
=\lp \widehat{f}_{\mathrm{tall},1}, \ldots, \widehat{f}_{\mathrm{tall},T} \rp^\T
=\sqrt{T}P_\mathrm{tall}, \nonumber \\
& \widehat{\Lambda}_\mathrm{tall}
=\lp \widehat{\lambda}_{\mathrm{tall},1}, \ldots, \widehat{\lambda}_{\mathrm{tall},N_0} \rp^\T
=\sqrt{N_0} Q_\mathrm{tall} D_\mathrm{tall}.
\end{align}

\item Perform a singular value decomposition of $\dfrac{Y_\mathrm{wide}}{%
\sqrt{T_0 N}}$, and let $D_\mathrm{wide}$ be an $r\times r$ diagonal matrix
with the $r$ largest singular values of $\dfrac{Y_\mathrm{wide}}{\sqrt{T_0 N}%
}$ on the diagonal in descending order. Then let $P_\mathrm{wide}$ and $Q_%
\mathrm{wide}$ be $T_0\times r$ and $N\times r$ matrices containing the left
and right singular vectors of $\dfrac{Y_\mathrm{wide}}{\sqrt{T_0 N}}$
respectively, corresponding to $D_\mathrm{wide}$. Compute
\begin{align}
&\widehat{F}_\mathrm{wide}
=\lp \widehat{f}_{\mathrm{wide},1}, \ldots, \widehat{f}_{\mathrm{wide},T_0} \rp^\T
=\sqrt{T_0}P_\mathrm{wide}, \nonumber \\
& \widehat{\Lambda}_\mathrm{wide}
=\lp \widehat{\lambda}_{\mathrm{wide},1}, \ldots, \widehat{\lambda}_{\mathrm{wide},N} \rp^\T
=\sqrt{N} Q_\mathrm{wide} D_\mathrm{wide}.
\end{align}

\item Let $\widehat{\Lambda}_{\mathrm{wide},0}$ be the first $N_0$ rows of $%
\widehat{\Lambda}_\mathrm{wide}$. Compute
\begin{align}
\widehat{H}_\mathrm{miss}=\widehat{\Lambda}_\mathrm{tall}^\mathsf{T}
\widehat{\Lambda}_{\mathrm{wide},0} \left( \widehat{\Lambda}_{\mathrm{wide}%
,0}^\mathsf{T} \widehat{\Lambda}_{\mathrm{wide},0} \right)^{-1} ,
\end{align}
and then let $\widehat{C}=\widehat{F}_\mathrm{tall}\widehat{H}_\mathrm{miss}%
\widehat{\Lambda}_\mathrm{wide}^\mathsf{T}$.

\item Let $\widehat{c}_{i,t}$ denote the $(t,i)$-th entry of $\widehat{C}$,
then compute the residuals $\widehat{e}_{i,t}=y_{i,t}-\widehat{c}_{i,t}$ for
$(i,t)\in \mathcal{I}\backslash\mathcal{I}_1$.

\item For $(i,t)\in\mathcal{I}_1$, the variance of $\widehat{c}_{i,t}$ is
estimated by
\begin{align}
\widehat{\mathbb{V}}_{i,t}&=\frac{1}{T_0} \widehat{f}_{\mathrm{tall},t}^%
\mathsf{T} \left( \frac{\widehat{F}_\mathrm{tall}^\mathsf{T} \widehat{F}_%
\mathrm{tall}}{T} \right)^{-1} \widehat{\Phi}_i \left( \frac{\widehat{F}_%
\mathrm{tall}^\mathsf{T} \widehat{F}_\mathrm{tall}}{T} \right)^{-1} \widehat{%
f}_{\mathrm{tall},t} \nonumber \\
&\phantom{=\:\:} +\frac{1}{N_0} \widehat{\lambda}_{\mathrm{wide},i}^\mathsf{T%
} \left( \frac{\widehat{\Lambda}_\mathrm{wide}^\mathsf{T} \widehat{\Lambda}_%
\mathrm{wide}}{N} \right)^{-1} \widehat{\Gamma}_t \left( \frac{\widehat{%
\Lambda}_\mathrm{wide}^\mathsf{T} \widehat{\Lambda}_\mathrm{wide}}{N}
\right)^{-1} \widehat{\lambda}_{\mathrm{wide},i},
\end{align}
where
\begin{align}
\widehat{\Gamma}_t &=\frac{1}{N_0}\sum_{j=1}^{N_0} \widehat{e}_{j,t}^2
\widehat{\lambda}_{\mathrm{wide},j} \widehat{\lambda}_{\mathrm{wide},j}^%
\mathsf{T}, \qquad L_{k,i}=\frac{1}{T_0}\sum_{s=k+1}^{T_0} \widehat{f}_{%
\mathrm{tall},s} \widehat{e}_{i,s} \widehat{e}_{i,s-k} \widehat{f}_{\mathrm{%
tall},s-k}^\mathsf{T}, \nonumber \\
\widehat{\Phi}_i&= L_{0,i}+\sum_{k=1}^K \left( 1-\frac{k}{K+1} \right)
\left( L_{k,i}+L_{k,i}^\mathsf{T} \right),
\end{align}
with $K\to\infty$ and $\dfrac{K}{T_0^{1/4}}\to 0$ as $T_0\to\infty$.

\item For every $N_0< i \le N$, compute $\displaystyle \widehat{\sigma}^2_i=%
\frac{1}{T_0} \sum_{s=1}^{T_0} \widehat{e}_{i,s}^2$.

\item For $(i,t)\in\mathcal{I}_1$, the estimated treatment effect is $%
\widehat{\Delta}_{i,t}=\mathscr{Y}_{i,t}-\widehat{c}_{i,t}$, and the
standard error of $\widehat{\Delta}_{i,t}$ is $\sqrt{\widehat{\mathbb{V}}%
_{i,t}+\widehat{\sigma}^2_i}$.
\end{enumerate}
\end{algo}

\begin{remark}
\label{remark:var estimate of chat} Since $\widehat{\mathbb{V}}_{i,t}=O_{%
\mathbb{P}}\left( \dfrac{1}{T_{0}}+\dfrac{1}{N_{0}}\right) $ for every $%
(i,t)\in \mathcal{I}_{1}$, it is asymptotically negligible in the standard
error of $\widehat{\Delta }_{i,t}$. Here we include $\widehat{\mathbb{V}}%
_{i,t}$ in the standard error of $\widehat{\Delta }_{i,t}$ to improve finite
sample performance.
\end{remark}

\begin{remark}
In the above algorithm, we discuss the estimation for individual treatment
effects for each treated unit. If the interest is the average treatment
effects (ATE) across treated units or across post-treatment periods, not individual treatment effects, see
 \citet{hsiao2022multiple} for discussion of the aggregation of multiple treatment effects across treated units,
 and see \citet{fujiki2015disentangling}
 and \citet{li2017estimation} for discussion of the average treatment effects over post-treatment periods.
\end{remark}

\subsection{Construction of Confidence Intervals}
When neither $N_1$ nor $T_1$ is large, the average treatment effects either across treated units or
 post-treatment periods may not be a good measure for inferential purpose since it is hard to establish
 the asymptotic properties of  such effects.
 In order to provide inferential procedure for individual treatment effects especially when
   neither $N_1$ nor $T_1$ is large, motivated by
 \citet{gonccalves2017bootstrap}, we adapt bootstrap method to
construct the confidence intervals of $\Delta_{i,t}$ for $(i,t)\in\mathcal{I}%
_1$, without a specific distributional assumption on the error term $e_{i,t}$ or particular quantity on $N_1$ or $T_1$ .

\begin{algo}
\label{algo:confidence interval} Confidence intervals of the treatment
effects via bootstrap (without covariates).

\begin{enumerate}[labelindent=\parindent, leftmargin=*, nosep]

\item Apply Algorithm \ref{algo:estimation} to the sample and obtain $%
\left\{ \widehat{c}_{i,t}: (i,t)\in\mathcal{I} \right\}$, $\left\{ \widehat{e%
}_{i,t}: (i,t)\in\mathcal{I} \right\}$, $\left\{ \widehat{\mathbb{V}}_{i,t}:
(i,t)\in\mathcal{I}_1 \right\}$, and $\left\{ \widehat{\sigma}^2_i: i=N_0+1,
\ldots, N \right\}$.

\item For $b=1,2,\ldots, B$:

\begin{enumerate}[nosep]

\item For $(i,t)\in\mathcal{I}\backslash\mathcal{I}_1$, let $%
e_{i,t}^*=u_{i,t}\widehat{e}_{i,t}$, where $\left\{ u_{i,t} \right\}$ are
i.i.d.\ or block i.i.d.\ from a standard normal distribution $\mathbb{N}
(0,1)$ and are independent of the raw sample.

\item For $(i,t)\in\mathcal{I}_1$, let $e_{i,t}^*$ be independently drawn
from a discrete uniform distribution on the set $\left\{ \widehat{e}_{i,1}-%
\overline{\widehat{e}}_i ,\widehat{e}_{i,2}-\overline{\widehat{e}}_i,
\ldots, \widehat{e}_{i,T_0}-\overline{\widehat{e}}_i \right\}$, where $%
\displaystyle\overline{\widehat{e}}_i=\frac{1}{T_0}\sum_{s=1}^{T_0}\widehat{e%
}_{i,s}$.

\item For $(i,t)\in\mathcal{I}$, let $y_{i,t}^*=\widehat{c}_{i,t}+e^*_{i,t}$%
. Use $\left\{ y_{i,t}^* \right\}$ to construct $Y_\mathrm{tall}^*$ and $Y_%
\mathrm{wide}^*$ in the same fashion as Equation \eqref{eq:submatrices}.

\item Apply Algorithm \ref{algo:estimation} to the bootstrapped sample $Y_%
\mathrm{tall}^*$ and $Y_\mathrm{wide}^*$, and obtain $\left\{ \widehat{c}%
_{i,t}^*: (i,t)\in\mathcal{I}_1 \right\}$, $\left\{ \widehat{\mathbb{V}}%
_{i,t}^*: (i,t)\in\mathcal{I}_1 \right\}$, and $\left\{ \left(\widehat{\sigma%
}^*_i\right)^2: i=N_0+1,\ldots, N \right\}$.

\item For $(i,t)\in\mathcal{I}_1$, compute
\begin{align}
s_{i,t}^*=\frac{\widehat{c}_{i,t}^*-y_{i,t}^*} {\sqrt{\widehat{\mathbb{V}}%
_{i,t}^*+\left(\widehat{\sigma}^*_i\right)^2}}.
\end{align}
\end{enumerate}

\item In the previous step, we generate $B$ statistics denoted by $%
s_{i,t}^*(1), s_{i,t}^*(2),\ldots, s_{i,t}^*(B)$ for every $(i,t)\in\mathcal{%
I}_1$. Let $q_{1-\alpha,i,t}$ be the $(1-\alpha)$ empirical quantile of $%
\left\{ s_{i,t}^*(1), s_{i,t}^*(2), \ldots, s_{i,t}^*(B)\right\} $, and let $%
p_{1-\alpha,i,t}$ be the $(1-\alpha)$ empirical quantile of $\left\{
\left\vert s_{i,t}^*(1)\right\vert, \left\vert s_{i,t}^*(2)\right\vert,
\ldots, \left\vert s_{i,t}^*(B) \right\vert \right\} $.

\item For $(i,t)\in\mathcal{I}_1$, the equal tailed $(1-\alpha)$ confidence
interval of $\Delta_{i,t}$ is
\begin{align}
\mathrm{EQ}_{1-\alpha,i,t}=\left[ \widehat{\Delta}_{i,t}+ q_{\alpha/2, i,t}
\sqrt{\widehat{\mathbb{V}}_{i,t}+\widehat{\sigma}^2_i}, \; \widehat{\Delta}%
_{i,t}+ q_{1-(\alpha/2),i,t} \sqrt{\widehat{\mathbb{V}}_{i,t}+\widehat{\sigma%
}^2_i} \right],  \label{eq:CI EQ}
\end{align}
and the symmetric $(1-\alpha)$ confidence interval of $\Delta_{i,t}$ is
\begin{align}
\mathrm{SY}_{1-\alpha,i,t}=\left[ \widehat{\Delta}_{i,t}- p_{1-\alpha,i,t}
\sqrt{\widehat{\mathbb{V}}_{i,t}+\widehat{\sigma}^2_i}, \; \widehat{\Delta}%
_{i,t}+ p_{1-\alpha, i,t} \sqrt{\widehat{\mathbb{V}}_{i,t}+\widehat{\sigma}%
^2_i} \right].  \label{eq:CI SY}
\end{align}
\end{enumerate}
\end{algo}

\begin{remark}
\label{remark:bootstrap for chat} As is shown in the proof of Theorem \ref%
{thry:coverage}, the distribution of $\left( \widehat{\Delta}%
_{i,t}-\Delta_{i,t} \right)$ is dominated by $e_{i,t}$ and the effect of $%
\left( \widehat{c}_{i,t}-c_{i,t} \right)$ is asymptotically negligible.
However, the bootstrap procedure above takes $\left( \widehat{c}%
_{i,t}-c_{i,t} \right)$ into consideration in order to improve the finite
sample performances.
\end{remark}

\begin{remark}
\label{remark:serial correlation} When the error terms $\left\{
e_{i,t}\right\} $ are suspected of having serial correlation, a block wild
bootstrap can be used to address this issue (\textit{e.g.},
\citealp{gonccalves2017bootstrap}). That is, a for block width $b\in
\mathbb{Z}_{+}$, we let $u_{i,(j-1)b+s}=\overline{u}_{i,j}$ for every $j\in
\mathbb{Z}_{+}$ and $s\in \left\{ 1,2,\cdots ,b\right\} $ such that $%
(j-1)b+s\leq T_{0}$ in Step 2(1).
\end{remark}

Under Assumptions \ref{ass:factors and loadings}--\ref{ass:order conditions}
in this paper, the asymptotic properties of the
bootstrapped confidence intervals \eqref{eq:CI EQ} and \eqref{eq:CI SY} are
summarized in the following theorem.

\begin{thry}
\label{thry:coverage} If Assumptions \ref{ass:factors and loadings}--\ref%
{ass:order conditions} hold, then
\begin{align}
\lim_{N_0,T_0\to\infty} \mathbb{P} \left( \Delta_{i,t}\in \mathrm{EQ}%
_{1-\alpha,i,t} \right) =1-\alpha\qquad \text{and}\qquad
\lim_{N_0,T_0\to\infty} \mathbb{P} \left( \Delta_{i,t}\in \mathrm{SY}%
_{1-\alpha,i,t} \right) =1-\alpha
\end{align}
for every $(i,t)\in\mathcal{I}_1$.
\end{thry}

From Theorem \ref{thry:coverage}, we can conclude that the bootstrap
confidence intervals will provide a correct coverage for the estimated treatment
effects in the post-treatment periods, and thus one could conduct
statistical inference about the treatment effects based on the bootstrap
confidence intervals. It also worths noting that even if our bootstrap
procedure follows the idea of \citet{gonccalves2017bootstrap}, the proof of bootstrap validity becomes
quite complicated due to the differences in purposes and model specifications.
Since \citet{gonccalves2017bootstrap} intend to
make inference on factor-augmented regression models \citep{bai2006confidence} where
the factors serves as intermediate variables,
they only require valid bootstrap approximations of factors.
In this paper, we intend to make inference on the factor model (with missing values) itself,
and the estimators involve products of estimated factors, their associated loadings,
and a rotation matrix.
This implies that we need valid bootstrap approximations of factors, loadings,
and the rotation matrix, which leads to laborious works.

\section{Estimation and Inference in a Model with Covariates}

\label{sec:Estimation and Inference in a Model with Covariates}

In the above section, we discussed the estimation of the treatment effects
and associated bootstrap confidence intervals for a pure factor model.
Besides the unobserved factors, in practice, the outcome variable could also
be affected by some exogenous regressors. To accommodate such a scenario, we
extend the factor-based approach to a model with covariates, \textit{i.e.},
Model \eqref{eq:model with covariates} in Section \ref{sec:The Model}. 

\subsection{Estimation of Treatment Effects}

For a factor model with covariates, \textit{i.e.}, a panel data model with
interactive fixed effect, our estimation strategies involve an interactive
fixed effect estimation (IFEE), which is described and studied in \cite%
{bai2009panel}. If $\left\{ y_{i,t}: (i,t)\in\mathcal{I} \right\}$ were
fully observed, we could directly use IFEE to estimate Model \eqref{eq:model with covariates}. 
The general framework of IFEE is summarised in the
algorithm to follow.

\begin{algo}
\label{algo:IFEE} Interactive fixed effect estimation.

\begin{enumerate}[labelindent=\parindent, leftmargin=*, nosep]

\item Input arguments. $\mathcal{Y}=\left( \mathcal{Y}_1, \ldots, \mathcal{Y}%
_\mathcal{N} \right)$, a $\mathcal{T}\times \mathcal{N}$ matrix. $\mathcal{X}%
=\left( \mathcal{X}_1, \ldots, \mathcal{X}_\mathcal{N} \right)$, a $\mathcal{%
T}\times (p\mathcal{N})$ matrix, where $\mathcal{X}_i$ is a $\mathcal{T}%
\times p$ matrix for every $i$.

\item Compute the starting value
\begin{align}
\beta^{(0)}=\left( \sum_{i=1}^\mathcal{N} \mathcal{X}_i^\mathsf{T} \mathcal{X%
}_i \right)^{-1} \sum_{i=1}^\mathcal{N} \mathcal{X}_i^\mathsf{T} \mathcal{Y}%
_i.
\end{align}

\item Perform the following iteration until $\left\Vert \beta^{(M)}-
\beta^{(M-1)} \right\Vert_2< \varepsilon$ for some $M\in\mathbb{Z}_+$, where
$\varepsilon$ is a sufficiently small positive number.

\begin{enumerate}[nosep]

\item Let
\begin{align}
\mathcal{R}^{(k)}= \dfrac{\mathcal{Y}-\mathcal{X}\left( I_\mathcal{N}
\otimes \beta^{(k-1)} \right)}{\sqrt{\mathcal{N}\mathcal{T}}}.
\end{align}

\item Perform a singular value decomposition of the matrix $\mathcal{R}%
^{(k)} $, and let $D^{(k)}$ be an $r\times r$ diagonal matrix with the $r$
largest singular values of $\mathcal{R}^{(k)}$ on the diagonal in descending
order. Then let $U^{(k)}$ and $V^{(k)}$ be $\mathcal{T}\times r$ and $%
\mathcal{N}\times r$ matrices containing the left and right singular vectors
of $\mathcal{R}^{(k)}$ corresponding to $D^{(k)}$.

\item Let $F^{(k)}=\sqrt{\mathcal{T}}U^{(k)}$ and $H^{(k)}= I_\mathcal{T}-%
\dfrac{F^{(k)} \left( F^{(k)} \right)^\mathsf{T}}{\mathcal{T}} $.

\item Let
\begin{align}
\beta^{(k)}=\left[ \sum_{i=1}^{\mathcal{N}} \mathcal{X}_i^\mathsf{T} H^{(k)}
\mathcal{X}_i \right]^{-1} \left[ \sum_{i=1}^{\mathcal{N}} \mathcal{X}_i^%
\mathsf{T} H^{(k)} \mathcal{Y}_i \right].
\end{align}
\end{enumerate}

\item Output arguments. $\beta^{(M)}$, $F^{(M)}$, and $\Lambda^{(M)}= \sqrt{%
\mathcal{N}}V^{(M)}D^{(M)}$.
\end{enumerate}
\end{algo}

Now we turn to the estimation of treatment effect when $\left\{
y_{i,t}:(i,t)\in\mathcal{I} \right\}$ are not observable. Let $Y$, $Y_%
\mathrm{tall}$, $Y_\mathrm{wide}$, $X_i$, $F$, ${F}_\mathrm{tall}$, ${F}_%
\mathrm{wide}$, $\Lambda$, ${\Lambda}_\mathrm{tall}$, and ${\Lambda}_\mathrm{%
wide}$ follow their definitions in Sections \ref{sec:The Model} and \ref%
{sec:Estimation and Inference in a Model without Covariates}. Let $%
X=\left( X_1, \ldots, X_N \right)$, $X_\mathrm{tall}= \left( X_1, \ldots,
X_{N_0} \right)$, and $X_\mathrm{wide}$ to be a sub-matrix formed by the
first $T_0$ rows of $X$.

Then we can use the algorithm below to estimate the treatment effects $%
\Delta_{i,t}$ for all $(i,t)\in\mathcal{I}_1$.

\begin{algo}
\label{algo:cov estimation} Treatment effect estimation via factor-based
matrix completion (with covariates).

\begin{enumerate}[labelindent=\parindent, leftmargin=*, nosep]

\item Apply Algorithm \ref{algo:IFEE} to $\left( Y_\mathrm{tall}, X_\mathrm{%
tall} \right)$, and obtain $\widehat{\beta}_\mathrm{tall}$, $\widehat{F}_\mathrm{tall}=
\lp \widehat{f}_{\mathrm{tall},1}, \ldots, \widehat{f}_{\mathrm{tall},T} \rp^\T$ and
$\widehat{\Lambda}_\mathrm{tall}=
\lp \widehat{\lambda}_{\mathrm{tall},1}, \ldots, \widehat{\lambda}_{\mathrm{tall},N_0} \rp^\T$.

\item Apply Algorithm \ref{algo:IFEE} to $\left( Y_\mathrm{wide}, X_\mathrm{%
wide} \right)$, and obtain $\widehat{F}_\mathrm{wide}
=\lp \widehat{f}_{\mathrm{wide},1}, \ldots, \widehat{f}_{\mathrm{wide},T_0} \rp^\T$
and $\widehat{\Lambda}_\mathrm{wide}
=\lp \widehat{\lambda}_{\mathrm{wide},1}, \ldots, \widehat{\lambda}_{\mathrm{wide},N} \rp^\T$.

\item Let $\widehat{\Lambda}_{\mathrm{wide},0}$ be the first $N_0$ rows of $%
\widehat{\Lambda}_\mathrm{wide}$. Compute
\begin{align}
\widehat{H}_\mathrm{miss}=\widehat{\Lambda}_\mathrm{tall}^\mathsf{T}
\widehat{\Lambda}_{\mathrm{wide},0} \left( \widehat{\Lambda}_{\mathrm{wide}%
,0}^\mathsf{T} \widehat{\Lambda}_{\mathrm{wide},0} \right)^{-1},
\end{align}
and then let $\widehat{C}=\widehat{F}_\mathrm{tall}\widehat{H}_\mathrm{miss}%
\widehat{\Lambda}_\mathrm{wide}^\mathsf{T}$.

\item Let $\widehat{c}_{i,t}$ denote the $(t,i)$-th entry of $\widehat{C}$,
then compute the residuals $\widehat{e}_{i,t}=y_{i,t}-x_{i,t}^\mathsf{T}%
\widehat{\beta}_\mathrm{tall} -\widehat{c}_{i,t}$ for $(i,t)\in \mathcal{I}%
\backslash\mathcal{I}_1$.

\item For $(i,t)\in\mathcal{I}_1$, the variance of $\widehat{c}_{i,t}$ is
estimated by
\begin{align}
\widehat{\mathbb{V}}_{i,t}&=\frac{1}{T_0} \widehat{f}_{\mathrm{tall},t}^%
\mathsf{T} \left( \frac{\widehat{F}_\mathrm{tall}^\mathsf{T} \widehat{F}_%
\mathrm{tall}}{T} \right)^{-1} \widehat{\Phi}_i \left( \frac{\widehat{F}_%
\mathrm{tall}^\mathsf{T} \widehat{F}_\mathrm{tall}}{T} \right)^{-1} \widehat{%
f}_{\mathrm{tall},t} \nonumber \\
&\phantom{=\:\:} +\frac{1}{N_0} \widehat{\lambda}_{\mathrm{wide},i}^\mathsf{T%
} \left( \frac{\widehat{\Lambda}_\mathrm{wide}^\mathsf{T} \widehat{\Lambda}_%
\mathrm{wide}}{N} \right)^{-1} \widehat{\Gamma}_t \left( \frac{\widehat{%
\Lambda}_\mathrm{wide}^\mathsf{T} \widehat{\Lambda}_\mathrm{wide}}{N}
\right)^{-1} \widehat{\lambda}_{\mathrm{wide},i},
\end{align}
where
\begin{align}
\widehat{\Gamma}_t &=\frac{1}{N_0}\sum_{j=1}^{N_0} \widehat{e}_{j,t}^2
\widehat{\lambda}_{\mathrm{wide},j} \widehat{\lambda}_{\mathrm{wide},j}^%
\mathsf{T}, \qquad L_{k,i}=\frac{1}{T_0}\sum_{s=k+1}^{T_0} \widehat{f}_{%
\mathrm{tall},s} \widehat{e}_{i,s} \widehat{e}_{i,s-k} \widehat{f}_{\mathrm{%
tall},s-k}^\mathsf{T}, \nonumber \\
\widehat{\Phi}_i&= L_{0,i}+\sum_{k=1}^K \left( 1-\frac{k}{K+1} \right)
\left( L_{k,i}+L_{k,i}^\mathsf{T} \right),
\end{align}
with $K\to\infty$ and $\dfrac{K}{T_0^{1/4}}\to 0$ as $T_0\to\infty$.

\item For every $N_0< i \le N$, compute $\displaystyle \widehat{\sigma}^2_i=%
\frac{1}{T_0} \sum_{s=1}^{T_0} \widehat{e}_{i,s}^2$.

\item For $(i,t)\in\mathcal{I}_1$, the estimated treatment effect is $%
\widehat{\Delta}_{i,t}=\mathscr{Y}_{i,t}-x_{i,t}^\mathsf{T} \widehat{\beta}_%
\mathrm{tall} -\widehat{c}_{i,t}$, and the standard error of $\widehat{\Delta%
}_{i,t}$ is $\sqrt{\widehat{\mathbb{V}}_{i,t}+\widehat{\sigma}^2_i}$.
\end{enumerate}
\end{algo}

\begin{remark}
\label{remark:standard error in cov model} Because
\begin{equation}
\widehat{\Delta }_{i,t}-\Delta _{i,t}=x_{i,t}^{\mathsf{T}}\left( \beta -%
\widehat{\beta }_{\mathrm{tall}}\right) +\left( c_{i,t}-\widehat{c}%
_{i,t}\right) +e_{i,t} \label{eq:decomposition of Deltahat in cov model}
\end{equation}%
for every $(i,t)\in \mathcal{I}_{1}$, the variance of $\widehat{\Delta }%
_{i,t}$ is comprised of $\mathrm{Var}\left( e_{i,t}\right) $, $\mathrm{Var}%
\left( \widehat{c}_{i,t}\right) $ and $\mathrm{Var}\left( \widehat{\beta }_{%
\mathrm{tall}}\right) $. In spirit of Remark \ref{remark:var estimate of
chat}, we should include an estimate of $\mathrm{Var}\left( \widehat{\beta }%
_{\mathrm{tall}}\right) $ in the standard error of $\widehat{\Delta }_{i,t}$
to improve the finite sample performances. However, by Theorem 3 of %
\citet{bai2009panel}, $\mathrm{Var}\left( \widehat{\beta }_{\mathrm{tall}%
}\right) =O\left( \dfrac{1}{N_{0}T}\right) $, which is of higher order than $%
\mathrm{Var}\left( \widehat{c}_{i,t}\right) =O\left( \dfrac{1}{N_{0}}+\dfrac{%
1}{T_{0}}\right) $ for $(i,t)\in \mathcal{I}_{1}$. Furthermore, a consistent
estimation of $\mathrm{Var}\left( \widehat{\beta }_{\mathrm{tall}}\right) $
is quite complicated. As a result of the cost-benefit trade-off, we do not
include an estimate of $\mathrm{Var}\left( \widehat{\beta }_{\mathrm{tall}%
}\right) $ in the standard error of $\widehat{\Delta }_{i,t}$. 
\end{remark}

\subsection{Construction of Confidence Intervals}

For the estimated $\widehat{\Delta}_{i,t}$ for $(i,t)\in\mathcal{I}_1$, we
shall again use the bootstrap procedure discussed above to construct their
confidence intervals.

\begin{algo}
\label{algo:cov confidence interval} Confidence intervals of the treatment
effects via bootstrap (with covariates).

\begin{enumerate}[labelindent=\parindent, leftmargin=*, nosep]

\item Apply Algorithm \ref{algo:cov estimation} to the sample and obtain $%
\left\{ \widehat{c}_{i,t}: (i,t)\in\mathcal{I} \right\}$, $\left\{ \widehat{e%
}_{i,t}: (i,t)\in\mathcal{I} \right\}$, $\left\{ \widehat{\mathbb{V}}_{i,t}:
(i,t)\in\mathcal{I}_1 \right\}$, and $\left\{ \widehat{\sigma}^2_i: i=N_0+1,
\ldots, N \right\}$.

\item For $b=1,2,\ldots, B$:

\begin{enumerate}[nosep]

\item For $(i,t)\in\mathcal{I}\backslash\mathcal{I}_1$, let $%
e_{i,t}^*=u_{i,t}\widehat{e}_{i,t}$, where $\left\{ u_{i,t} \right\}$ are
i.i.d.\ or block i.i.d.\ from a standard normal distribution $\mathbb{N}
(0,1)$ and are independent of the raw sample.

\item For $(i,t)\in\mathcal{I}_1$, let $e_{i,t}^*$ be independently drawn
from a discrete uniform distribution on the set $\left\{ \widehat{e}_{i,1}-%
\overline{\widehat{e}}_i ,\widehat{e}_{i,2}-\overline{\widehat{e}}_i,
\ldots, \widehat{e}_{i,T_0}-\overline{\widehat{e}}_i \right\}$, where $%
\displaystyle\overline{\widehat{e}}_i=\frac{1}{T_0}\sum_{s=1}^{T_0}\widehat{e%
}_{i,s}$.

\item For $(i,t)\in\mathcal{I}$, let $r_{i,t}^*=\widehat{c}_{i,t}+e^*_{i,t}$%
. Use $\left\{ r_{i,t}^* \right\}$ to construct $R_\mathrm{tall}^*$ and $R_%
\mathrm{wide}^*$ in the same fashion as Equation \eqref{eq:submatrices}.

\item Apply Algorithm \ref{algo:estimation} to the bootstrapped sample $R_%
\mathrm{tall}^*$ and $R_\mathrm{wide}^*$, and obtain $\left\{ \widehat{c}%
_{i,t}^*: (i,t)\in\mathcal{I}_1 \right\}$, $\left\{ \widehat{\mathbb{V}}%
_{i,t}^*: (i,t)\in\mathcal{I}_1 \right\}$, and $\left\{ \left(\widehat{\sigma%
}^*_i\right)^2: i=N_0+1,\ldots, N \right\}$.

\item For $(i,t)\in\mathcal{I}_1$, compute
\begin{align}
s_{i,t}^*=\frac{\widehat{c}_{i,t}^*-r_{i,t}^*} {\sqrt{\widehat{\mathbb{V}}%
_{i,t}^*+\left(\widehat{\sigma}^*_i\right)^2}}.
\end{align}
\end{enumerate}

\item In the previous step, we generate $B$ statistics denoted by $%
s_{i,t}^*(1), s_{i,t}^*(2),\ldots, s_{i,t}^*(B)$ for every $(i,t)\in\mathcal{%
I}_1$. Let $q_{1-\alpha,i,t}$ be the $(1-\alpha)$ empirical quantile of $%
\left\{ s_{i,t}^*(1), s_{i,t}^*(2), \ldots, s_{i,t}^*(B)\right\} $, and let $%
p_{1-\alpha,i,t}$ be the $(1-\alpha)$ empirical quantile of $\left\{
\left\vert s_{i,t}^*(1)\right\vert, \left\vert s_{i,t}^*(2)\right\vert,
\ldots, \left\vert s_{i,t}^*(B) \right\vert \right\} $.

\item For $(i,t)\in\mathcal{I}_1$, the equal tailed $(1-\alpha)$ confidence
interval of $\Delta_{i,t}$ is
\begin{align}
\mathrm{EQ}_{1-\alpha,i,t}=\left[ \widehat{\Delta}_{i,t}+ q_{\alpha/2, i,t}
\sqrt{\widehat{\mathbb{V}}_{i,t}+\widehat{\sigma}^2_i}, \; \widehat{\Delta}%
_{i,t}+ q_{1-(\alpha/2),i,t} \sqrt{\widehat{\mathbb{V}}_{i,t}+\widehat{\sigma%
}^2_i} \right],  \label{eq:cov CI EQ}
\end{align}
and the symmetric $(1-\alpha)$ confidence interval of $\Delta_{i,t}$ is
\begin{align}
\mathrm{SY}_{1-\alpha,i,t}=\left[ \widehat{\Delta}_{i,t}- p_{1-\alpha,i,t}
\sqrt{\widehat{\mathbb{V}}_{i,t}+\widehat{\sigma}^2_i}, \; \widehat{\Delta}%
_{i,t}+ p_{1-\alpha, i,t} \sqrt{\widehat{\mathbb{V}}_{i,t}+\widehat{\sigma}%
^2_i} \right].  \label{eq:cov CI SY}
\end{align}
\end{enumerate}
\end{algo}

\begin{remark}
\label{remark:pure factor bootstrap in sec4} By Equation 
\eqref{eq:decomposition of Deltahat in cov model} and in spirit of Remark \ref{remark:bootstrap for chat},
a bootstrap procedure should take the
distributions of all the 3 terms $x_{i,t}^\mathsf{T} \left( \beta-\widehat{%
\beta}_\mathrm{tall} \right)$, $\left( c_{i,t}-\widehat{c}_{i,t} \right)$
and $e_{i,t}$ into consideration to ensure the finite sample performances. But in
Algorithm \ref{algo:cov confidence interval}, resampled observations are
generated by a pure factor model and only the interactive fixed effects $%
\left\{ \widehat{c}_{i,t}^* \right\}$ are estimated in every bootstrapped
sample. Thus the bootstrap procedure only approximates the distribution of $%
\left( c_{i,t}-\widehat{c}_{i,t} \right)+e_{i,t}$, ignoring the effect of $%
x_{i,t}^\mathsf{T} \left( \beta-\widehat{\beta}_\mathrm{tall} \right)$. The
underlying rationale is also a cost-benefit trade-off mentioned in Remark %
\ref{remark:standard error in cov model}. On the one hand, by %
\citet{bai2009panel} and \citet{bai2021matrix}, we have $\left( c_{i,t}-%
\widehat{c}_{i,t} \right)=O_\mathbb{P}\left( \dfrac{1}{\sqrt{N_0\wedge T_0}}%
\right)$ and $x_{i,t}^\mathsf{T} \left( \beta-\widehat{\beta}_\mathrm{tall}
\right)=O_\mathbb{P} \left( \dfrac{1}{\sqrt{N_0 T}} \right)$ as $N_0,
T_0\to\infty$ for every $(i,t)\in\mathcal{I}_1$. On the other hand, the
computation of interactive fixed effect estimation is far more intensive
than that of estimating a pure factor model. These two facts motivate us to
ignore the effect of $x_{i,t}^\mathsf{T} \left( \beta-\widehat{\beta}_%
\mathrm{tall} \right)$ in the bootstrap procedure. Furthermore, evidence
from Section \ref{sec:Simulation Studies} shows the current bootstrap
procedure has already yielded satisfactory finite sample performances.
\end{remark}

\begin{remark}
If the error terms $\left\{ e_{i,t} \right\}$ are suspected of having serial
correlation, we suggest to use block wild bootstrap in Step 2(1). See Remark %
\ref{remark:serial correlation} for details.
\end{remark}

As above, we can establish the validity of the proposed confidence intervals
\eqref{eq:cov CI EQ} and \eqref{eq:cov CI SY} in the sense that they have
asymptotically correct coverage probabilities as $N_0,T_0\to \infty$.

\begin{thry}
\label{thry:cov coverage} If Assumptions \ref{ass:factors and loadings}--\ref%
{ass:error is independent of factor and x} hold, then
\begin{align}
\lim_{N_0,T_0\to\infty} \mathbb{P} \left( \Delta_{i,t}\in \mathrm{EQ}%
_{1-\alpha,i,t} \right) =1-\alpha\qquad \text{and}\qquad
\lim_{N_0,T_0\to\infty} \mathbb{P} \left( \Delta_{i,t}\in \mathrm{SY}%
_{1-\alpha,i,t} \right) =1-\alpha
\end{align}
for every $(i,t)\in\mathcal{I}_1$.
\end{thry}

Theorem \ref{thry:cov coverage} suggests that we can apply the proposed bootstrap procedure to conduct 
statistical inference for estimated treated effects from a panel with interactive fixed effects model.

\section{Simulation Studies}

\label{sec:Simulation Studies}

In this section, we conduct several Monte Carlo experiments to investigate
the finite sample properties of the proposed confidence intervals.
\footnote{MATLAB codes for simulation are available from the authors upon request.}

In the data generating processes below, we assume that the common factors $%
\left\{ f_{t}\right\} $ are i.i.d.\ as $\mathbb{N}\left( 0,I_{3}\right) $,
and the factor loadings $\left\{ \lambda _{i}\right\} $ are also i.i.d.\ as $%
\mathbb{N}\left( 0,I_{3}\right) $. For the model with covariates, we assume
that the covariates $\left\{ x_{i,t}\right\} $ are i.i.d.\ as $\mathbb{N}%
\left( 0,AA^{\mathsf{T}}\right) $, where each entry of the $2\times 2$
matrix $A$ is drawn from $\mathbb{N}(0,1)$. Let $\beta \sim \mathbb{N}\left(
0,I_{2}\right) $. We consider the following data generating processes.
\footnote{We also consider AR(1) factors, exponentially distributed errors, other variance
structures, and dependence between factors and covariates. These results are available
from the authors upon request.}

\begin{itemize}[labelindent=\parindent, leftmargin=*, nosep]

\item DGP1: Model without covariates, $y_{i,t}=f_t^\mathsf{T} \lambda_i+
e_{i,t}$.

\item DGP2: Model with covariates, $y_{i,t}=x_{i,t}^\mathsf{T} \beta +f_t^%
\mathsf{T} \lambda_i+ e_{i,t}$.
\end{itemize}

The error term is defined as
\begin{align}
e_{i,t}=v_{i,t}\sqrt{\dfrac{\sigma_{i}^2}{1-\rho_i^2}},
\end{align}
where $v_{i,t}=\rho_i v_{i,t-1}+\varepsilon_{i,t}$ and $\left\{
\varepsilon_{i,t} \right\}$ are i.i.d.\ drawn from $\mathbb{N} (0,1)$. We
specify two variance structures of $\left\{ e_{i,t} \right\}$.

\begin{itemize}[labelindent=\parindent, leftmargin=*, nosep]

\item Case 1: $\rho_i=0$ and $\sigma_{i}^2=1$.

\item Case 2: $\left\{ \rho_i \right\}$ are i.i.d.\ drawn from $\mathrm{Unif}%
\left( [-0.8,-0.2]\cup[0.2, 0.8] \right)$, and $\left\{ \sigma_{i}^2
\right\} $ are i.i.d.\ drawn from $\mathrm{logNormal}\left( 0,1 \right)$.
\end{itemize}

Moreover, we consider the following marginal distributions of $\left\{
v_{i,t}\right\}$.

\begin{itemize}[labelindent=\parindent, leftmargin=*, nosep]

\item Margin 1: $v_{i,t}\sim \left[ \chi^2(1)-1 \right]/\sqrt{2}$.

\item Margin 2: $v_{i,t}\sim \left( \mathrm{Unif}[-0.5,0.5] \right)/ \sqrt{12%
}$.
\end{itemize}

For demonstration, we assume there is only one treated unit $i=N$ and $5$ post-treatment periods. The
treatment effects are assumed to be constants equal to 1, \textit{i.e.}, $%
\Delta _{N,t}=1$ for $t=T_{0}+1,\ldots ,T_{0}+5$. The number of control
units $N_{0}\in \{30,50,100\}$ and the number of pre-treatment periods $%
T_{0}\in \{20,40\}$. We construct the $90\%$ and $95\%$, equal-tailed and
symmetric confidence intervals for $\Delta _{N,t}$, $t=T_{0}+1,\ldots
,T_{0}+5$. In Step 2(1) of Algorithms \ref{algo:confidence interval} and \ref%
{algo:cov confidence interval}, we use ordinary wild bootstrap procedure for
Case 1, and block wild bootstrap procedure with block width equal to 4 for
Case 2. The number of factors is either treated as known or estimated using
the method of \citet{bai2002determining}. For computational simplicity, a warp-speed method %
\citep{giacomini2013warp} is applied with 2000 replications for each
scenario. We report the coverage rates (in percent) of confidence intervals
in Tables \ref{tab:Pure factor model with homoscedastic i.i.d. chi-squared
errors}--\ref{tab:Factor model with covariates and heteroscedastic AR1
uniform errors}. In these tables, EQ stands for equal-tailed confidence
intervals and SY stands for symmetric confidence intervals.

Several interesting findings can be observed from the simulation results. On
the first hand, the results in Table \ref{tab:Pure factor model with
homoscedastic i.i.d. chi-squared errors}--\ref{tab:Pure factor model with
heteroscedastic AR1 uniform errors} clearly show that the empirical coverage
ratio for treatment effects from a pure factor model is quite close to the
nominal values (both 90\% and 95\% level) when the post-treatment period is
short, regardless of whether the idiosyncratic errors are heteroscedastic or
serially correlated, or whether the number of unobserved factors is known or
estimated from the data. On the other hand, when exogenous covariates are
included for treatment effects estimation, Tables \ref{tab:Factor model with
covariates and homoscedastic i.i.d. chi-squared errors}--\ref{tab:Factor
model with covariates and heteroscedastic AR1 uniform errors} also show that our
proposed bootstrapped confidence intervals are able to provide accurate
coverage ratios for the estimated treatment effects in a panel model with
exogenous regressors and with heteroscedastic or serially correlated errors. In
general, the simulation results confirm the validity of our proposed bootstrap procedure in
providing accurate and robust confidence intervals for estimated treatment effects
using a panel with interactive fixed effects.

\begin{table}[H] 
	\footnotesize\centering\setstretch{1.5} 
	\caption{Pure factor model with homoscedastic i.i.d.\ chi-squared errors} 
	\label{tab:Pure factor model with homoscedastic i.i.d. chi-squared errors} 
	\begin{tabular}{c|cc|cc|cc|cc|cc|cc} 
		\multicolumn{13}{c}{Known number of factors, $90\%$ CI} \\ 
		\hline\hline 
		\multirow{2}{*}{\diagbox{$t$}{$\lp T_0, N_0 \rp$}} &\multicolumn{2}{c|}{$ (20, 30) $} &\multicolumn{2}{c|}{$ (20, 50) $} &\multicolumn{2}{c|}{$ (20, 100) $} &\multicolumn{2}{c|}{$ (40, 30) $} &\multicolumn{2}{c|}{$ (40, 50) $} &\multicolumn{2}{c}{$ (40, 100) $} \\ 
		& EQ & SY & EQ & SY & EQ & SY & EQ & SY & EQ & SY & EQ & SY \\ 
		\hline 
		$ T_0+ 1 $&$ 90.30 $&$ 90.85 $&$ 91.55 $&$ 90.55 $&$ 90.70 $&$ 90.90 $&$ 93.05 $&$ 91.45 $&$ 91.60 $&$ 92.05 $&$ 91.85 $&$ 91.45 $ \\ 
		$ T_0+ 2 $&$ 91.35 $&$ 91.00 $&$ 90.85 $&$ 90.55 $&$ 92.55 $&$ 91.45 $&$ 91.15 $&$ 90.95 $&$ 92.85 $&$ 92.90 $&$ 91.35 $&$ 89.90 $ \\ 
		$ T_0+ 3 $&$ 90.50 $&$ 90.30 $&$ 92.95 $&$ 91.90 $&$ 93.30 $&$ 92.15 $&$ 91.50 $&$ 91.45 $&$ 92.75 $&$ 91.10 $&$ 91.90 $&$ 91.50 $ \\ 
		$ T_0+ 4 $&$ 93.15 $&$ 92.35 $&$ 92.05 $&$ 91.15 $&$ 91.00 $&$ 91.65 $&$ 91.80 $&$ 91.40 $&$ 92.40 $&$ 92.35 $&$ 91.85 $&$ 91.40 $ \\ 
		$ T_0+ 5 $&$ 91.55 $&$ 91.65 $&$ 89.85 $&$ 90.85 $&$ 90.20 $&$ 91.30 $&$ 92.35 $&$ 92.60 $&$ 92.55 $&$ 90.95 $&$ 92.05 $&$ 90.35 $ \\ 
		\hline\hline 
		\multicolumn{13}{c}{ }\\ 
		\multicolumn{13}{c}{ }\\ 
	\end{tabular} 
	
	\begin{tabular}{c|cc|cc|cc|cc|cc|cc} 
		\multicolumn{13}{c}{Known number of factors, $95\%$ CI} \\ 
		\hline\hline 
		\multirow{2}{*}{\diagbox{$t$}{$\lp T_0, N_0 \rp$}} &\multicolumn{2}{c|}{$ (20, 30) $} &\multicolumn{2}{c|}{$ (20, 50) $} &\multicolumn{2}{c|}{$ (20, 100) $} &\multicolumn{2}{c|}{$ (40, 30) $} &\multicolumn{2}{c|}{$ (40, 50) $} &\multicolumn{2}{c}{$ (40, 100) $} \\ 
		& EQ & SY & EQ & SY & EQ & SY & EQ & SY & EQ & SY & EQ & SY \\ 
		\hline 
		$ T_0+ 1 $&$ 94.85 $&$ 94.60 $&$ 95.25 $&$ 95.00 $&$ 95.60 $&$ 94.70 $&$ 96.40 $&$ 95.60 $&$ 95.75 $&$ 94.55 $&$ 96.35 $&$ 94.95 $ \\ 
		$ T_0+ 2 $&$ 94.90 $&$ 94.55 $&$ 95.25 $&$ 95.00 $&$ 95.70 $&$ 96.05 $&$ 95.35 $&$ 94.40 $&$ 96.20 $&$ 94.70 $&$ 95.60 $&$ 93.70 $ \\ 
		$ T_0+ 3 $&$ 95.00 $&$ 94.45 $&$ 95.80 $&$ 95.50 $&$ 95.85 $&$ 95.40 $&$ 95.35 $&$ 94.40 $&$ 95.90 $&$ 95.15 $&$ 95.95 $&$ 94.45 $ \\ 
		$ T_0+ 4 $&$ 96.25 $&$ 95.45 $&$ 94.85 $&$ 94.70 $&$ 95.15 $&$ 94.25 $&$ 95.20 $&$ 95.40 $&$ 96.65 $&$ 95.00 $&$ 96.05 $&$ 94.95 $ \\ 
		$ T_0+ 5 $&$ 96.10 $&$ 95.70 $&$ 94.65 $&$ 94.30 $&$ 95.00 $&$ 95.20 $&$ 96.65 $&$ 96.20 $&$ 96.70 $&$ 95.05 $&$ 96.35 $&$ 94.65 $ \\ 
		\hline\hline 
		\multicolumn{13}{c}{ }\\ 
		\multicolumn{13}{c}{ }\\ 
	\end{tabular} 
	
	\begin{tabular}{c|cc|cc|cc|cc|cc|cc} 
		\multicolumn{13}{c}{Estimated number of factors, $90\%$ CI} \\ 
		\hline\hline 
		\multirow{2}{*}{\diagbox{$t$}{$\lp T_0, N_0 \rp$}} &\multicolumn{2}{c|}{$ (20, 30) $} &\multicolumn{2}{c|}{$ (20, 50) $} &\multicolumn{2}{c|}{$ (20, 100) $} &\multicolumn{2}{c|}{$ (40, 30) $} &\multicolumn{2}{c|}{$ (40, 50) $} &\multicolumn{2}{c}{$ (40, 100) $} \\ 
		& EQ & SY & EQ & SY & EQ & SY & EQ & SY & EQ & SY & EQ & SY \\ 
		\hline 
		$ T_0+ 1 $&$ 90.30 $&$ 91.05 $&$ 91.35 $&$ 90.65 $&$ 90.65 $&$ 90.85 $&$ 93.30 $&$ 91.75 $&$ 91.65 $&$ 92.05 $&$ 91.85 $&$ 91.45 $ \\ 
		$ T_0+ 2 $&$ 91.35 $&$ 90.95 $&$ 90.65 $&$ 90.80 $&$ 92.35 $&$ 91.50 $&$ 91.10 $&$ 90.85 $&$ 92.90 $&$ 92.85 $&$ 91.35 $&$ 89.90 $ \\ 
		$ T_0+ 3 $&$ 90.60 $&$ 90.50 $&$ 92.35 $&$ 91.80 $&$ 93.30 $&$ 92.15 $&$ 91.35 $&$ 91.35 $&$ 92.70 $&$ 91.25 $&$ 91.90 $&$ 91.50 $ \\ 
		$ T_0+ 4 $&$ 92.55 $&$ 92.70 $&$ 91.80 $&$ 91.05 $&$ 91.00 $&$ 91.75 $&$ 91.70 $&$ 91.20 $&$ 92.45 $&$ 92.35 $&$ 91.85 $&$ 91.40 $ \\ 
		$ T_0+ 5 $&$ 91.70 $&$ 91.65 $&$ 90.65 $&$ 90.60 $&$ 90.20 $&$ 91.20 $&$ 92.35 $&$ 92.80 $&$ 92.50 $&$ 91.05 $&$ 92.05 $&$ 90.30 $ \\ 
		\hline\hline 
		\multicolumn{13}{c}{ }\\ 
		\multicolumn{13}{c}{ }\\ 
	\end{tabular} 
	
	\begin{tabular}{c|cc|cc|cc|cc|cc|cc} 
		\multicolumn{13}{c}{Estimated number of factors, $95\%$ CI} \\ 
		\hline\hline 
		\multirow{2}{*}{\diagbox{$t$}{$\lp T_0, N_0 \rp$}} &\multicolumn{2}{c|}{$ (20, 30) $} &\multicolumn{2}{c|}{$ (20, 50) $} &\multicolumn{2}{c|}{$ (20, 100) $} &\multicolumn{2}{c|}{$ (40, 30) $} &\multicolumn{2}{c|}{$ (40, 50) $} &\multicolumn{2}{c}{$ (40, 100) $} \\ 
		& EQ & SY & EQ & SY & EQ & SY & EQ & SY & EQ & SY & EQ & SY \\ 
		\hline 
		$ T_0+ 1 $&$ 94.85 $&$ 94.80 $&$ 95.35 $&$ 95.00 $&$ 95.55 $&$ 94.70 $&$ 96.50 $&$ 95.55 $&$ 95.65 $&$ 94.55 $&$ 96.35 $&$ 94.95 $ \\ 
		$ T_0+ 2 $&$ 95.05 $&$ 94.70 $&$ 95.40 $&$ 94.95 $&$ 95.65 $&$ 96.00 $&$ 95.35 $&$ 94.50 $&$ 96.15 $&$ 94.70 $&$ 95.60 $&$ 93.70 $ \\ 
		$ T_0+ 3 $&$ 94.85 $&$ 94.10 $&$ 96.15 $&$ 95.15 $&$ 95.85 $&$ 95.40 $&$ 95.35 $&$ 94.20 $&$ 95.95 $&$ 95.15 $&$ 95.95 $&$ 94.45 $ \\ 
		$ T_0+ 4 $&$ 95.90 $&$ 95.90 $&$ 94.85 $&$ 94.60 $&$ 95.15 $&$ 94.25 $&$ 95.20 $&$ 95.35 $&$ 96.55 $&$ 95.00 $&$ 96.05 $&$ 94.95 $ \\ 
		$ T_0+ 5 $&$ 95.60 $&$ 95.40 $&$ 94.75 $&$ 94.30 $&$ 95.05 $&$ 95.20 $&$ 96.75 $&$ 96.25 $&$ 96.80 $&$ 94.90 $&$ 96.35 $&$ 94.65 $ \\ 
		\hline\hline 
	\end{tabular} 
	
\end{table} 

\begin{table}[H] 
	\footnotesize\centering\setstretch{1.5} 
	\caption{Pure factor model with homoscedastic i.i.d.\ uniform errors} 
	\label{tab:Pure factor model with homoscedastic i.i.d. uniform errors} 
	\begin{tabular}{c|cc|cc|cc|cc|cc|cc} 
		\multicolumn{13}{c}{Known number of factors, $90\%$ CI} \\ 
		\hline\hline 
		\multirow{2}{*}{\diagbox{$t$}{$\lp T_0, N_0 \rp$}} &\multicolumn{2}{c|}{$ (20, 30) $} &\multicolumn{2}{c|}{$ (20, 50) $} &\multicolumn{2}{c|}{$ (20, 100) $} &\multicolumn{2}{c|}{$ (40, 30) $} &\multicolumn{2}{c|}{$ (40, 50) $} &\multicolumn{2}{c}{$ (40, 100) $} \\ 
		& EQ & SY & EQ & SY & EQ & SY & EQ & SY & EQ & SY & EQ & SY \\ 
		\hline 
		$ T_0+ 1 $&$ 91.00 $&$ 90.95 $&$ 89.10 $&$ 89.05 $&$ 92.40 $&$ 92.55 $&$ 91.55 $&$ 91.55 $&$ 92.05 $&$ 92.15 $&$ 91.15 $&$ 90.95 $ \\ 
		$ T_0+ 2 $&$ 91.15 $&$ 91.10 $&$ 90.95 $&$ 90.95 $&$ 90.80 $&$ 90.50 $&$ 91.60 $&$ 91.75 $&$ 92.85 $&$ 92.65 $&$ 92.00 $&$ 92.00 $ \\ 
		$ T_0+ 3 $&$ 91.80 $&$ 91.85 $&$ 91.45 $&$ 91.40 $&$ 91.45 $&$ 91.50 $&$ 91.60 $&$ 91.60 $&$ 91.60 $&$ 91.70 $&$ 90.20 $&$ 90.50 $ \\ 
		$ T_0+ 4 $&$ 90.95 $&$ 90.80 $&$ 91.35 $&$ 91.30 $&$ 92.00 $&$ 92.35 $&$ 92.10 $&$ 92.05 $&$ 91.60 $&$ 91.90 $&$ 91.75 $&$ 91.95 $ \\ 
		$ T_0+ 5 $&$ 92.85 $&$ 93.00 $&$ 90.10 $&$ 91.05 $&$ 91.95 $&$ 92.15 $&$ 91.80 $&$ 91.60 $&$ 90.20 $&$ 90.30 $&$ 90.80 $&$ 91.50 $ \\ 
		\hline\hline 
		\multicolumn{13}{c}{ }\\ 
		\multicolumn{13}{c}{ }\\ 
	\end{tabular} 
	
	\begin{tabular}{c|cc|cc|cc|cc|cc|cc} 
		\multicolumn{13}{c}{Known number of factors, $95\%$ CI} \\ 
		\hline\hline 
		\multirow{2}{*}{\diagbox{$t$}{$\lp T_0, N_0 \rp$}} &\multicolumn{2}{c|}{$ (20, 30) $} &\multicolumn{2}{c|}{$ (20, 50) $} &\multicolumn{2}{c|}{$ (20, 100) $} &\multicolumn{2}{c|}{$ (40, 30) $} &\multicolumn{2}{c|}{$ (40, 50) $} &\multicolumn{2}{c}{$ (40, 100) $} \\ 
		& EQ & SY & EQ & SY & EQ & SY & EQ & SY & EQ & SY & EQ & SY \\ 
		\hline 
		$ T_0+ 1 $&$ 96.35 $&$ 96.60 $&$ 94.75 $&$ 94.85 $&$ 96.80 $&$ 96.70 $&$ 96.15 $&$ 96.05 $&$ 97.20 $&$ 97.10 $&$ 96.40 $&$ 96.85 $ \\ 
		$ T_0+ 2 $&$ 96.90 $&$ 96.90 $&$ 97.20 $&$ 97.30 $&$ 96.45 $&$ 96.65 $&$ 95.85 $&$ 95.80 $&$ 97.45 $&$ 97.70 $&$ 96.60 $&$ 96.60 $ \\ 
		$ T_0+ 3 $&$ 96.85 $&$ 96.80 $&$ 96.65 $&$ 96.80 $&$ 97.20 $&$ 97.55 $&$ 96.05 $&$ 96.15 $&$ 97.25 $&$ 97.25 $&$ 96.75 $&$ 96.55 $ \\ 
		$ T_0+ 4 $&$ 97.25 $&$ 97.10 $&$ 96.85 $&$ 96.65 $&$ 96.65 $&$ 96.65 $&$ 96.75 $&$ 96.70 $&$ 97.10 $&$ 96.95 $&$ 97.20 $&$ 97.60 $ \\ 
		$ T_0+ 5 $&$ 97.55 $&$ 97.20 $&$ 96.20 $&$ 96.30 $&$ 96.45 $&$ 96.55 $&$ 96.00 $&$ 96.00 $&$ 96.65 $&$ 96.60 $&$ 97.30 $&$ 97.25 $ \\ 
		\hline\hline 
		\multicolumn{13}{c}{ }\\ 
		\multicolumn{13}{c}{ }\\ 
	\end{tabular} 
	
	\begin{tabular}{c|cc|cc|cc|cc|cc|cc} 
		\multicolumn{13}{c}{Estimated number of factors, $90\%$ CI} \\ 
		\hline\hline 
		\multirow{2}{*}{\diagbox{$t$}{$\lp T_0, N_0 \rp$}} &\multicolumn{2}{c|}{$ (20, 30) $} &\multicolumn{2}{c|}{$ (20, 50) $} &\multicolumn{2}{c|}{$ (20, 100) $} &\multicolumn{2}{c|}{$ (40, 30) $} &\multicolumn{2}{c|}{$ (40, 50) $} &\multicolumn{2}{c}{$ (40, 100) $} \\ 
		& EQ & SY & EQ & SY & EQ & SY & EQ & SY & EQ & SY & EQ & SY \\ 
		\hline 
		$ T_0+ 1 $&$ 90.95 $&$ 90.95 $&$ 89.10 $&$ 89.05 $&$ 92.40 $&$ 92.55 $&$ 91.55 $&$ 91.55 $&$ 92.05 $&$ 92.15 $&$ 91.15 $&$ 90.95 $ \\ 
		$ T_0+ 2 $&$ 90.70 $&$ 90.70 $&$ 90.95 $&$ 90.95 $&$ 90.80 $&$ 90.50 $&$ 91.60 $&$ 91.75 $&$ 92.85 $&$ 92.65 $&$ 92.00 $&$ 92.00 $ \\ 
		$ T_0+ 3 $&$ 91.75 $&$ 91.80 $&$ 91.45 $&$ 91.40 $&$ 91.45 $&$ 91.50 $&$ 91.60 $&$ 91.60 $&$ 91.60 $&$ 91.70 $&$ 90.20 $&$ 90.50 $ \\ 
		$ T_0+ 4 $&$ 90.85 $&$ 90.70 $&$ 91.35 $&$ 91.30 $&$ 92.00 $&$ 92.35 $&$ 92.10 $&$ 92.05 $&$ 91.60 $&$ 91.90 $&$ 91.75 $&$ 91.95 $ \\ 
		$ T_0+ 5 $&$ 92.75 $&$ 93.00 $&$ 90.10 $&$ 91.05 $&$ 91.95 $&$ 92.15 $&$ 91.80 $&$ 91.60 $&$ 90.20 $&$ 90.30 $&$ 90.80 $&$ 91.50 $ \\ 
		\hline\hline 
		\multicolumn{13}{c}{ }\\ 
		\multicolumn{13}{c}{ }\\ 
	\end{tabular} 
	
	\begin{tabular}{c|cc|cc|cc|cc|cc|cc} 
		\multicolumn{13}{c}{Estimated number of factors, $95\%$ CI} \\ 
		\hline\hline 
		\multirow{2}{*}{\diagbox{$t$}{$\lp T_0, N_0 \rp$}} &\multicolumn{2}{c|}{$ (20, 30) $} &\multicolumn{2}{c|}{$ (20, 50) $} &\multicolumn{2}{c|}{$ (20, 100) $} &\multicolumn{2}{c|}{$ (40, 30) $} &\multicolumn{2}{c|}{$ (40, 50) $} &\multicolumn{2}{c}{$ (40, 100) $} \\ 
		& EQ & SY & EQ & SY & EQ & SY & EQ & SY & EQ & SY & EQ & SY \\ 
		\hline 
		$ T_0+ 1 $&$ 96.30 $&$ 96.60 $&$ 94.75 $&$ 94.85 $&$ 96.80 $&$ 96.70 $&$ 96.15 $&$ 96.05 $&$ 97.20 $&$ 97.10 $&$ 96.40 $&$ 96.85 $ \\ 
		$ T_0+ 2 $&$ 96.85 $&$ 96.90 $&$ 97.20 $&$ 97.30 $&$ 96.45 $&$ 96.65 $&$ 95.85 $&$ 95.80 $&$ 97.45 $&$ 97.70 $&$ 96.60 $&$ 96.60 $ \\ 
		$ T_0+ 3 $&$ 96.85 $&$ 96.80 $&$ 96.65 $&$ 96.80 $&$ 97.20 $&$ 97.55 $&$ 96.05 $&$ 96.15 $&$ 97.25 $&$ 97.25 $&$ 96.75 $&$ 96.55 $ \\ 
		$ T_0+ 4 $&$ 97.20 $&$ 97.05 $&$ 96.85 $&$ 96.65 $&$ 96.65 $&$ 96.65 $&$ 96.75 $&$ 96.70 $&$ 97.10 $&$ 96.95 $&$ 97.20 $&$ 97.60 $ \\ 
		$ T_0+ 5 $&$ 97.50 $&$ 97.20 $&$ 96.20 $&$ 96.30 $&$ 96.45 $&$ 96.55 $&$ 96.00 $&$ 96.00 $&$ 96.65 $&$ 96.60 $&$ 97.30 $&$ 97.25 $ \\ 
		\hline\hline 
	\end{tabular} 
	
\end{table} 

\begin{table}[H] 
	\footnotesize\centering\setstretch{1.5} 
	\caption{Pure factor model with heteroscedastic AR(1) chi-squared errors} 
	\label{tab:Pure factor model with heteroscedastic AR1 chi-squared errors} 
	\begin{tabular}{c|cc|cc|cc|cc|cc|cc} 
		\multicolumn{13}{c}{Known number of factors, $90\%$ CI} \\ 
		\hline\hline 
		\multirow{2}{*}{\diagbox{$t$}{$\lp T_0, N_0 \rp$}} &\multicolumn{2}{c|}{$ (20, 30) $} &\multicolumn{2}{c|}{$ (20, 50) $} &\multicolumn{2}{c|}{$ (20, 100) $} &\multicolumn{2}{c|}{$ (40, 30) $} &\multicolumn{2}{c|}{$ (40, 50) $} &\multicolumn{2}{c}{$ (40, 100) $} \\ 
		& EQ & SY & EQ & SY & EQ & SY & EQ & SY & EQ & SY & EQ & SY \\ 
		\hline 
		$ T_0+ 1 $&$ 92.90 $&$ 92.90 $&$ 93.25 $&$ 93.15 $&$ 92.85 $&$ 92.00 $&$ 92.70 $&$ 92.75 $&$ 90.60 $&$ 91.30 $&$ 92.90 $&$ 93.80 $ \\ 
		$ T_0+ 2 $&$ 94.55 $&$ 94.70 $&$ 91.15 $&$ 91.15 $&$ 92.45 $&$ 92.20 $&$ 92.35 $&$ 92.65 $&$ 91.10 $&$ 92.20 $&$ 94.45 $&$ 93.50 $ \\ 
		$ T_0+ 3 $&$ 93.20 $&$ 92.90 $&$ 90.20 $&$ 90.60 $&$ 93.00 $&$ 93.35 $&$ 91.90 $&$ 91.85 $&$ 91.75 $&$ 91.20 $&$ 91.95 $&$ 91.10 $ \\ 
		$ T_0+ 4 $&$ 92.15 $&$ 92.20 $&$ 91.15 $&$ 91.15 $&$ 93.40 $&$ 93.15 $&$ 93.30 $&$ 93.15 $&$ 91.35 $&$ 91.75 $&$ 92.60 $&$ 92.40 $ \\ 
		$ T_0+ 5 $&$ 92.20 $&$ 92.00 $&$ 90.10 $&$ 90.55 $&$ 92.65 $&$ 92.25 $&$ 91.15 $&$ 90.95 $&$ 91.10 $&$ 91.50 $&$ 93.05 $&$ 92.30 $ \\ 
		\hline\hline 
		\multicolumn{13}{c}{ }\\ 
		\multicolumn{13}{c}{ }\\ 
	\end{tabular} 
	
	\begin{tabular}{c|cc|cc|cc|cc|cc|cc} 
		\multicolumn{13}{c}{Known number of factors, $95\%$ CI} \\ 
		\hline\hline 
		\multirow{2}{*}{\diagbox{$t$}{$\lp T_0, N_0 \rp$}} &\multicolumn{2}{c|}{$ (20, 30) $} &\multicolumn{2}{c|}{$ (20, 50) $} &\multicolumn{2}{c|}{$ (20, 100) $} &\multicolumn{2}{c|}{$ (40, 30) $} &\multicolumn{2}{c|}{$ (40, 50) $} &\multicolumn{2}{c}{$ (40, 100) $} \\ 
		& EQ & SY & EQ & SY & EQ & SY & EQ & SY & EQ & SY & EQ & SY \\ 
		\hline 
		$ T_0+ 1 $&$ 96.80 $&$ 96.70 $&$ 96.45 $&$ 97.20 $&$ 96.55 $&$ 95.75 $&$ 97.00 $&$ 96.80 $&$ 95.45 $&$ 96.30 $&$ 96.85 $&$ 95.95 $ \\ 
		$ T_0+ 2 $&$ 96.95 $&$ 96.80 $&$ 94.70 $&$ 94.55 $&$ 96.20 $&$ 95.60 $&$ 95.90 $&$ 96.10 $&$ 95.05 $&$ 95.40 $&$ 97.65 $&$ 96.40 $ \\ 
		$ T_0+ 3 $&$ 96.65 $&$ 96.20 $&$ 94.80 $&$ 94.55 $&$ 97.10 $&$ 96.10 $&$ 96.50 $&$ 96.20 $&$ 94.85 $&$ 95.25 $&$ 96.75 $&$ 95.10 $ \\ 
		$ T_0+ 4 $&$ 96.40 $&$ 96.35 $&$ 94.75 $&$ 94.90 $&$ 97.00 $&$ 96.35 $&$ 96.85 $&$ 96.65 $&$ 95.65 $&$ 95.70 $&$ 96.85 $&$ 95.20 $ \\ 
		$ T_0+ 5 $&$ 96.50 $&$ 96.35 $&$ 94.95 $&$ 94.50 $&$ 97.45 $&$ 96.15 $&$ 95.75 $&$ 95.75 $&$ 94.55 $&$ 94.05 $&$ 96.25 $&$ 95.30 $ \\ 
		\hline\hline 
		\multicolumn{13}{c}{ }\\ 
		\multicolumn{13}{c}{ }\\ 
	\end{tabular} 
	
	\begin{tabular}{c|cc|cc|cc|cc|cc|cc} 
		\multicolumn{13}{c}{Estimated number of factors, $90\%$ CI} \\ 
		\hline\hline 
		\multirow{2}{*}{\diagbox{$t$}{$\lp T_0, N_0 \rp$}} &\multicolumn{2}{c|}{$ (20, 30) $} &\multicolumn{2}{c|}{$ (20, 50) $} &\multicolumn{2}{c|}{$ (20, 100) $} &\multicolumn{2}{c|}{$ (40, 30) $} &\multicolumn{2}{c|}{$ (40, 50) $} &\multicolumn{2}{c}{$ (40, 100) $} \\ 
		& EQ & SY & EQ & SY & EQ & SY & EQ & SY & EQ & SY & EQ & SY \\ 
		\hline 
		$ T_0+ 1 $&$ 92.95 $&$ 92.55 $&$ 91.85 $&$ 92.80 $&$ 92.25 $&$ 91.95 $&$ 92.60 $&$ 93.05 $&$ 90.90 $&$ 91.10 $&$ 92.70 $&$ 93.40 $ \\ 
		$ T_0+ 2 $&$ 92.65 $&$ 92.75 $&$ 90.20 $&$ 90.30 $&$ 92.00 $&$ 91.95 $&$ 92.80 $&$ 92.35 $&$ 91.35 $&$ 91.70 $&$ 93.95 $&$ 93.25 $ \\ 
		$ T_0+ 3 $&$ 92.55 $&$ 92.55 $&$ 88.80 $&$ 89.35 $&$ 93.20 $&$ 93.45 $&$ 91.80 $&$ 91.70 $&$ 91.25 $&$ 90.35 $&$ 92.75 $&$ 91.80 $ \\ 
		$ T_0+ 4 $&$ 91.75 $&$ 91.70 $&$ 89.60 $&$ 89.60 $&$ 93.50 $&$ 93.30 $&$ 93.20 $&$ 93.10 $&$ 91.25 $&$ 92.35 $&$ 92.65 $&$ 92.00 $ \\ 
		$ T_0+ 5 $&$ 92.30 $&$ 92.30 $&$ 89.60 $&$ 89.00 $&$ 91.85 $&$ 92.20 $&$ 91.95 $&$ 91.80 $&$ 91.10 $&$ 91.25 $&$ 92.20 $&$ 91.80 $ \\ 
		\hline\hline 
		\multicolumn{13}{c}{ }\\ 
		\multicolumn{13}{c}{ }\\ 
	\end{tabular} 
	
	\begin{tabular}{c|cc|cc|cc|cc|cc|cc} 
		\multicolumn{13}{c}{Estimated number of factors, $95\%$ CI} \\ 
		\hline\hline 
		\multirow{2}{*}{\diagbox{$t$}{$\lp T_0, N_0 \rp$}} &\multicolumn{2}{c|}{$ (20, 30) $} &\multicolumn{2}{c|}{$ (20, 50) $} &\multicolumn{2}{c|}{$ (20, 100) $} &\multicolumn{2}{c|}{$ (40, 30) $} &\multicolumn{2}{c|}{$ (40, 50) $} &\multicolumn{2}{c}{$ (40, 100) $} \\ 
		& EQ & SY & EQ & SY & EQ & SY & EQ & SY & EQ & SY & EQ & SY \\ 
		\hline 
		$ T_0+ 1 $&$ 96.35 $&$ 96.25 $&$ 96.10 $&$ 96.15 $&$ 95.95 $&$ 95.75 $&$ 96.00 $&$ 95.75 $&$ 95.70 $&$ 95.90 $&$ 96.35 $&$ 95.95 $ \\ 
		$ T_0+ 2 $&$ 96.05 $&$ 95.70 $&$ 94.30 $&$ 94.35 $&$ 95.15 $&$ 94.95 $&$ 96.35 $&$ 96.10 $&$ 94.95 $&$ 95.25 $&$ 97.60 $&$ 96.25 $ \\ 
		$ T_0+ 3 $&$ 96.85 $&$ 96.95 $&$ 93.45 $&$ 93.55 $&$ 96.55 $&$ 96.05 $&$ 95.85 $&$ 95.55 $&$ 94.40 $&$ 94.75 $&$ 96.55 $&$ 95.75 $ \\ 
		$ T_0+ 4 $&$ 95.80 $&$ 95.20 $&$ 93.50 $&$ 93.45 $&$ 97.20 $&$ 96.40 $&$ 96.85 $&$ 96.40 $&$ 95.40 $&$ 95.55 $&$ 96.40 $&$ 95.75 $ \\ 
		$ T_0+ 5 $&$ 95.55 $&$ 95.60 $&$ 93.80 $&$ 94.15 $&$ 95.80 $&$ 95.40 $&$ 95.70 $&$ 95.55 $&$ 94.85 $&$ 94.70 $&$ 95.35 $&$ 95.15 $ \\ 
		\hline\hline 
	\end{tabular} 
	
\end{table} 

\begin{table}[H] 
	\footnotesize\centering\setstretch{1.5} 
	\caption{Pure factor model with heteroscedastic AR(1) uniform errors} 
	\label{tab:Pure factor model with heteroscedastic AR1 uniform errors} 
	\begin{tabular}{c|cc|cc|cc|cc|cc|cc} 
		\multicolumn{13}{c}{Known number of factors, $90\%$ CI} \\ 
		\hline\hline 
		\multirow{2}{*}{\diagbox{$t$}{$\lp T_0, N_0 \rp$}} &\multicolumn{2}{c|}{$ (20, 30) $} &\multicolumn{2}{c|}{$ (20, 50) $} &\multicolumn{2}{c|}{$ (20, 100) $} &\multicolumn{2}{c|}{$ (40, 30) $} &\multicolumn{2}{c|}{$ (40, 50) $} &\multicolumn{2}{c}{$ (40, 100) $} \\ 
		& EQ & SY & EQ & SY & EQ & SY & EQ & SY & EQ & SY & EQ & SY \\ 
		\hline 
		$ T_0+ 1 $&$ 91.30 $&$ 91.50 $&$ 91.65 $&$ 91.95 $&$ 95.30 $&$ 95.20 $&$ 91.50 $&$ 91.30 $&$ 91.90 $&$ 91.85 $&$ 92.50 $&$ 92.50 $ \\ 
		$ T_0+ 2 $&$ 91.70 $&$ 91.75 $&$ 91.50 $&$ 91.00 $&$ 95.15 $&$ 94.90 $&$ 92.70 $&$ 92.65 $&$ 91.70 $&$ 91.75 $&$ 92.55 $&$ 92.90 $ \\ 
		$ T_0+ 3 $&$ 90.25 $&$ 90.50 $&$ 91.45 $&$ 91.75 $&$ 94.80 $&$ 94.95 $&$ 90.85 $&$ 90.65 $&$ 91.55 $&$ 91.20 $&$ 94.70 $&$ 94.75 $ \\ 
		$ T_0+ 4 $&$ 91.70 $&$ 91.65 $&$ 90.55 $&$ 90.65 $&$ 94.50 $&$ 94.35 $&$ 90.40 $&$ 90.55 $&$ 90.15 $&$ 90.00 $&$ 92.90 $&$ 92.60 $ \\ 
		$ T_0+ 5 $&$ 92.40 $&$ 92.45 $&$ 92.10 $&$ 91.95 $&$ 94.35 $&$ 93.95 $&$ 93.10 $&$ 92.60 $&$ 91.35 $&$ 91.10 $&$ 93.35 $&$ 93.35 $ \\ 
		\hline\hline 
		\multicolumn{13}{c}{ }\\ 
		\multicolumn{13}{c}{ }\\ 
	\end{tabular} 
	
	\begin{tabular}{c|cc|cc|cc|cc|cc|cc} 
		\multicolumn{13}{c}{Known number of factors, $95\%$ CI} \\ 
		\hline\hline 
		\multirow{2}{*}{\diagbox{$t$}{$\lp T_0, N_0 \rp$}} &\multicolumn{2}{c|}{$ (20, 30) $} &\multicolumn{2}{c|}{$ (20, 50) $} &\multicolumn{2}{c|}{$ (20, 100) $} &\multicolumn{2}{c|}{$ (40, 30) $} &\multicolumn{2}{c|}{$ (40, 50) $} &\multicolumn{2}{c}{$ (40, 100) $} \\ 
		& EQ & SY & EQ & SY & EQ & SY & EQ & SY & EQ & SY & EQ & SY \\ 
		\hline 
		$ T_0+ 1 $&$ 96.75 $&$ 96.80 $&$ 96.90 $&$ 97.05 $&$ 98.65 $&$ 98.65 $&$ 96.20 $&$ 96.35 $&$ 97.30 $&$ 97.15 $&$ 97.85 $&$ 97.25 $ \\ 
		$ T_0+ 2 $&$ 95.65 $&$ 95.60 $&$ 97.35 $&$ 97.25 $&$ 98.60 $&$ 98.60 $&$ 96.50 $&$ 96.45 $&$ 95.80 $&$ 96.15 $&$ 97.80 $&$ 97.75 $ \\ 
		$ T_0+ 3 $&$ 95.35 $&$ 95.30 $&$ 96.00 $&$ 96.05 $&$ 98.90 $&$ 98.90 $&$ 95.80 $&$ 95.75 $&$ 96.35 $&$ 96.65 $&$ 98.60 $&$ 98.60 $ \\ 
		$ T_0+ 4 $&$ 95.85 $&$ 95.75 $&$ 95.10 $&$ 95.15 $&$ 98.45 $&$ 98.50 $&$ 95.55 $&$ 95.05 $&$ 96.35 $&$ 96.40 $&$ 97.30 $&$ 97.10 $ \\ 
		$ T_0+ 5 $&$ 96.25 $&$ 96.20 $&$ 96.75 $&$ 96.70 $&$ 97.80 $&$ 98.85 $&$ 96.35 $&$ 96.55 $&$ 96.30 $&$ 96.05 $&$ 97.85 $&$ 98.00 $ \\ 
		\hline\hline 
		\multicolumn{13}{c}{ }\\ 
		\multicolumn{13}{c}{ }\\ 
	\end{tabular} 
	
	\begin{tabular}{c|cc|cc|cc|cc|cc|cc} 
		\multicolumn{13}{c}{Estimated number of factors, $90\%$ CI} \\ 
		\hline\hline 
		\multirow{2}{*}{\diagbox{$t$}{$\lp T_0, N_0 \rp$}} &\multicolumn{2}{c|}{$ (20, 30) $} &\multicolumn{2}{c|}{$ (20, 50) $} &\multicolumn{2}{c|}{$ (20, 100) $} &\multicolumn{2}{c|}{$ (40, 30) $} &\multicolumn{2}{c|}{$ (40, 50) $} &\multicolumn{2}{c}{$ (40, 100) $} \\ 
		& EQ & SY & EQ & SY & EQ & SY & EQ & SY & EQ & SY & EQ & SY \\ 
		\hline 
		$ T_0+ 1 $&$ 91.50 $&$ 91.65 $&$ 90.50 $&$ 90.15 $&$ 93.95 $&$ 93.90 $&$ 92.60 $&$ 92.35 $&$ 91.45 $&$ 92.15 $&$ 92.05 $&$ 91.70 $ \\ 
		$ T_0+ 2 $&$ 91.95 $&$ 91.75 $&$ 90.70 $&$ 90.60 $&$ 93.65 $&$ 93.60 $&$ 93.15 $&$ 93.00 $&$ 91.40 $&$ 91.70 $&$ 91.90 $&$ 92.45 $ \\ 
		$ T_0+ 3 $&$ 91.75 $&$ 91.65 $&$ 90.90 $&$ 90.55 $&$ 93.80 $&$ 94.15 $&$ 90.35 $&$ 90.60 $&$ 90.85 $&$ 90.55 $&$ 94.00 $&$ 94.00 $ \\ 
		$ T_0+ 4 $&$ 91.90 $&$ 92.20 $&$ 89.45 $&$ 89.80 $&$ 93.75 $&$ 93.90 $&$ 91.55 $&$ 91.70 $&$ 90.90 $&$ 91.20 $&$ 92.95 $&$ 92.85 $ \\ 
		$ T_0+ 5 $&$ 90.85 $&$ 91.20 $&$ 89.85 $&$ 89.85 $&$ 93.50 $&$ 93.10 $&$ 92.95 $&$ 93.20 $&$ 92.00 $&$ 91.90 $&$ 92.50 $&$ 92.30 $ \\ 
		\hline\hline 
		\multicolumn{13}{c}{ }\\ 
		\multicolumn{13}{c}{ }\\ 
	\end{tabular} 
	
	\begin{tabular}{c|cc|cc|cc|cc|cc|cc} 
		\multicolumn{13}{c}{Estimated number of factors, $95\%$ CI} \\ 
		\hline\hline 
		\multirow{2}{*}{\diagbox{$t$}{$\lp T_0, N_0 \rp$}} &\multicolumn{2}{c|}{$ (20, 30) $} &\multicolumn{2}{c|}{$ (20, 50) $} &\multicolumn{2}{c|}{$ (20, 100) $} &\multicolumn{2}{c|}{$ (40, 30) $} &\multicolumn{2}{c|}{$ (40, 50) $} &\multicolumn{2}{c}{$ (40, 100) $} \\ 
		& EQ & SY & EQ & SY & EQ & SY & EQ & SY & EQ & SY & EQ & SY \\ 
		\hline 
		$ T_0+ 1 $&$ 95.95 $&$ 96.00 $&$ 96.55 $&$ 96.40 $&$ 98.30 $&$ 98.25 $&$ 96.55 $&$ 96.70 $&$ 97.15 $&$ 97.10 $&$ 97.75 $&$ 97.45 $ \\ 
		$ T_0+ 2 $&$ 96.25 $&$ 96.20 $&$ 96.10 $&$ 96.25 $&$ 97.50 $&$ 97.50 $&$ 97.40 $&$ 97.05 $&$ 96.50 $&$ 96.35 $&$ 97.80 $&$ 98.05 $ \\ 
		$ T_0+ 3 $&$ 96.35 $&$ 96.00 $&$ 95.40 $&$ 95.30 $&$ 98.35 $&$ 98.25 $&$ 94.90 $&$ 95.30 $&$ 95.70 $&$ 96.30 $&$ 98.20 $&$ 98.25 $ \\ 
		$ T_0+ 4 $&$ 96.25 $&$ 96.00 $&$ 95.35 $&$ 95.65 $&$ 98.10 $&$ 98.10 $&$ 95.80 $&$ 95.70 $&$ 95.70 $&$ 95.90 $&$ 97.00 $&$ 97.10 $ \\ 
		$ T_0+ 5 $&$ 96.45 $&$ 96.55 $&$ 95.85 $&$ 95.85 $&$ 97.65 $&$ 97.50 $&$ 96.85 $&$ 96.80 $&$ 96.50 $&$ 96.30 $&$ 97.55 $&$ 97.60 $ \\ 
		\hline\hline 
	\end{tabular} 
	
\end{table} 

\begin{table}[H] 
	\footnotesize\centering\setstretch{1.5} 
	\caption{Factor model with covariates and homoscedastic i.i.d.\ chi-squared errors} 
	\label{tab:Factor model with covariates and homoscedastic i.i.d. chi-squared errors} 
	\begin{tabular}{c|cc|cc|cc|cc|cc|cc} 
		\multicolumn{13}{c}{Known number of factors, $90\%$ CI} \\ 
		\hline\hline 
		\multirow{2}{*}{\diagbox{$t$}{$\lp T_0, N_0 \rp$}} &\multicolumn{2}{c|}{$ (20, 30) $} &\multicolumn{2}{c|}{$ (20, 50) $} &\multicolumn{2}{c|}{$ (20, 100) $} &\multicolumn{2}{c|}{$ (40, 30) $} &\multicolumn{2}{c|}{$ (40, 50) $} &\multicolumn{2}{c}{$ (40, 100) $} \\ 
		& EQ & SY & EQ & SY & EQ & SY & EQ & SY & EQ & SY & EQ & SY \\ 
		\hline 
		$ T_0+ 1 $&$ 92.00 $&$ 91.80 $&$ 91.90 $&$ 91.75 $&$ 91.75 $&$ 90.60 $&$ 91.65 $&$ 90.30 $&$ 92.95 $&$ 92.30 $&$ 93.55 $&$ 92.35 $ \\ 
		$ T_0+ 2 $&$ 93.35 $&$ 91.95 $&$ 91.35 $&$ 91.40 $&$ 91.40 $&$ 91.00 $&$ 93.10 $&$ 91.20 $&$ 92.65 $&$ 91.30 $&$ 92.40 $&$ 91.50 $ \\ 
		$ T_0+ 3 $&$ 91.75 $&$ 91.35 $&$ 90.75 $&$ 90.60 $&$ 90.70 $&$ 91.20 $&$ 92.70 $&$ 92.00 $&$ 91.60 $&$ 91.00 $&$ 92.60 $&$ 92.25 $ \\ 
		$ T_0+ 4 $&$ 91.90 $&$ 91.60 $&$ 92.15 $&$ 92.15 $&$ 92.00 $&$ 92.10 $&$ 93.40 $&$ 92.65 $&$ 91.85 $&$ 92.05 $&$ 93.15 $&$ 92.20 $ \\ 
		$ T_0+ 5 $&$ 91.65 $&$ 91.55 $&$ 90.45 $&$ 90.70 $&$ 91.60 $&$ 90.70 $&$ 91.85 $&$ 91.20 $&$ 93.70 $&$ 91.90 $&$ 93.30 $&$ 90.95 $ \\ 
		\hline\hline 
		\multicolumn{13}{c}{ }\\ 
		\multicolumn{13}{c}{ }\\ 
	\end{tabular} 
	
	\begin{tabular}{c|cc|cc|cc|cc|cc|cc} 
		\multicolumn{13}{c}{Known number of factors, $95\%$ CI} \\ 
		\hline\hline 
		\multirow{2}{*}{\diagbox{$t$}{$\lp T_0, N_0 \rp$}} &\multicolumn{2}{c|}{$ (20, 30) $} &\multicolumn{2}{c|}{$ (20, 50) $} &\multicolumn{2}{c|}{$ (20, 100) $} &\multicolumn{2}{c|}{$ (40, 30) $} &\multicolumn{2}{c|}{$ (40, 50) $} &\multicolumn{2}{c}{$ (40, 100) $} \\ 
		& EQ & SY & EQ & SY & EQ & SY & EQ & SY & EQ & SY & EQ & SY \\ 
		\hline 
		$ T_0+ 1 $&$ 95.80 $&$ 94.80 $&$ 94.80 $&$ 94.50 $&$ 95.55 $&$ 94.80 $&$ 96.40 $&$ 94.70 $&$ 96.50 $&$ 95.70 $&$ 96.60 $&$ 95.50 $ \\ 
		$ T_0+ 2 $&$ 96.05 $&$ 95.40 $&$ 96.10 $&$ 94.55 $&$ 95.20 $&$ 94.75 $&$ 96.30 $&$ 95.60 $&$ 96.30 $&$ 94.85 $&$ 96.35 $&$ 95.00 $ \\ 
		$ T_0+ 3 $&$ 96.80 $&$ 95.05 $&$ 95.20 $&$ 94.50 $&$ 94.65 $&$ 93.85 $&$ 95.90 $&$ 95.50 $&$ 95.95 $&$ 94.75 $&$ 96.70 $&$ 94.40 $ \\ 
		$ T_0+ 4 $&$ 95.85 $&$ 94.60 $&$ 95.25 $&$ 94.65 $&$ 96.50 $&$ 95.05 $&$ 97.15 $&$ 96.10 $&$ 95.85 $&$ 94.65 $&$ 97.15 $&$ 94.90 $ \\ 
		$ T_0+ 5 $&$ 95.75 $&$ 94.65 $&$ 94.10 $&$ 94.30 $&$ 95.55 $&$ 94.85 $&$ 96.25 $&$ 95.45 $&$ 97.25 $&$ 95.65 $&$ 96.60 $&$ 95.45 $ \\ 
		\hline\hline 
		\multicolumn{13}{c}{ }\\ 
		\multicolumn{13}{c}{ }\\ 
	\end{tabular} 
	
	\begin{tabular}{c|cc|cc|cc|cc|cc|cc} 
		\multicolumn{13}{c}{Estimated number of factors, $90\%$ CI} \\ 
		\hline\hline 
		\multirow{2}{*}{\diagbox{$t$}{$\lp T_0, N_0 \rp$}} &\multicolumn{2}{c|}{$ (20, 30) $} &\multicolumn{2}{c|}{$ (20, 50) $} &\multicolumn{2}{c|}{$ (20, 100) $} &\multicolumn{2}{c|}{$ (40, 30) $} &\multicolumn{2}{c|}{$ (40, 50) $} &\multicolumn{2}{c}{$ (40, 100) $} \\ 
		& EQ & SY & EQ & SY & EQ & SY & EQ & SY & EQ & SY & EQ & SY \\ 
		\hline 
		$ T_0+ 1 $&$ 92.40 $&$ 92.70 $&$ 91.80 $&$ 91.45 $&$ 91.75 $&$ 90.60 $&$ 91.60 $&$ 90.50 $&$ 93.05 $&$ 92.25 $&$ 93.55 $&$ 92.35 $ \\ 
		$ T_0+ 2 $&$ 92.30 $&$ 91.80 $&$ 91.45 $&$ 91.35 $&$ 91.40 $&$ 90.90 $&$ 92.80 $&$ 91.45 $&$ 92.65 $&$ 91.35 $&$ 92.55 $&$ 91.50 $ \\ 
		$ T_0+ 3 $&$ 92.20 $&$ 92.10 $&$ 90.95 $&$ 91.30 $&$ 90.75 $&$ 91.20 $&$ 92.55 $&$ 92.25 $&$ 91.75 $&$ 91.00 $&$ 92.65 $&$ 92.35 $ \\ 
		$ T_0+ 4 $&$ 92.25 $&$ 91.75 $&$ 92.25 $&$ 92.10 $&$ 92.00 $&$ 92.10 $&$ 93.45 $&$ 92.70 $&$ 91.95 $&$ 92.10 $&$ 93.15 $&$ 92.20 $ \\ 
		$ T_0+ 5 $&$ 91.60 $&$ 91.35 $&$ 90.70 $&$ 90.80 $&$ 91.40 $&$ 90.70 $&$ 91.70 $&$ 91.15 $&$ 93.75 $&$ 92.05 $&$ 93.30 $&$ 90.95 $ \\ 
		\hline\hline 
		\multicolumn{13}{c}{ }\\ 
		\multicolumn{13}{c}{ }\\ 
	\end{tabular} 
	
	\begin{tabular}{c|cc|cc|cc|cc|cc|cc} 
		\multicolumn{13}{c}{Estimated number of factors, $95\%$ CI} \\ 
		\hline\hline 
		\multirow{2}{*}{\diagbox{$t$}{$\lp T_0, N_0 \rp$}} &\multicolumn{2}{c|}{$ (20, 30) $} &\multicolumn{2}{c|}{$ (20, 50) $} &\multicolumn{2}{c|}{$ (20, 100) $} &\multicolumn{2}{c|}{$ (40, 30) $} &\multicolumn{2}{c|}{$ (40, 50) $} &\multicolumn{2}{c}{$ (40, 100) $} \\ 
		& EQ & SY & EQ & SY & EQ & SY & EQ & SY & EQ & SY & EQ & SY \\ 
		\hline 
		$ T_0+ 1 $&$ 96.25 $&$ 95.05 $&$ 94.60 $&$ 94.60 $&$ 95.55 $&$ 94.80 $&$ 96.75 $&$ 94.70 $&$ 96.45 $&$ 95.90 $&$ 96.60 $&$ 95.50 $ \\ 
		$ T_0+ 2 $&$ 96.15 $&$ 95.45 $&$ 95.95 $&$ 94.60 $&$ 95.15 $&$ 94.75 $&$ 96.35 $&$ 95.55 $&$ 96.30 $&$ 94.80 $&$ 96.35 $&$ 95.00 $ \\ 
		$ T_0+ 3 $&$ 96.75 $&$ 95.60 $&$ 94.90 $&$ 94.45 $&$ 94.70 $&$ 93.85 $&$ 96.25 $&$ 95.30 $&$ 96.05 $&$ 94.75 $&$ 96.65 $&$ 94.45 $ \\ 
		$ T_0+ 4 $&$ 95.50 $&$ 94.75 $&$ 95.10 $&$ 94.55 $&$ 96.55 $&$ 95.05 $&$ 97.05 $&$ 96.00 $&$ 95.95 $&$ 94.70 $&$ 97.15 $&$ 94.90 $ \\ 
		$ T_0+ 5 $&$ 95.50 $&$ 94.85 $&$ 94.40 $&$ 94.35 $&$ 95.55 $&$ 94.85 $&$ 96.15 $&$ 95.20 $&$ 97.25 $&$ 95.65 $&$ 96.60 $&$ 95.45 $ \\ 
		\hline\hline 
	\end{tabular} 
	
\end{table} 

\begin{table}[H] 
	\footnotesize\centering\setstretch{1.5} 
	\caption{Factor model with covariates and homoscedastic i.i.d.\ uniform errors} 
	\label{tab:Factor model with covariates and homoscedastic i.i.d. uniform errors} 
	\begin{tabular}{c|cc|cc|cc|cc|cc|cc} 
		\multicolumn{13}{c}{Known number of factors, $90\%$ CI} \\ 
		\hline\hline 
		\multirow{2}{*}{\diagbox{$t$}{$\lp T_0, N_0 \rp$}} &\multicolumn{2}{c|}{$ (20, 30) $} &\multicolumn{2}{c|}{$ (20, 50) $} &\multicolumn{2}{c|}{$ (20, 100) $} &\multicolumn{2}{c|}{$ (40, 30) $} &\multicolumn{2}{c|}{$ (40, 50) $} &\multicolumn{2}{c}{$ (40, 100) $} \\ 
		& EQ & SY & EQ & SY & EQ & SY & EQ & SY & EQ & SY & EQ & SY \\ 
		\hline 
		$ T_0+ 1 $&$ 93.15 $&$ 92.75 $&$ 93.20 $&$ 93.15 $&$ 91.90 $&$ 91.55 $&$ 90.95 $&$ 91.05 $&$ 92.40 $&$ 92.20 $&$ 90.25 $&$ 90.25 $ \\ 
		$ T_0+ 2 $&$ 90.40 $&$ 91.10 $&$ 92.65 $&$ 92.85 $&$ 91.10 $&$ 90.60 $&$ 91.60 $&$ 91.60 $&$ 90.95 $&$ 91.10 $&$ 90.55 $&$ 90.60 $ \\ 
		$ T_0+ 3 $&$ 93.35 $&$ 93.25 $&$ 92.00 $&$ 91.90 $&$ 91.60 $&$ 91.35 $&$ 91.85 $&$ 92.85 $&$ 91.65 $&$ 91.70 $&$ 92.00 $&$ 91.95 $ \\ 
		$ T_0+ 4 $&$ 92.30 $&$ 93.65 $&$ 88.90 $&$ 88.85 $&$ 91.55 $&$ 91.55 $&$ 90.70 $&$ 90.85 $&$ 92.05 $&$ 92.15 $&$ 91.35 $&$ 91.35 $ \\ 
		$ T_0+ 5 $&$ 90.75 $&$ 90.90 $&$ 92.15 $&$ 92.05 $&$ 89.55 $&$ 89.55 $&$ 91.35 $&$ 91.35 $&$ 89.75 $&$ 90.10 $&$ 91.60 $&$ 91.40 $ \\ 
		\hline\hline 
		\multicolumn{13}{c}{ }\\ 
		\multicolumn{13}{c}{ }\\ 
	\end{tabular} 
	
	\begin{tabular}{c|cc|cc|cc|cc|cc|cc} 
		\multicolumn{13}{c}{Known number of factors, $95\%$ CI} \\ 
		\hline\hline 
		\multirow{2}{*}{\diagbox{$t$}{$\lp T_0, N_0 \rp$}} &\multicolumn{2}{c|}{$ (20, 30) $} &\multicolumn{2}{c|}{$ (20, 50) $} &\multicolumn{2}{c|}{$ (20, 100) $} &\multicolumn{2}{c|}{$ (40, 30) $} &\multicolumn{2}{c|}{$ (40, 50) $} &\multicolumn{2}{c}{$ (40, 100) $} \\ 
		& EQ & SY & EQ & SY & EQ & SY & EQ & SY & EQ & SY & EQ & SY \\ 
		\hline 
		$ T_0+ 1 $&$ 97.65 $&$ 97.70 $&$ 97.35 $&$ 97.50 $&$ 96.70 $&$ 96.70 $&$ 97.30 $&$ 97.35 $&$ 96.85 $&$ 96.75 $&$ 96.65 $&$ 96.75 $ \\ 
		$ T_0+ 2 $&$ 96.65 $&$ 97.10 $&$ 96.95 $&$ 96.80 $&$ 95.60 $&$ 95.70 $&$ 96.50 $&$ 96.50 $&$ 96.60 $&$ 96.55 $&$ 95.55 $&$ 95.45 $ \\ 
		$ T_0+ 3 $&$ 96.90 $&$ 97.05 $&$ 96.80 $&$ 97.05 $&$ 96.75 $&$ 96.90 $&$ 97.00 $&$ 97.05 $&$ 97.50 $&$ 97.65 $&$ 96.90 $&$ 97.05 $ \\ 
		$ T_0+ 4 $&$ 97.05 $&$ 98.00 $&$ 96.10 $&$ 96.05 $&$ 96.10 $&$ 96.25 $&$ 96.80 $&$ 96.75 $&$ 97.35 $&$ 97.60 $&$ 96.40 $&$ 96.25 $ \\ 
		$ T_0+ 5 $&$ 96.45 $&$ 96.65 $&$ 97.30 $&$ 97.30 $&$ 96.30 $&$ 96.25 $&$ 96.35 $&$ 96.05 $&$ 96.15 $&$ 96.25 $&$ 96.95 $&$ 96.80 $ \\ 
		\hline\hline 
		\multicolumn{13}{c}{ }\\ 
		\multicolumn{13}{c}{ }\\ 
	\end{tabular} 
	
	\begin{tabular}{c|cc|cc|cc|cc|cc|cc} 
		\multicolumn{13}{c}{Estimated number of factors, $90\%$ CI} \\ 
		\hline\hline 
		\multirow{2}{*}{\diagbox{$t$}{$\lp T_0, N_0 \rp$}} &\multicolumn{2}{c|}{$ (20, 30) $} &\multicolumn{2}{c|}{$ (20, 50) $} &\multicolumn{2}{c|}{$ (20, 100) $} &\multicolumn{2}{c|}{$ (40, 30) $} &\multicolumn{2}{c|}{$ (40, 50) $} &\multicolumn{2}{c}{$ (40, 100) $} \\ 
		& EQ & SY & EQ & SY & EQ & SY & EQ & SY & EQ & SY & EQ & SY \\ 
		\hline 
		$ T_0+ 1 $&$ 93.15 $&$ 92.75 $&$ 93.20 $&$ 93.15 $&$ 91.90 $&$ 91.55 $&$ 90.95 $&$ 91.05 $&$ 92.40 $&$ 92.20 $&$ 90.25 $&$ 90.25 $ \\ 
		$ T_0+ 2 $&$ 90.35 $&$ 91.10 $&$ 92.65 $&$ 92.85 $&$ 91.10 $&$ 90.60 $&$ 91.60 $&$ 91.60 $&$ 90.95 $&$ 91.10 $&$ 90.55 $&$ 90.60 $ \\ 
		$ T_0+ 3 $&$ 93.20 $&$ 92.95 $&$ 92.00 $&$ 91.90 $&$ 91.60 $&$ 91.35 $&$ 91.85 $&$ 92.85 $&$ 91.65 $&$ 91.70 $&$ 92.00 $&$ 91.95 $ \\ 
		$ T_0+ 4 $&$ 92.30 $&$ 93.60 $&$ 88.90 $&$ 88.85 $&$ 91.55 $&$ 91.55 $&$ 90.70 $&$ 90.85 $&$ 92.05 $&$ 92.15 $&$ 91.35 $&$ 91.35 $ \\ 
		$ T_0+ 5 $&$ 90.75 $&$ 90.80 $&$ 92.15 $&$ 92.05 $&$ 89.55 $&$ 89.55 $&$ 91.35 $&$ 91.35 $&$ 89.75 $&$ 90.10 $&$ 91.60 $&$ 91.40 $ \\ 
		\hline\hline 
		\multicolumn{13}{c}{ }\\ 
		\multicolumn{13}{c}{ }\\ 
	\end{tabular} 
	
	\begin{tabular}{c|cc|cc|cc|cc|cc|cc} 
		\multicolumn{13}{c}{Estimated number of factors, $95\%$ CI} \\ 
		\hline\hline 
		\multirow{2}{*}{\diagbox{$t$}{$\lp T_0, N_0 \rp$}} &\multicolumn{2}{c|}{$ (20, 30) $} &\multicolumn{2}{c|}{$ (20, 50) $} &\multicolumn{2}{c|}{$ (20, 100) $} &\multicolumn{2}{c|}{$ (40, 30) $} &\multicolumn{2}{c|}{$ (40, 50) $} &\multicolumn{2}{c}{$ (40, 100) $} \\ 
		& EQ & SY & EQ & SY & EQ & SY & EQ & SY & EQ & SY & EQ & SY \\ 
		\hline 
		$ T_0+ 1 $&$ 97.60 $&$ 97.65 $&$ 97.35 $&$ 97.50 $&$ 96.70 $&$ 96.70 $&$ 97.30 $&$ 97.35 $&$ 96.85 $&$ 96.75 $&$ 96.65 $&$ 96.75 $ \\ 
		$ T_0+ 2 $&$ 96.65 $&$ 97.10 $&$ 96.95 $&$ 96.80 $&$ 95.60 $&$ 95.70 $&$ 96.50 $&$ 96.50 $&$ 96.60 $&$ 96.55 $&$ 95.55 $&$ 95.45 $ \\ 
		$ T_0+ 3 $&$ 96.90 $&$ 97.00 $&$ 96.75 $&$ 97.05 $&$ 96.75 $&$ 96.90 $&$ 97.00 $&$ 97.05 $&$ 97.50 $&$ 97.65 $&$ 96.90 $&$ 97.05 $ \\ 
		$ T_0+ 4 $&$ 97.05 $&$ 98.00 $&$ 96.10 $&$ 96.05 $&$ 96.10 $&$ 96.25 $&$ 96.80 $&$ 96.75 $&$ 97.35 $&$ 97.60 $&$ 96.40 $&$ 96.25 $ \\ 
		$ T_0+ 5 $&$ 96.55 $&$ 96.70 $&$ 97.30 $&$ 97.30 $&$ 96.30 $&$ 96.25 $&$ 96.35 $&$ 96.05 $&$ 96.15 $&$ 96.25 $&$ 96.95 $&$ 96.80 $ \\ 
		\hline\hline 
	\end{tabular} 
	
\end{table} 

\begin{table}[H] 
	\footnotesize\centering\setstretch{1.5} 
	\caption{Factor model with covariates and heteroscedastic AR(1) chi-squared errors} 
	\label{tab:Factor model with covariates and heteroscedastic AR1 chi-squared errors} 
	\begin{tabular}{c|cc|cc|cc|cc|cc|cc} 
		\multicolumn{13}{c}{Known number of factors, $90\%$ CI} \\ 
		\hline\hline 
		\multirow{2}{*}{\diagbox{$t$}{$\lp T_0, N_0 \rp$}} &\multicolumn{2}{c|}{$ (20, 30) $} &\multicolumn{2}{c|}{$ (20, 50) $} &\multicolumn{2}{c|}{$ (20, 100) $} &\multicolumn{2}{c|}{$ (40, 30) $} &\multicolumn{2}{c|}{$ (40, 50) $} &\multicolumn{2}{c}{$ (40, 100) $} \\ 
		& EQ & SY & EQ & SY & EQ & SY & EQ & SY & EQ & SY & EQ & SY \\ 
		\hline 
		$ T_0+ 1 $&$ 92.55 $&$ 91.75 $&$ 89.50 $&$ 89.70 $&$ 90.20 $&$ 90.35 $&$ 92.15 $&$ 91.60 $&$ 90.20 $&$ 90.40 $&$ 92.50 $&$ 93.45 $ \\ 
		$ T_0+ 2 $&$ 92.45 $&$ 91.90 $&$ 88.80 $&$ 88.80 $&$ 90.75 $&$ 90.95 $&$ 92.20 $&$ 91.95 $&$ 91.25 $&$ 91.80 $&$ 91.70 $&$ 92.25 $ \\ 
		$ T_0+ 3 $&$ 91.40 $&$ 91.25 $&$ 87.65 $&$ 87.75 $&$ 91.25 $&$ 90.90 $&$ 91.90 $&$ 91.05 $&$ 91.80 $&$ 91.85 $&$ 92.20 $&$ 91.85 $ \\ 
		$ T_0+ 4 $&$ 89.45 $&$ 89.50 $&$ 89.25 $&$ 89.45 $&$ 91.50 $&$ 91.25 $&$ 90.80 $&$ 91.15 $&$ 89.70 $&$ 89.95 $&$ 91.20 $&$ 91.20 $ \\ 
		$ T_0+ 5 $&$ 91.05 $&$ 90.90 $&$ 88.35 $&$ 88.30 $&$ 90.30 $&$ 90.25 $&$ 92.15 $&$ 92.35 $&$ 90.80 $&$ 90.45 $&$ 90.95 $&$ 91.00 $ \\ 
		\hline\hline 
		\multicolumn{13}{c}{ }\\ 
		\multicolumn{13}{c}{ }\\ 
	\end{tabular} 
	
	\begin{tabular}{c|cc|cc|cc|cc|cc|cc} 
		\multicolumn{13}{c}{Known number of factors, $95\%$ CI} \\ 
		\hline\hline 
		\multirow{2}{*}{\diagbox{$t$}{$\lp T_0, N_0 \rp$}} &\multicolumn{2}{c|}{$ (20, 30) $} &\multicolumn{2}{c|}{$ (20, 50) $} &\multicolumn{2}{c|}{$ (20, 100) $} &\multicolumn{2}{c|}{$ (40, 30) $} &\multicolumn{2}{c|}{$ (40, 50) $} &\multicolumn{2}{c}{$ (40, 100) $} \\ 
		& EQ & SY & EQ & SY & EQ & SY & EQ & SY & EQ & SY & EQ & SY \\ 
		\hline 
		$ T_0+ 1 $&$ 96.45 $&$ 96.05 $&$ 94.90 $&$ 94.65 $&$ 94.75 $&$ 93.95 $&$ 96.25 $&$ 95.75 $&$ 95.70 $&$ 95.10 $&$ 96.85 $&$ 96.85 $ \\ 
		$ T_0+ 2 $&$ 95.20 $&$ 95.20 $&$ 94.90 $&$ 94.50 $&$ 95.90 $&$ 95.05 $&$ 95.95 $&$ 95.35 $&$ 95.90 $&$ 96.40 $&$ 96.20 $&$ 95.85 $ \\ 
		$ T_0+ 3 $&$ 95.20 $&$ 95.10 $&$ 93.75 $&$ 94.15 $&$ 94.95 $&$ 95.25 $&$ 95.60 $&$ 95.25 $&$ 96.05 $&$ 95.75 $&$ 96.00 $&$ 96.15 $ \\ 
		$ T_0+ 4 $&$ 93.40 $&$ 93.55 $&$ 95.05 $&$ 95.15 $&$ 95.40 $&$ 95.30 $&$ 95.50 $&$ 95.20 $&$ 94.90 $&$ 94.20 $&$ 96.20 $&$ 95.30 $ \\ 
		$ T_0+ 5 $&$ 95.35 $&$ 95.15 $&$ 93.80 $&$ 93.80 $&$ 94.40 $&$ 94.25 $&$ 95.70 $&$ 95.10 $&$ 95.10 $&$ 95.25 $&$ 95.70 $&$ 95.20 $ \\ 
		\hline\hline 
		\multicolumn{13}{c}{ }\\ 
		\multicolumn{13}{c}{ }\\ 
	\end{tabular} 
	
	\begin{tabular}{c|cc|cc|cc|cc|cc|cc} 
		\multicolumn{13}{c}{Estimated number of factors, $90\%$ CI} \\ 
		\hline\hline 
		\multirow{2}{*}{\diagbox{$t$}{$\lp T_0, N_0 \rp$}} &\multicolumn{2}{c|}{$ (20, 30) $} &\multicolumn{2}{c|}{$ (20, 50) $} &\multicolumn{2}{c|}{$ (20, 100) $} &\multicolumn{2}{c|}{$ (40, 30) $} &\multicolumn{2}{c|}{$ (40, 50) $} &\multicolumn{2}{c}{$ (40, 100) $} \\ 
		& EQ & SY & EQ & SY & EQ & SY & EQ & SY & EQ & SY & EQ & SY \\ 
		\hline 
		$ T_0+ 1 $&$ 92.15 $&$ 92.05 $&$ 89.75 $&$ 89.85 $&$ 89.35 $&$ 90.65 $&$ 92.45 $&$ 92.40 $&$ 90.55 $&$ 90.45 $&$ 93.00 $&$ 93.50 $ \\ 
		$ T_0+ 2 $&$ 91.85 $&$ 92.40 $&$ 89.25 $&$ 89.35 $&$ 90.35 $&$ 91.10 $&$ 92.00 $&$ 92.45 $&$ 90.85 $&$ 92.00 $&$ 92.40 $&$ 92.10 $ \\ 
		$ T_0+ 3 $&$ 92.15 $&$ 92.05 $&$ 88.00 $&$ 88.00 $&$ 90.65 $&$ 90.85 $&$ 92.80 $&$ 92.20 $&$ 91.25 $&$ 91.55 $&$ 92.10 $&$ 91.55 $ \\ 
		$ T_0+ 4 $&$ 89.70 $&$ 89.85 $&$ 88.70 $&$ 89.20 $&$ 90.20 $&$ 89.95 $&$ 92.75 $&$ 92.95 $&$ 89.75 $&$ 90.30 $&$ 91.05 $&$ 90.55 $ \\ 
		$ T_0+ 5 $&$ 91.45 $&$ 91.40 $&$ 86.85 $&$ 87.30 $&$ 89.90 $&$ 89.80 $&$ 91.10 $&$ 91.55 $&$ 90.60 $&$ 90.80 $&$ 91.10 $&$ 91.05 $ \\ 
		\hline\hline 
		\multicolumn{13}{c}{ }\\ 
		\multicolumn{13}{c}{ }\\ 
	\end{tabular} 
	
	\begin{tabular}{c|cc|cc|cc|cc|cc|cc} 
		\multicolumn{13}{c}{Estimated number of factors, $95\%$ CI} \\ 
		\hline\hline 
		\multirow{2}{*}{\diagbox{$t$}{$\lp T_0, N_0 \rp$}} &\multicolumn{2}{c|}{$ (20, 30) $} &\multicolumn{2}{c|}{$ (20, 50) $} &\multicolumn{2}{c|}{$ (20, 100) $} &\multicolumn{2}{c|}{$ (40, 30) $} &\multicolumn{2}{c|}{$ (40, 50) $} &\multicolumn{2}{c}{$ (40, 100) $} \\ 
		& EQ & SY & EQ & SY & EQ & SY & EQ & SY & EQ & SY & EQ & SY \\ 
		\hline 
		$ T_0+ 1 $&$ 96.60 $&$ 95.95 $&$ 95.15 $&$ 95.25 $&$ 94.70 $&$ 94.75 $&$ 95.95 $&$ 95.45 $&$ 94.95 $&$ 94.65 $&$ 97.15 $&$ 96.55 $ \\ 
		$ T_0+ 2 $&$ 95.75 $&$ 95.80 $&$ 94.85 $&$ 94.80 $&$ 95.05 $&$ 94.45 $&$ 95.30 $&$ 95.30 $&$ 96.50 $&$ 96.00 $&$ 95.95 $&$ 95.90 $ \\ 
		$ T_0+ 3 $&$ 95.60 $&$ 95.55 $&$ 93.40 $&$ 93.25 $&$ 95.05 $&$ 94.70 $&$ 95.60 $&$ 95.35 $&$ 95.90 $&$ 95.90 $&$ 96.35 $&$ 96.30 $ \\ 
		$ T_0+ 4 $&$ 93.85 $&$ 93.80 $&$ 93.55 $&$ 93.95 $&$ 94.95 $&$ 94.90 $&$ 95.95 $&$ 95.95 $&$ 95.75 $&$ 94.70 $&$ 95.70 $&$ 95.05 $ \\ 
		$ T_0+ 5 $&$ 95.10 $&$ 94.95 $&$ 92.00 $&$ 92.10 $&$ 94.25 $&$ 93.65 $&$ 95.40 $&$ 95.20 $&$ 95.05 $&$ 94.00 $&$ 95.40 $&$ 95.25 $ \\ 
		\hline\hline 
	\end{tabular} 
	
\end{table} 

\begin{table}[H] 
	\footnotesize\centering\setstretch{1.5} 
	\caption{Factor model with covariates and heteroscedastic AR(1) uniform errors} 
	\label{tab:Factor model with covariates and heteroscedastic AR1 uniform errors} 
	\begin{tabular}{c|cc|cc|cc|cc|cc|cc} 
		\multicolumn{13}{c}{Known number of factors, $90\%$ CI} \\ 
		\hline\hline 
		\multirow{2}{*}{\diagbox{$t$}{$\lp T_0, N_0 \rp$}} &\multicolumn{2}{c|}{$ (20, 30) $} &\multicolumn{2}{c|}{$ (20, 50) $} &\multicolumn{2}{c|}{$ (20, 100) $} &\multicolumn{2}{c|}{$ (40, 30) $} &\multicolumn{2}{c|}{$ (40, 50) $} &\multicolumn{2}{c}{$ (40, 100) $} \\ 
		& EQ & SY & EQ & SY & EQ & SY & EQ & SY & EQ & SY & EQ & SY \\ 
		\hline 
		$ T_0+ 1 $&$ 93.70 $&$ 93.95 $&$ 92.25 $&$ 92.60 $&$ 89.55 $&$ 90.60 $&$ 93.05 $&$ 93.10 $&$ 92.45 $&$ 92.55 $&$ 90.60 $&$ 90.50 $ \\ 
		$ T_0+ 2 $&$ 93.30 $&$ 93.45 $&$ 92.80 $&$ 93.20 $&$ 89.50 $&$ 89.55 $&$ 93.35 $&$ 93.30 $&$ 92.10 $&$ 92.05 $&$ 92.30 $&$ 92.30 $ \\ 
		$ T_0+ 3 $&$ 94.35 $&$ 94.35 $&$ 93.95 $&$ 94.15 $&$ 89.75 $&$ 89.85 $&$ 93.80 $&$ 94.00 $&$ 93.45 $&$ 93.65 $&$ 91.30 $&$ 91.25 $ \\ 
		$ T_0+ 4 $&$ 94.25 $&$ 94.25 $&$ 90.85 $&$ 91.00 $&$ 92.65 $&$ 92.55 $&$ 93.30 $&$ 92.80 $&$ 93.25 $&$ 93.05 $&$ 93.25 $&$ 93.30 $ \\ 
		$ T_0+ 5 $&$ 94.95 $&$ 94.95 $&$ 90.60 $&$ 90.60 $&$ 90.30 $&$ 90.25 $&$ 92.80 $&$ 93.00 $&$ 91.40 $&$ 91.40 $&$ 92.00 $&$ 91.80 $ \\ 
		\hline\hline 
		\multicolumn{13}{c}{ }\\ 
		\multicolumn{13}{c}{ }\\ 
	\end{tabular} 
	
	\begin{tabular}{c|cc|cc|cc|cc|cc|cc} 
		\multicolumn{13}{c}{Known number of factors, $95\%$ CI} \\ 
		\hline\hline 
		\multirow{2}{*}{\diagbox{$t$}{$\lp T_0, N_0 \rp$}} &\multicolumn{2}{c|}{$ (20, 30) $} &\multicolumn{2}{c|}{$ (20, 50) $} &\multicolumn{2}{c|}{$ (20, 100) $} &\multicolumn{2}{c|}{$ (40, 30) $} &\multicolumn{2}{c|}{$ (40, 50) $} &\multicolumn{2}{c}{$ (40, 100) $} \\ 
		& EQ & SY & EQ & SY & EQ & SY & EQ & SY & EQ & SY & EQ & SY \\ 
		\hline 
		$ T_0+ 1 $&$ 97.50 $&$ 97.50 $&$ 96.95 $&$ 96.95 $&$ 95.70 $&$ 95.85 $&$ 98.20 $&$ 98.20 $&$ 96.95 $&$ 96.85 $&$ 95.90 $&$ 95.60 $ \\ 
		$ T_0+ 2 $&$ 98.10 $&$ 98.00 $&$ 97.40 $&$ 97.65 $&$ 96.05 $&$ 95.70 $&$ 97.50 $&$ 97.50 $&$ 96.80 $&$ 96.80 $&$ 96.50 $&$ 96.50 $ \\ 
		$ T_0+ 3 $&$ 97.90 $&$ 97.85 $&$ 97.25 $&$ 97.50 $&$ 95.95 $&$ 95.95 $&$ 96.95 $&$ 97.00 $&$ 97.25 $&$ 97.55 $&$ 96.60 $&$ 96.65 $ \\ 
		$ T_0+ 4 $&$ 98.15 $&$ 98.10 $&$ 96.35 $&$ 96.35 $&$ 96.85 $&$ 96.75 $&$ 97.50 $&$ 97.35 $&$ 97.80 $&$ 97.70 $&$ 97.40 $&$ 97.50 $ \\ 
		$ T_0+ 5 $&$ 97.95 $&$ 98.00 $&$ 96.70 $&$ 96.70 $&$ 96.00 $&$ 96.05 $&$ 97.65 $&$ 97.60 $&$ 97.20 $&$ 97.05 $&$ 96.70 $&$ 96.55 $ \\ 
		\hline\hline 
		\multicolumn{13}{c}{ }\\ 
		\multicolumn{13}{c}{ }\\ 
	\end{tabular} 
	
	\begin{tabular}{c|cc|cc|cc|cc|cc|cc} 
		\multicolumn{13}{c}{Estimated number of factors, $90\%$ CI} \\ 
		\hline\hline 
		\multirow{2}{*}{\diagbox{$t$}{$\lp T_0, N_0 \rp$}} &\multicolumn{2}{c|}{$ (20, 30) $} &\multicolumn{2}{c|}{$ (20, 50) $} &\multicolumn{2}{c|}{$ (20, 100) $} &\multicolumn{2}{c|}{$ (40, 30) $} &\multicolumn{2}{c|}{$ (40, 50) $} &\multicolumn{2}{c}{$ (40, 100) $} \\ 
		& EQ & SY & EQ & SY & EQ & SY & EQ & SY & EQ & SY & EQ & SY \\ 
		\hline 
		$ T_0+ 1 $&$ 92.10 $&$ 92.30 $&$ 91.45 $&$ 90.85 $&$ 91.10 $&$ 91.20 $&$ 93.15 $&$ 93.30 $&$ 91.80 $&$ 91.80 $&$ 91.10 $&$ 90.75 $ \\ 
		$ T_0+ 2 $&$ 92.40 $&$ 92.30 $&$ 93.30 $&$ 93.35 $&$ 89.60 $&$ 89.55 $&$ 92.10 $&$ 92.10 $&$ 90.25 $&$ 90.30 $&$ 92.50 $&$ 92.85 $ \\ 
		$ T_0+ 3 $&$ 92.75 $&$ 92.85 $&$ 91.85 $&$ 92.25 $&$ 88.90 $&$ 88.95 $&$ 92.55 $&$ 92.55 $&$ 93.25 $&$ 93.25 $&$ 90.50 $&$ 90.20 $ \\ 
		$ T_0+ 4 $&$ 93.50 $&$ 93.40 $&$ 90.25 $&$ 90.20 $&$ 91.45 $&$ 91.45 $&$ 93.55 $&$ 93.90 $&$ 93.10 $&$ 93.10 $&$ 92.85 $&$ 92.80 $ \\ 
		$ T_0+ 5 $&$ 92.25 $&$ 92.35 $&$ 90.35 $&$ 90.45 $&$ 90.20 $&$ 90.00 $&$ 93.45 $&$ 93.80 $&$ 91.50 $&$ 91.50 $&$ 91.85 $&$ 91.65 $ \\ 
		\hline\hline 
		\multicolumn{13}{c}{ }\\ 
		\multicolumn{13}{c}{ }\\ 
	\end{tabular} 
	
	\begin{tabular}{c|cc|cc|cc|cc|cc|cc} 
		\multicolumn{13}{c}{Estimated number of factors, $95\%$ CI} \\ 
		\hline\hline 
		\multirow{2}{*}{\diagbox{$t$}{$\lp T_0, N_0 \rp$}} &\multicolumn{2}{c|}{$ (20, 30) $} &\multicolumn{2}{c|}{$ (20, 50) $} &\multicolumn{2}{c|}{$ (20, 100) $} &\multicolumn{2}{c|}{$ (40, 30) $} &\multicolumn{2}{c|}{$ (40, 50) $} &\multicolumn{2}{c}{$ (40, 100) $} \\ 
		& EQ & SY & EQ & SY & EQ & SY & EQ & SY & EQ & SY & EQ & SY \\ 
		\hline 
		$ T_0+ 1 $&$ 97.20 $&$ 97.20 $&$ 95.25 $&$ 95.60 $&$ 96.25 $&$ 96.25 $&$ 97.90 $&$ 97.90 $&$ 95.90 $&$ 95.90 $&$ 96.15 $&$ 96.10 $ \\ 
		$ T_0+ 2 $&$ 97.80 $&$ 98.15 $&$ 96.90 $&$ 96.85 $&$ 95.85 $&$ 95.70 $&$ 96.85 $&$ 96.80 $&$ 95.85 $&$ 95.90 $&$ 96.80 $&$ 96.95 $ \\ 
		$ T_0+ 3 $&$ 98.10 $&$ 98.25 $&$ 96.80 $&$ 96.95 $&$ 94.70 $&$ 94.70 $&$ 96.85 $&$ 96.80 $&$ 96.65 $&$ 96.85 $&$ 95.95 $&$ 96.40 $ \\ 
		$ T_0+ 4 $&$ 97.70 $&$ 97.60 $&$ 95.85 $&$ 95.80 $&$ 95.75 $&$ 96.00 $&$ 97.85 $&$ 97.55 $&$ 97.10 $&$ 97.10 $&$ 97.25 $&$ 97.35 $ \\ 
		$ T_0+ 5 $&$ 96.80 $&$ 96.80 $&$ 96.75 $&$ 96.85 $&$ 95.80 $&$ 95.75 $&$ 97.40 $&$ 97.45 $&$ 96.50 $&$ 96.50 $&$ 96.55 $&$ 96.60 $ \\ 
		\hline\hline 
	\end{tabular} 
	
\end{table} 

\section{Empirical Applications}

\label{sec:Empirical Application}
In this section, we re-evaluate the impacts of Hong Kong's Political and Economic Integration with Mainland
China as well as the effects of California's Tobacco Control Program using our proposed bootstrap procedure.
\footnote{MATLAB, Python and R codes for application are available from the authors upon request.}

\subsection{Hong Kong's Political and Economic Integration with Mainland
	China Revisited}

In this subsection, we revisit the impacts of political and economic integration of Hong
Kong with Mainland China, which has been analysed in \citet{hsiao2012panel}.
Since \citet{hsiao2012panel} specify a pure factor model without covariates,
we apply the methods in Section \ref{sec:Estimation and Inference in a Model
	without Covariates} of this paper to the dataset of \citet{hsiao2012panel}. 
For the results about political integration, quarterly real GDP growth rates
from 1993Q1 to 1997Q2 of 10 countries and districts are used to form the
counter-factual path of Hong Kong from 1997Q3 up to 2003Q4. The 10 countries
and districts are Mainland China, Indonesia, Japan, Korea, Malaysia,
Philippines, Singapore, Taiwan, Thailand and US. For the analysis of
economic integration, quarterly real GDP growth rates of 24 countries and
districts from 1993Q1 to 2003Q4 are used to form counter-factual path of
Hong Kong from 2004Q1 to 2008Q1.

\begin{figure}[ht]
	\centering
	\includegraphics[width=7.5cm]{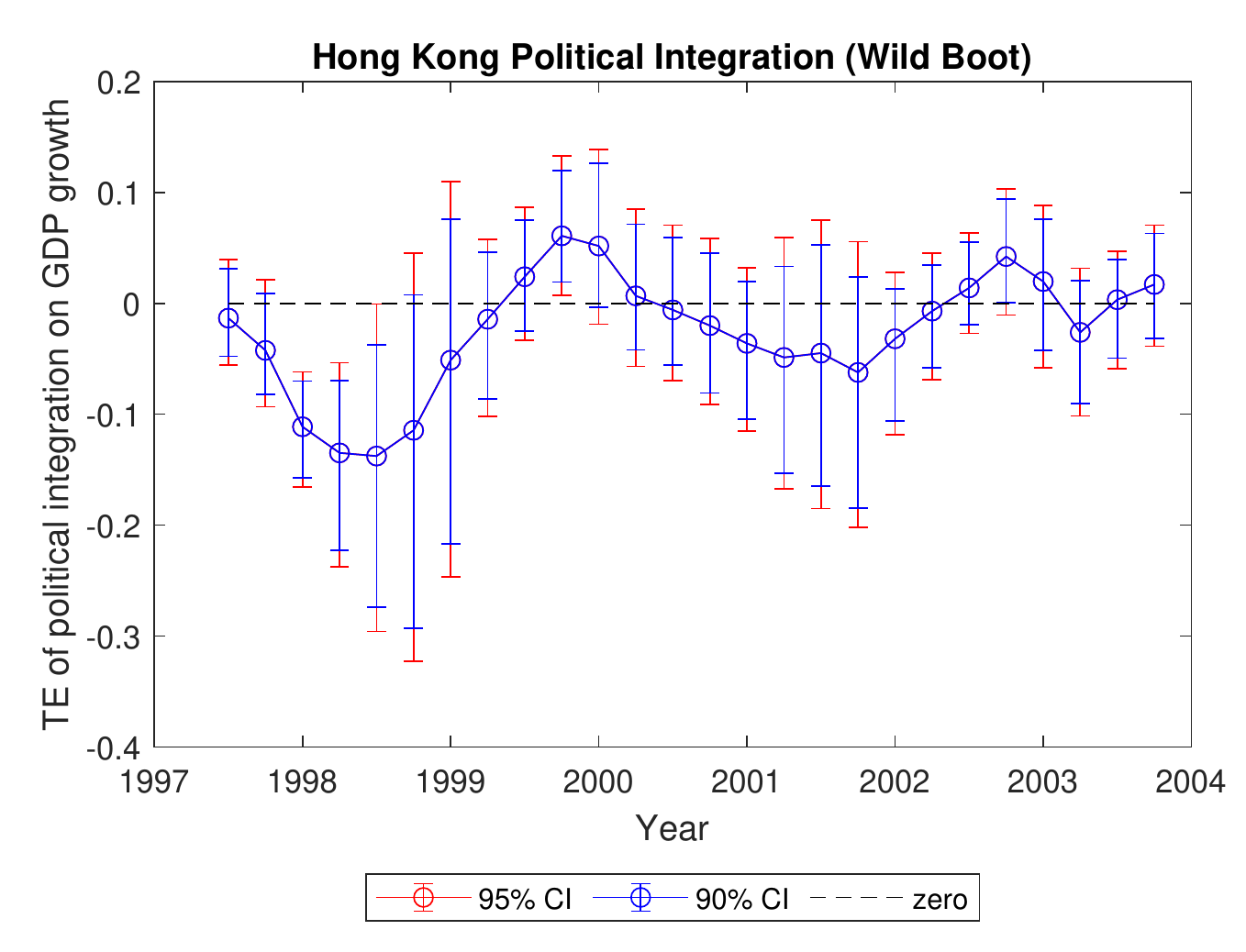} %
	\includegraphics[width=7.5cm]{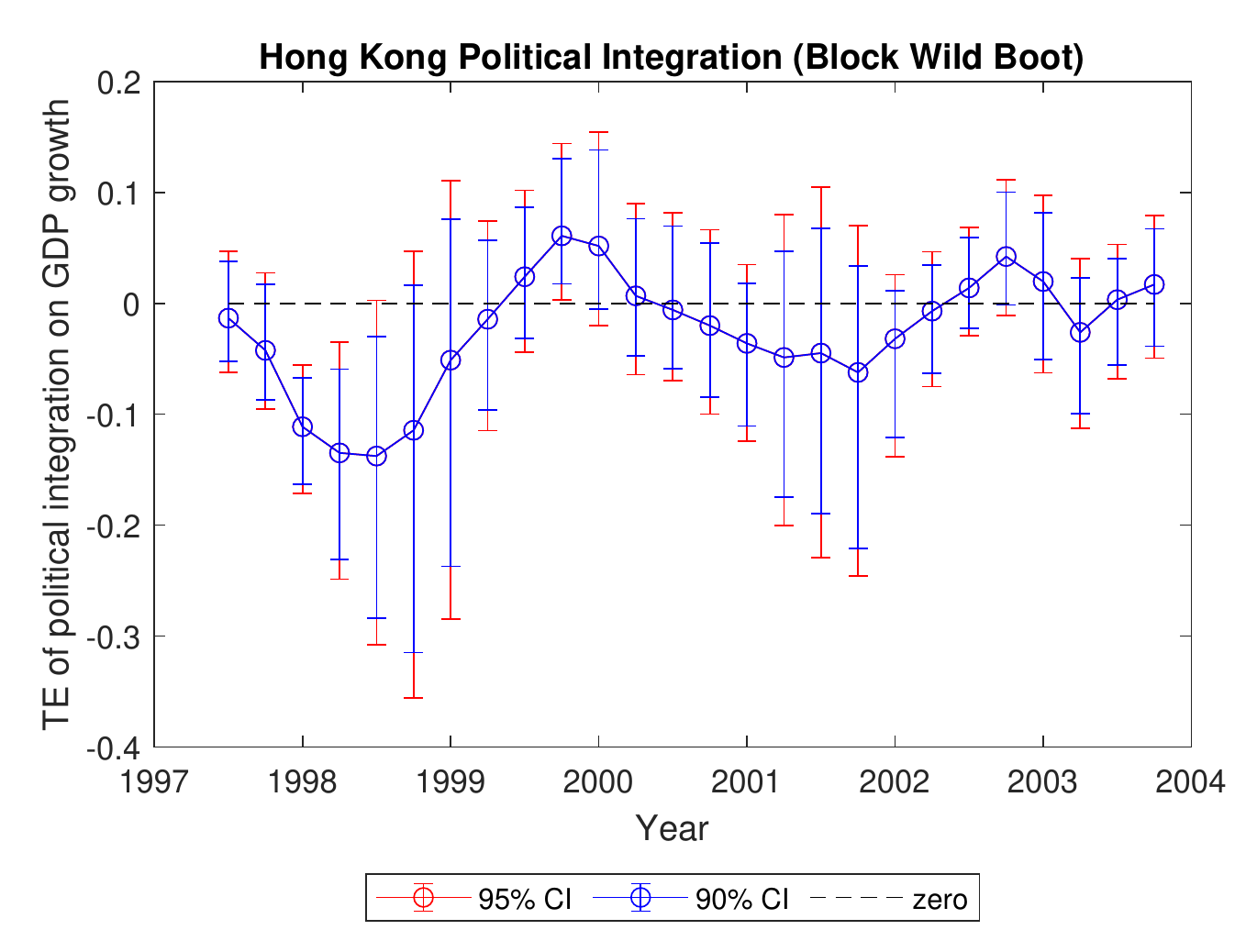}
	\caption{Impact of Political Integration with Mainland China on Hong Kong
		Economic Growth}
	\label{fig:HKPI}
\end{figure}

\begin{figure}[ht]
	\centering
	\includegraphics[width=7.5cm]{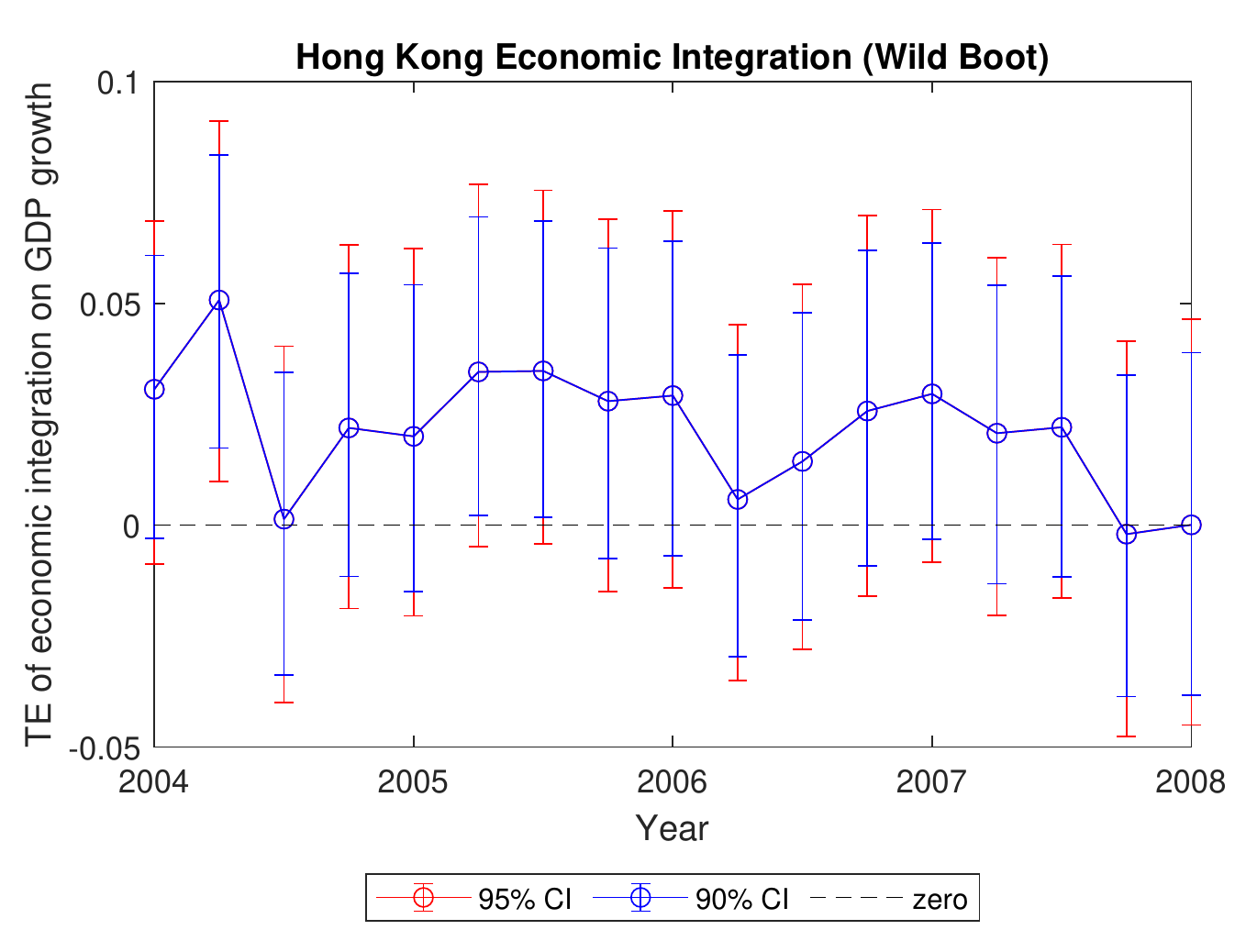} %
	\includegraphics[width=7.5cm]{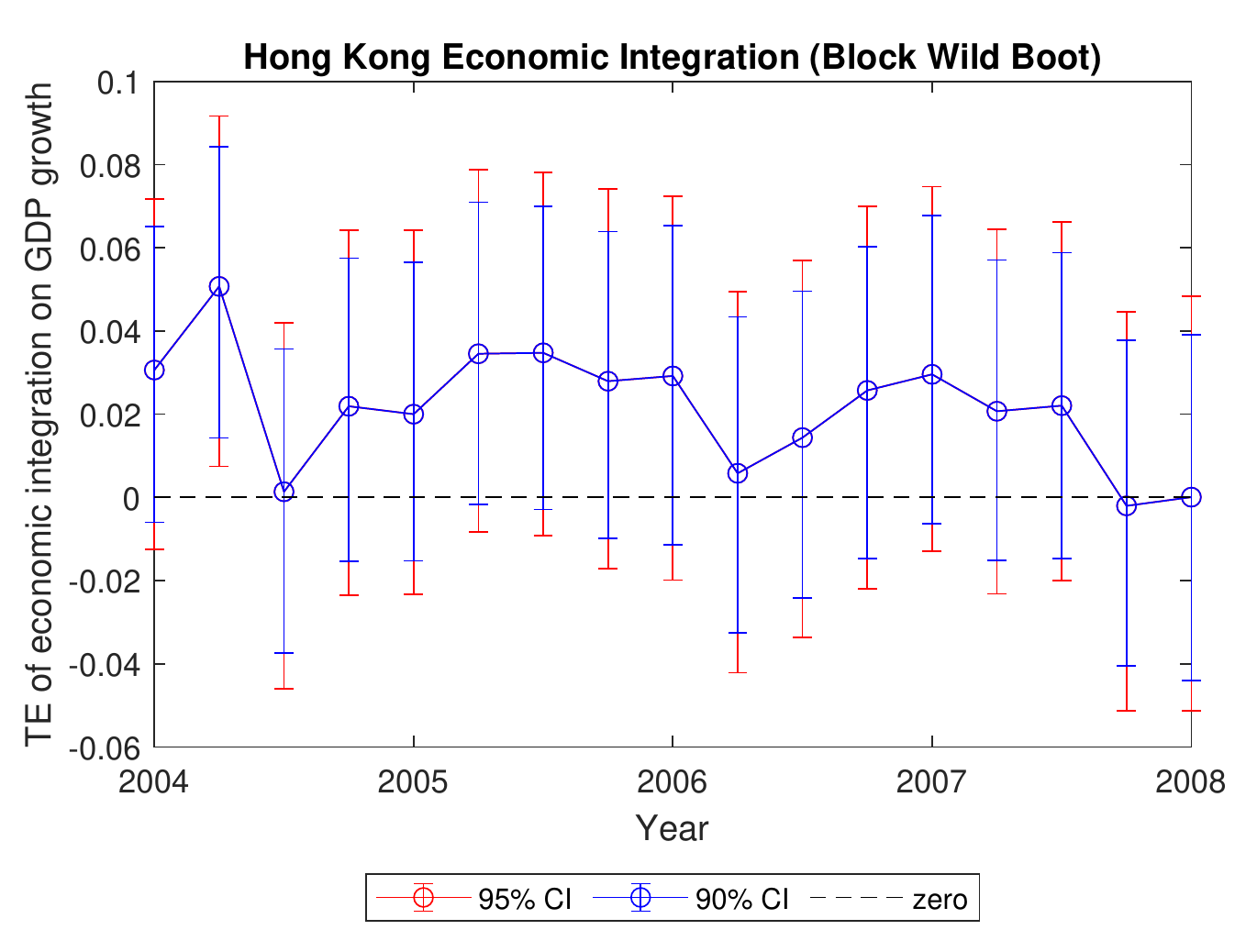}
	\caption{Impact of Economic Integration with Mainland China on Hong Kong
		Economic Growth}
	\label{fig:HKEI}
\end{figure}

The results for the impact of political integration and economic integration with Mainland China
on Hong Kong's economic growth are provided in \ref{fig:HKPI} and \ref{fig:HKEI}, respectively.

As is shown in Figure \ref{fig:HKPI}, the estimated treatment effects of
Hong Kong political integration are of similar magnitudes and patterns as
those in \citet{hsiao2012panel}. In Figure \ref{fig:HKEI}, the estimated
treatment effects of Hong Kong economic integration are positive, as in %
\citet{hsiao2012panel}, and confidence intervals cover 0 in most of the
periods. This suggests that the impact of political and economic integration
of Hong Kong with Mainland China is significant in the first few years after
integration, and vanishes afterward. This observation is in general
consistent with the insignificant average treatment effects across
post-treatment periods in \citet{hsiao2012panel}.

\subsection{California's Tobacco Control Program Revisited}
Now we revisit the effectiveness of CTCP on per capita cigarettes consumption and
personal healthcare expenditures using the methods discussed in Section 4 of this article.
In November 1988, California passed the Proposition 99, which increased
California's cigarettes tax by 25 cents per pack and earmarked the tax
revenue for health and anti-smoking measures. Proposition 99 triggered a
wave of local clean-air ordinances in California. \citet{abadie2010synthetic}
used the synthetic control method for the period 1970-2000 to show that the
California Tobacco Control Program had a significant impact on per capita
cigarette consumption for the period 1989-2000, and that its impact
continued to be enhanced over time.

Note that in the model and dataset of \citet{abadie2010synthetic}, covariates do not change over time, which
violates Assumption \ref{ass:covariates x}(2) of this study. To accommodate
time-variant covariates, we use the dataset of \citet{hsiao2019panel}, who
also revisit the impact of CTCP on per capita cigarettes consumption and
personal healthcare expenditures but use a set of time-variant covariates:
per capita GDP obtained from \citet{abadie2010synthetic};
poverty rates obtained from the National Census Bureau;
educational attainment, defined as the percentage of obtaining college
degree of population 25 years and over, obtained from the National Census
Bureau.

\begin{figure}[ht]
\centering
\includegraphics[width=7.5cm]{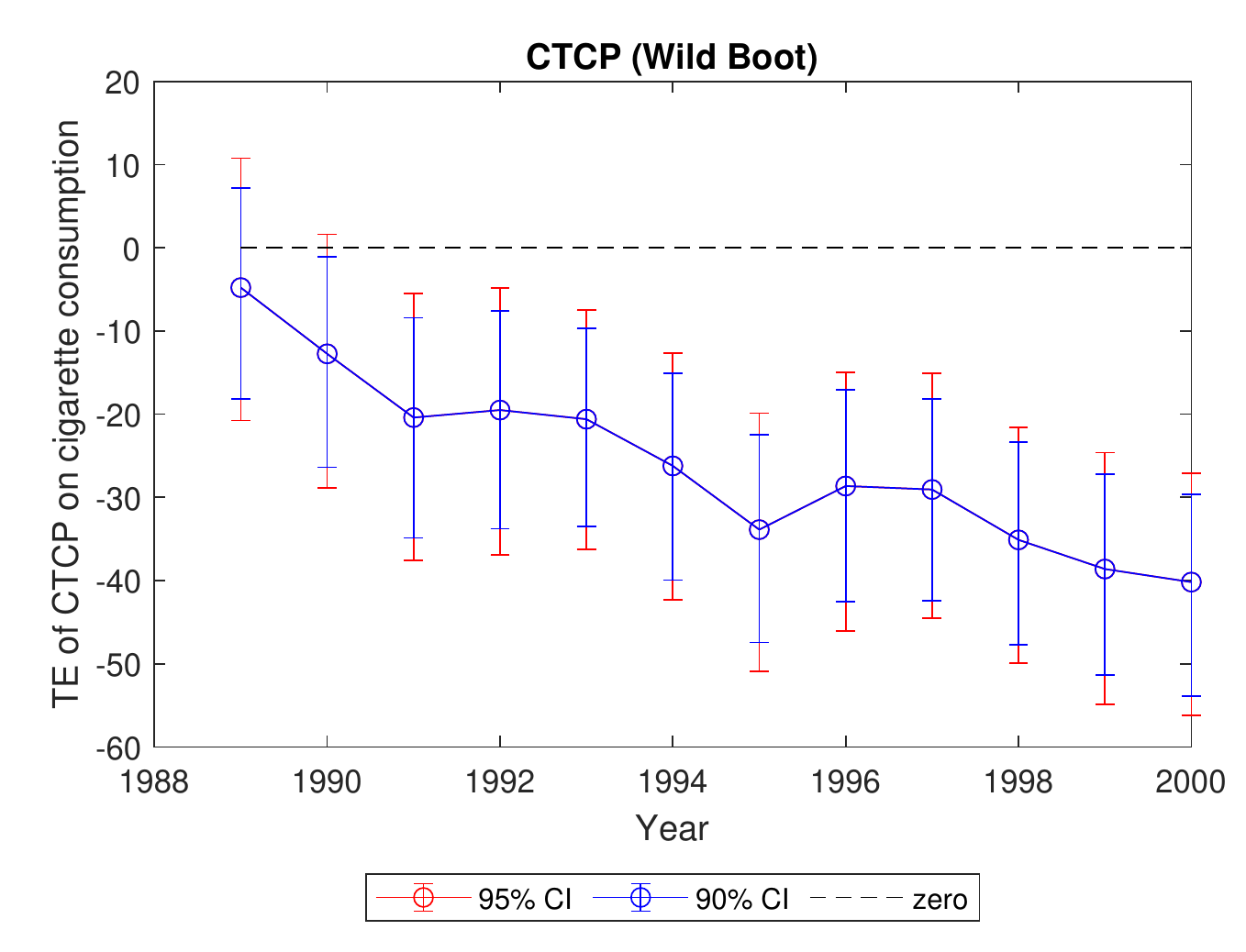} %
\includegraphics[width=7.5cm]{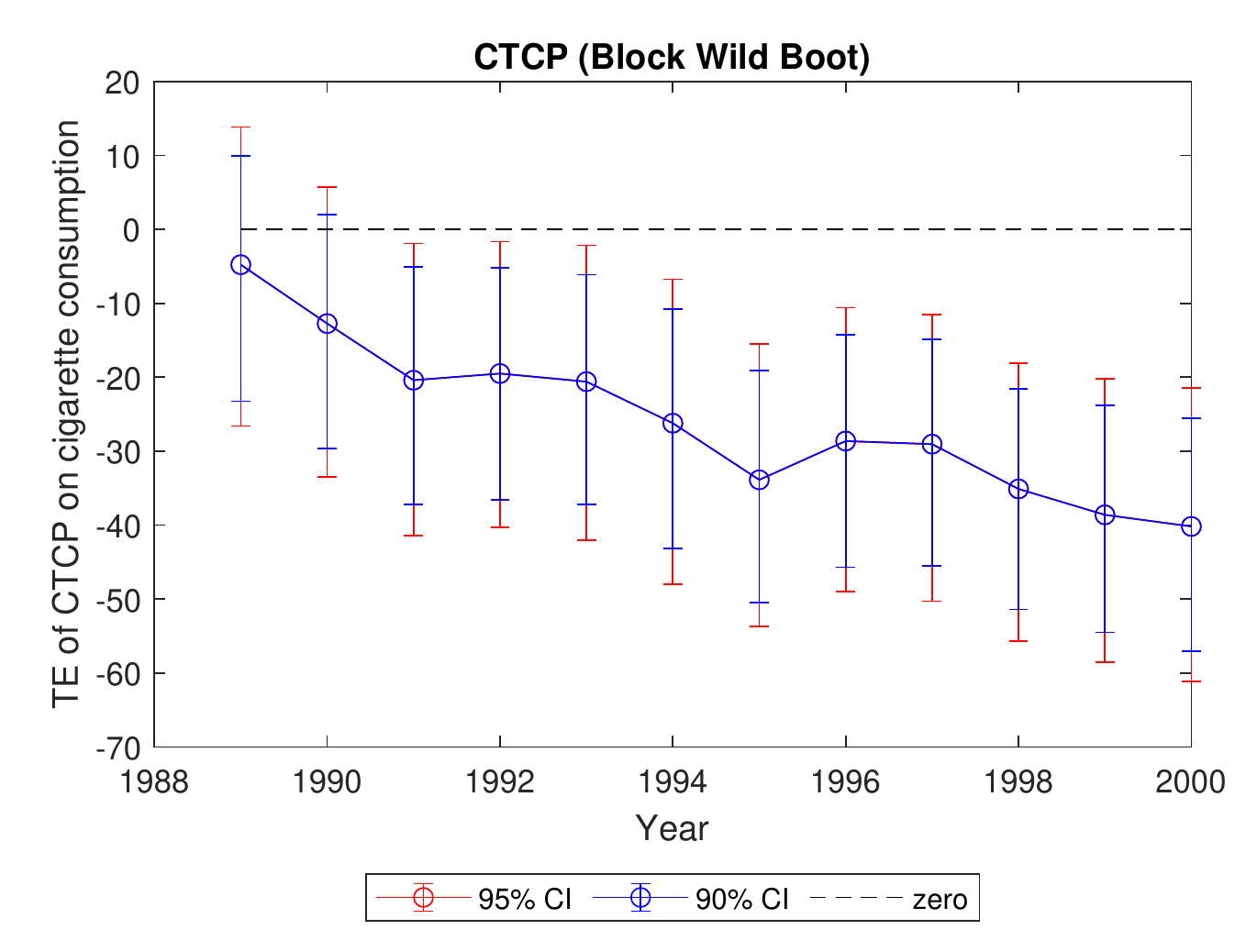}
\caption{Impact of CTCP on per capita cigarette consumption}
\label{fig:CTCP cig}
\end{figure}

\begin{figure}[ht]
\centering
\includegraphics[width=7.5cm]{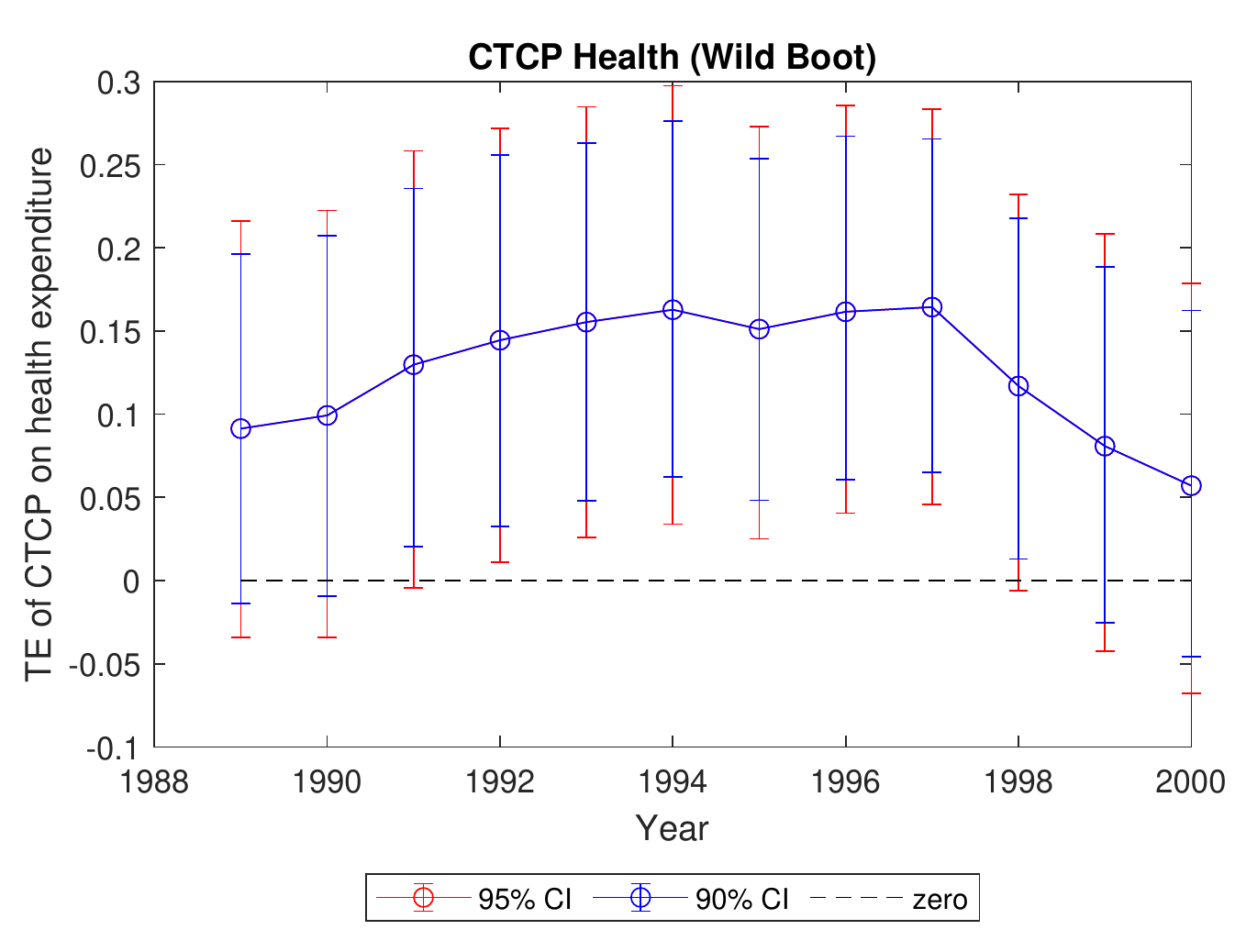} %
\includegraphics[width=7.5cm]{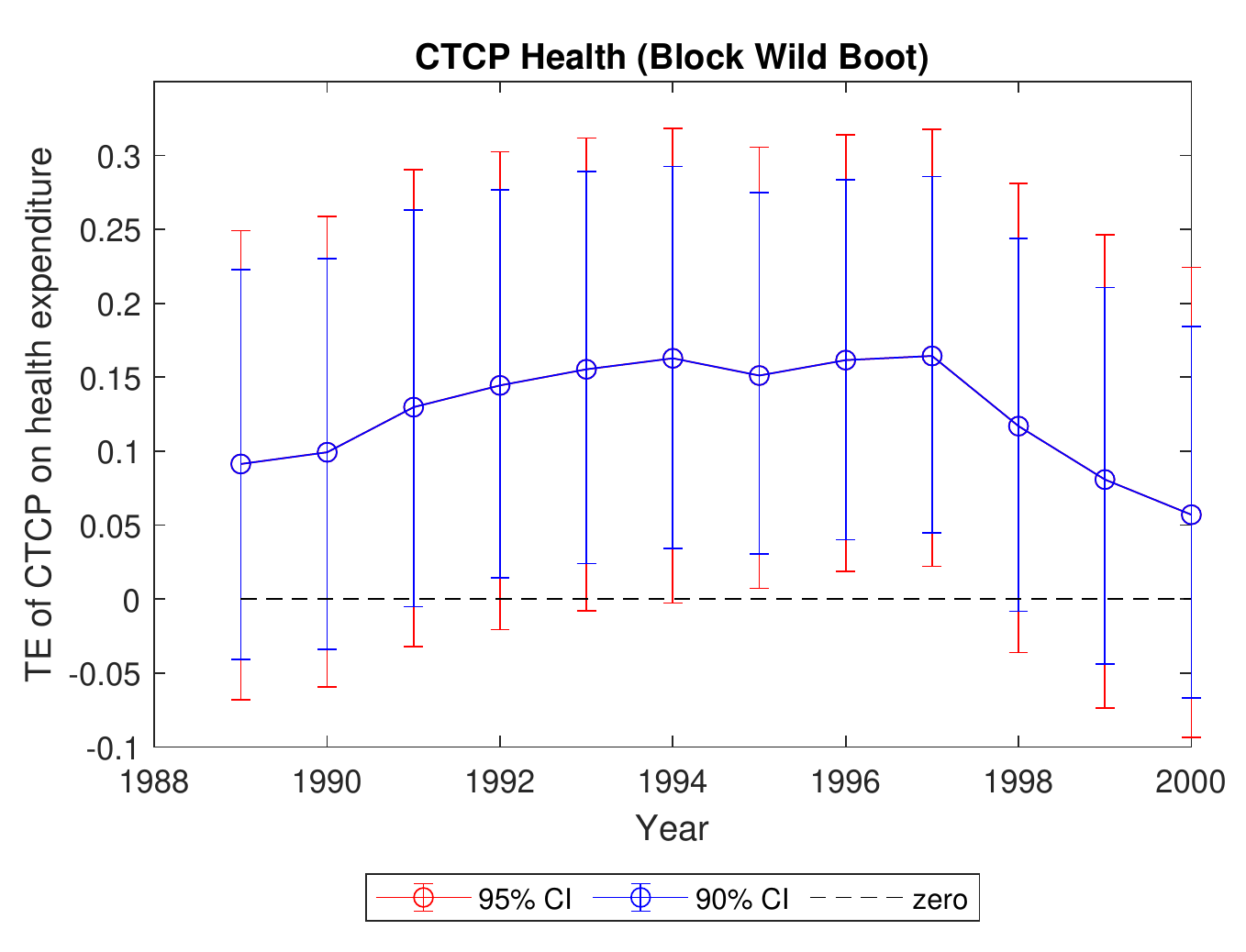}
\caption{Impact of CTCP on personal healthcare expenditures}
\label{fig:CTCP health}
\end{figure}

In our analysis, the number of factors is estimated using the method proposed by %
\citet{alessi2010improved}, which shows better performance than %
\citet{bai2002determining} in this and the next applications. Both ordinary
wild bootstrap procedure and block wild bootstrap procedure with block width
equal to 3 are considered, and equal tailed confidence intervals are
reported. The results for the impact of CTCP on per capita cigarette consumption and personal 
healthcare expenditures are provided in \ref{fig:CTCP cig} and \ref{fig:CTCP health}, respectively.

Figure \ref{fig:CTCP cig} shows that the estimated treatment
effects are of similar magnitudes as those in \citet{abadie2010synthetic};
and that the confidence intervals indicate negative and significant impacts over
time, consistent with their findings using permutation tests. Figure \ref{fig:CTCP health} shows that the estimated treatment effects of CTCP on
health expenditure are of similar magnitudes as reported in %
\citet{hsiao2019panel}. Consistent with their findings, confidence intervals
indicate that the effects are short-lived. In both figures, we can see that
the confidence intervals based on ordinary and block wild bootstrap
procedures produce similar results.

\section{Conclusion}

\label{sec:Conclusion}

In this paper, we consider the construction of confidence intervals for
treatment effects estimated in panel models with interactive fixed effects,
which serves as an alternative inferential approach. We first use the
factor-based matrix completion technique proposed by \citet{bai2021matrix}
for panel models to estimate the treatment effects, and then use bootstrap
method to construct confidence intervals of the treatment effects for
treated units at each post-treatment period. Our construction of confidence
intervals requires neither specific distributional assumptions on the error
terms nor large number of post-treatment periods. We also establish the
validity of proposed bootstrap procedure that these confidence intervals
have asymptotically correct coverage probabilities. Simulation studies show
that these confidence intervals have satisfactory finite sample
performances, and empirical applications using classical datasets yield
treatment effect estimates of similar magnitude and reliable confidence
intervals.

\bibliography{references}


\pagebreak

\begin{center}		
	\Large{\bf Confidence Intervals of Treatment Effects in Panel Data	Models 
		with Interactive Fixed Effects\\ Online Supplementary Appendices}\\
	[0.75cm]
	
	\large{	\begin{tabular}{cc}
			Xingyu Li  & Yan Shen \\
			National School of Development  & National School of Development \\
			Peking University  & Peking University \\
			x.y@pku.edu.cn  & yshen@nsd.pku.edu.cn \\
		\end{tabular}
	}\\	
	[0.75cm]
		
	\large{Qiankun Zhou\\Department of Economics\\Louisiana State University\\qzhou@lsu.edu} \\
	\bigskip
	\today
	
\end{center}
\bigskip

\begin{appendices}

The online appendices include auxiliary lemmas, technical notes,
and the proofs of Theorems \ref{thry:coverage} and \ref{thry:cov coverage}.
We first list some results for estimators and the associated residuals in panel models with interactive
effects, and then provide the technical properties of bootstrap procedure.
Finally, we use the above mentioned results to prove the theorems in the main paper.

\section{Properties of Estimators and Residuals}

The lemmas in this section are cited or derived from some of the results in %
\citet{bai2003inferential}, \citet{bai2009panel}, %
\citet{gonccalves2014bootstrapping}, and \citet{bai2021matrix}. Note that
Assumptions \ref{ass:factors and loadings}--\ref{ass:error is independent of
factor and x} of this paper are sufficient for Assumptions A--G of %
\citet{bai2003inferential}, Assumptions A--E of \citet{bai2009panel}, panel
factor model relevant conditions of Assumptions 1--5 of %
\citet{gonccalves2014bootstrapping}, and Assumptions A--D of %
\citet{bai2021matrix}.

\subsection{Estimators and Residuals in Section \ref{sec:Estimation
and Inference in a Model without Covariates}}

\label{sec:Properties of the Estimators and Residuals pure factor}

We introduce some notations. Let $\Vhattall=\Dtall^2$ and $\Vhatwide=\Dwide^2
$. Define rotation matrices
\begin{align*}
\Htall=\left( \frac{ \Ltall^\mathsf{T} \Ltall}{N_0} \right) \left( \frac{%
\Ftall^\mathsf{T} \Fhattall}{T} \right) \Vhattall^{-1}, \qquad \Hwide=\left(
\frac{ \Lwide^\mathsf{T} \Lwide}{N} \right) \left( \frac{\Fwide^\mathsf{T} %
\Fhatwide}{T_0} \right) \Vhatwide^{-1}.
\end{align*}

\begin{lemma}
\label{lemma:pointwise convergence of factors and loadings in sec3} If
Assumptions \ref{ass:factors and loadings}--\ref{ass:order conditions} hold,
then as $N_0, T_0\to\infty$,

\begin{enumerate}[label=(\arabic*), labelindent=\parindent, leftmargin=*, nosep]

\item $\Vhattall\xrightarrow{\mathbb{P}} V$ and $\Vhatwide%
\xrightarrow{\mathbb{P}} V$, where $V$ is the diagonal matrix consisting of
the eigenvalues of $\Sigma_{\Lambda}\Sigma_{F}$, or equivalently, the
eigenvalues of $\Sigma_{\Lambda}^{1/2}\Sigma_{F}\Sigma_{\Lambda}^{1/2}$.

\item $\Htall\xrightarrow{\mathbb{P}} \Sigma_{\Lambda}^{1/2} U V^{-1/2}$, $%
\Hwide\xrightarrow{\mathbb{P}} \Sigma_{\Lambda}^{1/2} U V^{-1/2}$, $\Htall%
^{-1} \xrightarrow{\mathbb{P}} V^{1/2} U^\mathsf{T} \Sigma_{\Lambda}^{-1/2}$%
, and $\Hwide^{-1} \xrightarrow{\mathbb{P}} V^{1/2} U^\mathsf{T}
\Sigma_{\Lambda}^{-1/2}$, where $V$ is the diagonal matrix consisting of the
eigenvalues of $\Sigma_{\Lambda}^{1/2}\Sigma_{F}\Sigma_{\Lambda}^{1/2}$, and
$U$ is the corresponding eigenvector matrix such that $\Sigma_{%
\Lambda}^{1/2}\Sigma_{F}\Sigma_{\Lambda}^{1/2} U =UV$.

\item $\fhattallt-\Htall^\mathsf{T} f_t = O_\mathbb{P}\left( \dfrac{1}{\sqrt{%
N_0}} \right)$ for every $t\in \{1,\ldots, T\}$.

\item $\lhatwidei-\Hwide^{-1}\lambda_{i} = O_\mathbb{P} \left( \dfrac{1}{%
\sqrt{T_0}} \right)$ for every $i\in \{ 1,\ldots, N\}$.

\item $\Hhatmiss-\Htall^{-1}\Hwide=O_\mathbb{P} \left( \dfrac{1}{N_0\wedge
T_0} \right)$, and $\Hhatmiss\xrightarrow{\mathbb{P}} I_r $.

\item
\begin{align*}
\frac{1}{T}\sum_{t=1}^T \left\Vert \fhattallt-\Htall^\mathsf{T} f_t
\right\Vert^2 =O_\mathbb{P}\left( \frac{1}{N_0 \wedge T} \right), \qquad
\frac{1}{T_0}\sum_{t=1}^{T_0} \left\Vert \fhatwidet-\Hwide^\mathsf{T} f_t
\right\Vert^2 =O_\mathbb{P}\left( \frac{1}{N\wedge T_0} \right).
\end{align*}
\end{enumerate}
\end{lemma}

\begin{pf}{}
Claim (1) is obtained by Lemma A.3(1) of \citet{bai2003inferential} (applied
to the control subsample and pre-treatment subsample) and Assumption \ref%
{ass:factors and loadings}(3)(4) of this paper. Claim (2) follows from
Proposition of \citet{bai2003inferential} (applied to the control subsample
and pre-treatment subsample), Assumption \ref{ass:factors and loadings}%
(3)(4) of this paper, and the continuous mapping theorem. Note that
Assumption \ref{ass:order conditions}(2) implies $\left. \sqrt{N_0} \right/
T \to 0$ and $\left. \sqrt{T_0} \right/ N \to 0$ as $N_0, T_0\to\infty$.
Then applying Theorem 1(i) to the control subsample yields Claim (3), and
applying Theorem 2(i) to the pre-treatment subsample yields Claim (4). The
first part of Claim (5) is borrowed from Lemma A.1(i) of %
\citet{bai2021matrix}, and the second part is by the first part and Claim
(2). Claim (6) is obtained by applying Lemma A.1 of %
\citet{bai2003inferential} to the control subsample and pre-treatment
subsample.
\end{pf}

For ease of citation, we state the $C_p$ inequality here. For $k,m\in \bZ_+$
and $\left\{ z_j \right\}_{j=1}^k\subset \mathbb{R}$, we have
\begin{align*}
\left\vert \sum_{j=1}^k z_j \right\vert^m \le k^{m-1} \sum_{j=1}^k
\left\vert z_j \right\vert^m.
\end{align*}

\begin{lemma}
\label{lemma:MSE between tall and wide loadings in sec3} If Assumptions \ref%
{ass:factors and loadings}--\ref{ass:order conditions} hold, then as $N_0,
T_0\to\infty$,
\begin{align*}
\frac{1}{N_0}\sum_{i=1}^{N_0} \left\Vert \lhattalli-\Htall^{-1}\Hwide%
\lhatwidei \right\Vert^2 = O_\mathbb{P} \left( \frac{1}{T_0} \right).
\end{align*}
\end{lemma}

\begin{pf}{}
Let $\ewidei=\left( e_{i,1}, \ldots, e_{i,T_0} \right)^\mathsf{T}$ for every $%
i\in\{1,\ldots, N\}$. By the proof of Theorem 2 of \citet{bai2003inferential}
(applied to the pre-treatment subsample), for every $i\in\{1,\ldots, N\}$,
we have the following decomposition
\begin{align*}
\lhatwidei=\Hwide^{-1}\lambda_{i}+b_{1,i}+b_{2,i}+b_{3,i},
\end{align*}
where
\begin{align*}
b_{1,i} &= \frac{1}{T_0} \left( \Fhatwide -\Fwide \Hwide \right)^\mathsf{T} %
\ewidei, \qquad \qquad b_{2,i} = \frac{1}{T_0} \Hwide \Fwide^\mathsf{T} %
\ewidei, \\
b_{3,i} &= \frac{1}{T_0} \Fhatwide^\mathsf{T} \left( \Fwide-\Fhatwide \Hwide%
^{-1} \right) \lambda_{i}.
\end{align*}

By the triangle inequality and Cauchy-Schwartz inequality,
\begin{align*}
\left\Vert b_{1,i} \right\Vert &=\left\Vert \frac{1}{T_0}\sum_{s=1}^{T_0}
\left( \fhatwides-\Hwide^\mathsf{T} f_s \right) e_{i,s} \right\Vert \le
\frac{1}{T_0} \sum_{s=1}^{T_0} \left\Vert \fhatwides-\Hwide^\mathsf{T} f_s
\right\Vert \left\vert e_{i,s} \right\vert \\
& \le \sqrt{\frac{1}{T_0}\sum_{s=1}^{T_0} \left\Vert \fhatwides-\Hwide^%
\mathsf{T} f_s \right\Vert^2} \sqrt{\frac{1}{T_0}\sum_{s=1}^{T_0} e_{i,s}^2}.
\end{align*}
By Assumption \ref{ass:moments of idiosyncratic errors}(1),
\begin{align*}
\mathbb{E} \left( \left\vert \frac{1}{N T_0} \sum_{i=1}^N \sum_{s=1}^{T_0}
e_{i,s}^2 \right\vert \right) =\frac{1}{N T_0} \sum_{i=1}^N \sum_{s=1}^{T_0}
\mathbb{E} \left( e_{i,s}^2 \right) \le M+1.
\end{align*}
Then Markov's inequality implies that $\displaystyle\frac{1}{N T_0}
\sum_{i=1}^N \sum_{s=1}^{T_0} e_{i,s}^2 = O_\mathbb{P} (1)$. By Lemma \ref%
{lemma:pointwise convergence of factors and loadings in sec3}(6) and
Assumption \ref{ass:order conditions}(2),
\begin{align*}
\frac{1}{N}\sum_{i=1}^N \left\Vert b_{1,i} \right\Vert^2 \le \left( \frac{1}{%
T_0}\sum_{s=1}^{T_0} \left\Vert \fhatwides-\Hwide^\mathsf{T} f_s
\right\Vert^2 \right) \left( \frac{1}{N T_0} \sum_{i=1}^N \sum_{s=1}^{T_0}
e_{i,s}^2 \right) =O_\mathbb{P}\left( \frac{1}{T_0} \right).
\end{align*}

From Assumption \ref{ass:moments of idiosyncratic errors}(5),
\begin{align*}
\mathbb{E} \left( \left\vert \frac{1}{N}\sum_{i=1}^N \left\Vert \frac{1}{T_0}%
\sum_{s=1}^{T_0} f_s e_{i,s} \right\Vert^2 \right\vert \right) =\frac{1}{T_0}
\mathbb{E} \left( \frac{1}{N} \sum_{i=1}^N \left\Vert \frac{1}{\sqrt{T_0}}%
\sum_{s=1}^{T_0} f_s e_{i,s} \right\Vert^2 \right) \le \frac{M}{T_0}.
\end{align*}
By the properties of matrix norms, the triangle inequality, Lemma \ref%
{lemma:pointwise convergence of factors and loadings in sec3}(2) and
Markov's inequality,
\begin{align*}
\frac{1}{N}\sum_{i=1}^N \left\Vert b_{2,i} \right\Vert^2 = \frac{1}{N}
\sum_{i=1}^N \left\Vert \frac{1}{T_0} \Hwide \Fwide^\mathsf{T} \ewidei%
\right\Vert^2 \le \left\Vert \Hwide \right\Vert^2 \left( \frac{1}{N}
\sum_{i=1}^N \left\Vert \frac{1}{T_0}\sum_{s=1}^{T_0} f_s e_{i,s}
\right\Vert^2 \right)= O_\mathbb{P} \left( \frac{1}{T_0} \right).
\end{align*}

Applying Lemma B.3 of \citet{bai2003inferential} to the pre-treatment
subsample yields
\begin{align*}
\frac{1}{T_0} \Fhatwide^\mathsf{T} \left( \Fwide \Hwide-\Fhatwide \right) =O_%
\mathbb{P} \left( \frac{1}{N\wedge T_0} \right).
\end{align*}
Then by the definition and property of Frobenius norm, Lemma \ref%
{lemma:pointwise convergence of factors and loadings in sec3}(2), Assumption %
\ref{ass:factors and loadings}(4) and \ref{ass:order conditions}(2),
\begin{align*}
\frac{1}{N}\sum_{i=1}^N \left\Vert b_{3,i} \right\Vert^2 & \le \frac{1}{N}%
\sum_{i=1}^N \left\Vert \frac{1}{T_0} \Fhatwide^\mathsf{T} \left( \Fwide %
\Hwide-\Fhatwide \right) \right\Vert^2 \left\Vert \Hwide^{-1} \right\Vert^2
\left\Vert \lambda_i \right\Vert^2 \\
&=\left\Vert \frac{1}{T_0}\Fhatwide^\mathsf{T} \left( \Fwide \Hwide-%
\Fhatwide \right) \right\Vert^2 \tr \left( \frac{\Lambda^\mathsf{T} \Lambda}{%
N} \right)=O_\mathbb{P} \left( \frac{1}{T_0^2} \right).
\end{align*}

Therefore, by the properties of matrix norms, the $C_p$ inequality and Lemma %
\ref{lemma:pointwise convergence of factors and loadings in sec3}(2),
\begin{align*}
\frac{1}{N}\sum_{i=1}^N \left\Vert \Hwide \lhatwidei -\lambda_{i}
\right\Vert^2 &\le \left\Vert \Hwide \right\Vert^2 \left( \frac{1}{N}%
\sum_{i=1}^N \left\Vert b_{1,i}+b_{2,i} +b_{3,i} \right\Vert^2 \right) \\
& \le \left\Vert \Hwide \right\Vert^2 \left( \frac{3}{N} \sum_{i=1}^N
\left\Vert b_{1,i} \right\Vert^2 +\left\Vert b_{2,i} \right\Vert^2
+\left\Vert b_{3,i} \right\Vert^2 \right) =O_\mathbb{P} \left( \frac{1}{T_0}
\right).
\end{align*}
Similarly we can show that
\begin{align*}
\frac{1}{N_0}\sum_{i=1}^{N_0} \left\Vert \Htall \lhattalli -\lambda_{i}
\right\Vert^2 =O_\mathbb{P} \left( \frac{1}{T_0} \right).
\end{align*}
By the properties of matrix norms, the $C_p$ inequality, Lemma \ref%
{lemma:pointwise convergence of factors and loadings in sec3}(2) and
Assumption \ref{ass:order conditions}(2),
\begin{align*}
&\frac{1}{N_0}\sum_{i=1}^{N_0} \left\Vert \lhattalli-\Htall^{-1}\Hwide %
\lhatwidei \right\Vert^2 \\
& \le 2 \left\Vert \Htall^{-1} \right\Vert^2 \left[ \left( \frac{N}{N_0}
\frac{1}{N}\sum_{i=1}^N \left\Vert \Hwide\lhatwidei -\lambda_{i}
\right\Vert^2 \right)+ \left( \frac{1}{N_0}\sum_{i=1}^{N_0} \left\Vert \Htall%
\lhattalli-\lambda_{i} \right\Vert^2 \right) \right] =O_\mathbb{P} \left(
\frac{1}{T_0} \right),
\end{align*}
which proves the conclusion.
\end{pf}

\begin{lemma}
\label{lemma:p order sample mean is bounded} If Assumptions \ref{ass:factors
and loadings}--\ref{ass:order conditions} hold, then for every $%
m\in\{1,2,\ldots,8 \}$, as $N_0, T_0\to\infty$,

\begin{enumerate}[label=(\arabic*), labelindent=\parindent, leftmargin=*, nosep]

\item
\begin{align*}
\frac{1}{T}\sum_{t=1}^T \left\Vert \fhattallt-\Htall^\mathsf{T} f_t
\right\Vert^m =O_\mathbb{P}(1), \qquad \frac{1}{T_0}\sum_{t=1}^{T_0}
\left\Vert \fhatwidet-\Hwide^\mathsf{T} f_t \right\Vert^m =O_\mathbb{P}(1).
\end{align*}

\item
\begin{align*}
\frac{1}{N_0}\sum_{i=1}^{N_0} \left\Vert \lhattalli-\Htall^{-1}\lambda_{i}
\right\Vert^m =O_\mathbb{P}(1), \qquad \frac{1}{N}\sum_{i=1}^N \left\Vert %
\lhatwidei-\Hwide^{-1}\lambda_{i} \right\Vert^m =O_\mathbb{P}(1).
\end{align*}

\item
\begin{align*}
\frac{1}{T}\sum_{t=1}^T \left\Vert \fhattallt \right\Vert^m =O_\mathbb{P}%
(1), \qquad \frac{1}{T_0}\sum_{t=1}^{T_0} \left\Vert \fhatwidet %
\right\Vert^m =O_\mathbb{P}(1).
\end{align*}

\item
\begin{align*}
\frac{1}{N_0}\sum_{i=1}^{N_0} \left\Vert \lhattalli \right\Vert^m =O_\mathbb{%
P}(1), \qquad \frac{1}{N}\sum_{i=1}^N \left\Vert \lhatwidei \right\Vert^m =O_%
\mathbb{P}(1).
\end{align*}

\item
\begin{align*}
\frac{1}{T}\sum_{t=1}^T \left\vert \ehat_{i,t} \right\vert^m =O_\mathbb{P}%
(1) \quad \text{for every } i\in \{1,\ldots, N_0\}.
\end{align*}

\item
\begin{align*}
\frac{1}{T_0}\sum_{t=1}^{T_0} \left\vert \ehat_{i,t} \right\vert^m =O_%
\mathbb{P}(1) \quad \text{for every } i\in \{1,\ldots, N\}.
\end{align*}

\item
\begin{align*}
\frac{1}{N}\sum_{i=1}^{N} \left\vert \ehat_{i,t} \right\vert^m =O_\mathbb{P}%
(1) \quad \text{for every } t\in \{1,\ldots, T_0\}.
\end{align*}

\item
\begin{align*}
\frac{1}{N_0}\sum_{i=1}^{N_0} \left\vert \ehat_{i,t} \right\vert^m =O_%
\mathbb{P}(1) \quad \text{for every } t\in \{1,\ldots, T\}.
\end{align*}

\item
\begin{align*}
\frac{1}{T N_0}\sum_{t=1}^T \sum_{i=1}^{N_0} \left\vert \ehat_{i,t}
\right\vert^m =O_\mathbb{P}(1).
\end{align*}

\item
\begin{align*}
\frac{1}{T_0 N}\sum_{t=1}^{T_0} \sum_{i=1}^{N} \left\vert \ehat_{i,t}
\right\vert^m =O_\mathbb{P}(1).
\end{align*}
\end{enumerate}
\end{lemma}

\begin{pf}{}
Firstly, we prove Claims (1) and (2) with $m=1$.
By Assumptions \ref{ass:factors and loadings}(1)(2) and \ref{ass:moments of
idiosyncratic errors}(1), we have $\mathbb{E}\left( \left\Vert f_t
\right\Vert \right) \le M+1$, $\left\Vert \lambda_{i} \right\Vert \le
M$, and $\mathbb{E}\left( e_{i,t}^{2} \right) \le M+1$ for every $t\in \{1,\ldots, T\}$ 
and every $i\in\{1,\ldots, N\}$. Note that although Lemma C.1 of %
\citet{gonccalves2014bootstrapping} requires their Assumptions 1--5, only
the panel factor model relevant conditions of their Assumptions 1--5 are
actually used, and these panel factor model relevant conditions
automatically hold under Assumptions \ref{ass:factors and loadings}--\ref{ass:order conditions} 
of this paper. Then Claims (1) and (2) with $m=1$ are obtained by applying Lemma C.1 (i) and (ii) of %
\citet{gonccalves2014bootstrapping} to the control subsample and
pre-treatment subsample.

Claim (1) with $m=2$ follows from Lemma \ref{lemma:pointwise convergence of factors and loadings in sec3}(6)
and the fact that $o_\p(1)$ is trivially $O_\p(1)$. 
For $m>2$, we use Lemma \ref{lemma:pointwise convergence of factors and loadings in sec3}(6)
and Assumption \ref{ass:order conditions}(2) to conclude that \begin{align*}
	\sum_{t=1}^T \left\Vert \fhattallt-\Htall^\mathsf{T} f_t
	\right\Vert^2 =O_\mathbb{P}(1) \qquad \text{and} \qquad
	\sum_{t=1}^{T_0}
	\left\Vert \fhatwidet-\Hwide^\mathsf{T} f_t \right\Vert^2 =O_\p (1),
\end{align*}
which in turn implies that \begin{align*}
	\sum_{t=1}^T \left\Vert \fhattallt-\Htall^\mathsf{T} f_t
	\right\Vert^m =O_\mathbb{P}(1) \qquad \text{and} \qquad
	\sum_{t=1}^{T_0}
	\left\Vert \fhatwidet-\Hwide^\mathsf{T} f_t \right\Vert^m =O_\p (1)
\end{align*}
holds for any integer $m>2$. Then Claim (1) with $m>2$ follows from
the relationship between $o_\p(1)$ and $O_\p(1)$. Claim (2) with $m\ge 2$
can be proved analogously using some facts in the proof of
Lemma \ref{lemma:MSE between tall and wide loadings in sec3}.

For every $m\in \{1,2,\ldots, 8\}$, Assumption \ref{ass:factors and loadings}(1)
implies that
\begin{align*}
\mathbb{E}\left( \left\vert \frac{1}{T}\sum_{t=1}^T \left\Vert f_t
\right\Vert^m \right\vert \right) =\frac{1}{T} \sum_{t=1}^T \mathbb{E}
\left( \left\Vert f_t \right\Vert^m \right) \le M+1.
\end{align*}
By the $C_p$ inequality, the properties of matrix norms, Markov's
inequality, Lemma \ref{lemma:pointwise convergence of factors and loadings
in sec3}(2) and the first part of Claim (1),
\begin{align*}
\frac{1}{T}\sum_{t=1}^T \left\Vert \fhattallt \right\Vert^m \le 2^{m-1} %
\left[ \left\Vert \Htall \right\Vert^m \left( \frac{1}{T} \sum_{t=1}^T
\left\Vert f_t \right\Vert^m \right) + \frac{1}{T}\sum_{t=1}^T \left\Vert %
\fhattallt-\Htall^\mathsf{T} f_t \right\Vert^m \right) =O_\mathbb{P}(1),
\end{align*}
which establishes the first part of Claim (3). The second part and Claim (4)
can be proved analogously.

To prove Claims (5)--(10), we perform the following decomposition of $\ehat%
_{i,t}$ for every $(i,t)\in \mI\backslash\mI_1$.
\begin{align}
\ehat_{i,t} &=e_{i,t}-\left( \chat_{i,t}-c_{i,t} \right)  \notag \\
& = e_{i,t}- \left( \fhattallt-\Htall^\mathsf{T} f_t \right)^\mathsf{T} %
\Hhatmiss \left( \lhatwidei-\Hwide^{-1}\lambda_{i} \right) - f_t^\mathsf{T} %
\Htall \Hhatmiss \left( \lhatwidei-\Hwide^{-1} \lambda_{i} \right)  \notag \\
&\phantom{=\;\;} -\left( \fhattallt-\Htall^\mathsf{T} f_t \right)^\mathsf{T} %
\Hhatmiss \Hwide^{-1} \lambda_{i} -f_t^\mathsf{T} \left( \Htall \Hhatmiss %
\Hwide^{-1} -I_r \right) \lambda_{i} .  \label{eq:decomposition of ehat}
\end{align}
By the $C_p$ inequality and properties of matrix norms,
\begin{align*}
\left\vert \ehat_{i,t} \right\vert^m & \le 5^{m-1} \left( \left\vert e_{i,t}
\right\vert^m +\left\Vert \Hhatmiss \right\Vert^m \left\Vert \fhattallt-%
\Htall^\mathsf{T} f_t \right\Vert^m \left\Vert \lhatwidei-\Hwide^{-1}
\lambda_{i} \right\Vert^m \right. \\
& \phantom{=\;\;} +\left\Vert \Htall \right\Vert^m \left\Vert \Hhatmiss %
\right\Vert^m \left\Vert f_t \right\Vert^m \left\Vert \lhatwidei-\Hwide^{-1}
\lambda_{i} \right\Vert^m \\
& \phantom{=\;\;} + \left\Vert \Hhatmiss \right\Vert^m \left\Vert \Hwide%
^{-1} \right\Vert^m \left\Vert \fhattallt-\Htall^\mathsf{T} f_t
\right\Vert^m \left\Vert \lambda_{i} \right\Vert^m \\
& \phantom{=\;\;} \left. + \left\Vert \Htall \Hhatmiss \Hwide^{-1}-I_r
\right\Vert^m \left\Vert f_t \right\Vert^m \left\Vert \lambda_{i}
\right\Vert^m \right).
\end{align*}
Lemma \ref{lemma:pointwise convergence of factors and loadings in sec3}(5)
implies that $\Hhatmiss=O_\mathbb{P}(1)$. By Assumptions \ref{ass:factors
and loadings}(1)(2), \ref{ass:moments of idiosyncratic errors}(1) and
Markov's inequality,
\begin{align*}
& \frac{1}{T}\sum_{t=1}^T \left\Vert f_t \right\Vert^m=O_\mathbb{P}(1),
\qquad \frac{1}{N}\sum_{i=1}^N \left\Vert \lambda_{i} \right\Vert^m =O_%
\mathbb{P}(1), \\
& \frac{1}{T}\sum_{t=1}^T \left\vert e_{i,t} \right\vert^m=O_\mathbb{P}(1),
\qquad \frac{1}{N}\sum_{i=1}^N \left\vert e_{i,t} \right\vert^m=O_\mathbb{P}%
(1), \qquad \frac{1}{TN}\sum_{t=1}^T\sum_{i=1}^N \left\vert e_{i,t}
\right\vert^m =O_\mathbb{P}(1).
\end{align*}
Then Claims (5)--(10) follow from Claims (1)(2), Lemma \ref{lemma:pointwise
convergence of factors and loadings in sec3}(2)(5), Assumption \ref%
{ass:order conditions}(2), and above facts.
\end{pf}

\begin{lemma}
\label{lemma:MSE of ehat with respect to e} If Assumptions \ref{ass:factors
and loadings}--\ref{ass:order conditions} hold, then for every $%
i\in\{1,\ldots, N\}$ and every integer $m\ge 2$, as $N_0, T_0 \to\infty$,
\begin{align*}
\frac{1}{T_0}\sum_{t=1}^{T_0} \left\vert \ehat_{i,t} -e_{i,t}
\right\vert^m=O_\mathbb{P} \left( \frac{1}{T_0} \right).
\end{align*}
\end{lemma}

\begin{pf}{}
Firstly consider the case of $m=2$ By Equation \eqref{eq:decomposition of ehat}, the $C_p$ inequality and the properties of matrix norms,
\begin{align*}
\frac{1}{T_0}\sum_{t=1}^{T_0}\left( \ehat_{i,t}-e_{i,t} \right)^2 &\le 4
\left\Vert \Hhatmiss \right\Vert^2 \left\Vert \lhatwidei-\Hwide^{-1}
\lambda_{i} \right\Vert^2 \frac{1}{T_0} \sum_{t=1}^{T_0} \left\Vert %
\fhattallt-\Htall^\mathsf{T} f_s \right\Vert^2 \\
& \phantom{=\;\;} +4 \left\Vert \Htall \right\Vert^2 \left\Vert \Hhatmiss %
\right\Vert^2 \left\Vert \lhatwidei-\Hwide^{-1} \lambda_{i} \right\Vert^2
\frac{1}{T_0} \sum_{t=1}^{T_0} \left\Vert f_t \right\Vert^2 \\
& \phantom{=\;\;} +4 \left\Vert \Hhatmiss \right\Vert^2 \left\Vert \Hwide%
^{-1} \right\Vert^2 \left\Vert \lambda_{i} \right\Vert^2 \frac{1}{T_0}%
\sum_{t=1}^{T_0} \left\Vert \fhattallt-\Htall^\mathsf{T} f_t \right\Vert^2 \\
& \phantom{=\;\;} +4 \left\Vert \Htall\Hhatmiss\Hwide^{-1}-I_r \right\Vert^2
\left\Vert \lambda_{i} \right\Vert^2 \frac{1}{T_0} \sum_{t=1}^{T_0}
\left\Vert f_t \right\Vert^2.
\end{align*}
And the conclusion follows from Assumptions \ref{ass:factors and loadings}%
(1)(2), \ref{ass:order conditions}(2), Markov's inequality, and Lemma \ref%
{lemma:pointwise convergence of factors and loadings in sec3}(2)(4)(5)(6).

For the case of $m>2$, the result for the case of $m=2$ implies that
\begin{align*}
\sum_{t=1}^{T_0} \left( \ehat_{i,t} -e_{i,t} \right)^2=O_\mathbb{P}(1),
\end{align*}
which in turn implies that
\begin{align*}
\sum_{t=1}^{T_0} \left\vert \ehat_{i,t} -e_{i,t} \right\vert^m=O_\mathbb{P}%
(1)
\end{align*}
for every integer $m>2$.
\end{pf}

\subsection{Estimators and Residuals in Section \ref{sec:Estimation
and Inference in a Model with Covariates}}

In this subsection, we show that adding covariates into the panel factor
model does not alter the conclusions in Subsection \ref{sec:Properties of
the Estimators and Residuals pure factor}. Now $\Fhattall$, $\Fhatwide$, $%
\Lhattall$ and $\Lhatwide$ are estimated by Algorithm \ref{algo:IFEE}
(interactive fixed effect estimation), and
\begin{align*}
\Vhattall=\frac{1}{N_0}\Lhattall^\mathsf{T} \Lhattall, \qquad \Vhatwide=%
\frac{1}{N}\Lhatwide^\mathsf{T} \Lhatwide, \qquad \ehat%
_{i,t}=y_{i,t}-x_{i,t}^\mathsf{T} \betahattall-\chat_{i,t}.
\end{align*}
Other quantities follow their definitions in Subsection \ref{sec:Properties
of the Estimators and Residuals pure factor}.

\begin{lemma}
\label{lemma:properties of estimators with covariates} If Assumptions \ref%
{ass:factors and loadings}--\ref{ass:error is independent of factor and x}
hold, then the results in Lemmas \ref{lemma:pointwise convergence of factors
and loadings in sec3}, \ref{lemma:MSE between tall and wide loadings in sec3}
and \ref{lemma:p order sample mean is bounded}(1)--(4) are true for the
estimators in Section \ref{sec:Estimation and Inference in a Model with
Covariates}.
\end{lemma}

\begin{pf}{}
By Remark 5 of \citet{bai2009panel}, estimation of $\beta$ does not affect
the rates of convergence and the limiting distributions of the estimated
factors and loadings, so they are the same as those of a pure factor model.
Since Lemmas \ref{lemma:pointwise convergence of factors and loadings in
sec3}, \ref{lemma:MSE between tall and wide loadings in sec3} and \ref%
{lemma:p order sample mean is bounded}(1)--(4) are asymptotic properties of
the estimated factors and loadings in a pure factor model, they can be
naturally extended to a factor model with covariates.
\end{pf}

\begin{lemma}
\label{lemma:properties of residuals with covariates} If Assumptions \ref%
{ass:factors and loadings}--\ref{ass:error is independent of factor and x}
hold, then the results in Lemmas \ref{lemma:p order sample mean is bounded}%
(5)--(10) and \ref{lemma:MSE of ehat with respect to e} are true for the
residuals in Section \ref{sec:Estimation and Inference in a Model with
Covariates}.
\end{lemma}

\begin{pf}{}
Applying Theorem 1 of \citet{bai2009panel} to the control subsample yields
\begin{align}
\betahattall-\beta =O_\mathbb{P} \left( \frac{1}{\sqrt{N_0 T}} \right).
\label{eq:rate of convergence of betahat}
\end{align}
Now for every $(i,t)\in\mI\backslash\mI_1$, $e_{i,t}$ admits a decomposition
\begin{align*}
\ehat_{i,t} &=e_{i,t}-\left( \chat_{i,t}-c_{i,t} \right) -x_{i,t}^\mathsf{T}
\left( \betahattall-\beta \right)  \notag \\
& = e_{i,t}- \left( \fhattallt-\Htall^\mathsf{T} f_t \right)^\mathsf{T} %
\Hhatmiss \left( \lhatwidei-\Hwide^{-1}\lambda_{i} \right) - f_t^\mathsf{T} %
\Htall \Hhatmiss \left( \lhatwidei-\Hwide^{-1} \lambda_{i} \right)  \notag \\
&\phantom{=\;\;} -\left( \fhattallt-\Htall^\mathsf{T} f_t \right)^\mathsf{T} %
\Hhatmiss \Hwide^{-1} \lambda_{i} -f_t^\mathsf{T} \left( \Htall \Hhatmiss %
\Hwide^{-1} -I_r \right) \lambda_{i} -x_{i,t}^\mathsf{T} \left( \betahattall%
-\beta \right).
\end{align*}
For every $m\in\{1,2, \ldots,8\}$, by the $C_p$ inequality and properties of
matrix norms,
\begin{align*}
\left\vert \ehat_{i,t} \right\vert^m & \le 6^{m-1} \left( \left\vert e_{i,t}
\right\vert^m +\left\Vert \Hhatmiss \right\Vert^m \left\Vert \fhattallt-%
\Htall^\mathsf{T} f_t \right\Vert^m \left\Vert \lhatwidei-\Hwide^{-1}
\lambda_{i} \right\Vert^m \right. \\
& \phantom{=\;\;} +\left\Vert \Htall \right\Vert^m \left\Vert \Hhatmiss %
\right\Vert^m \left\Vert f_t \right\Vert^m \left\Vert \lhatwidei-\Hwide^{-1}
\lambda_{i} \right\Vert^m \\
& \phantom{=\;\;} + \left\Vert \Hhatmiss \right\Vert^m \left\Vert \Hwide%
^{-1} \right\Vert^m \left\Vert \fhattallt-\Htall^\mathsf{T} f_t
\right\Vert^m \left\Vert \lambda_{i} \right\Vert^m \\
& \phantom{=\;\;} \left. + \left\Vert \Htall \Hhatmiss \Hwide^{-1}-I_r
\right\Vert^m \left\Vert f_t \right\Vert^m \left\Vert \lambda_{i}
\right\Vert^m + \left\Vert x_{i,t} \right\Vert^m \left\Vert \betahattall%
-\beta \right\Vert^m \right).
\end{align*}
Furthermore,
\begin{align*}
\frac{1}{T_0}\sum_{t=1}^{T_0}\left( \ehat_{i,t}-e_{i,t} \right)^2 &\le 5
\left\Vert \Hhatmiss \right\Vert^2 \left\Vert \lhatwidei-\Hwide^{-1}
\lambda_{i} \right\Vert^2 \frac{1}{T_0} \sum_{t=1}^{T_0} \left\Vert %
\fhattallt-\Htall^\mathsf{T} f_s \right\Vert^2 \\
& \phantom{=\;\;} +5 \left\Vert \Htall \right\Vert^2 \left\Vert \Hhatmiss %
\right\Vert^2 \left\Vert \lhatwidei-\Hwide^{-1} \lambda_{i} \right\Vert^2
\frac{1}{T_0} \sum_{t=1}^{T_0} \left\Vert f_t \right\Vert^2 \\
& \phantom{=\;\;} +5 \left\Vert \Hhatmiss \right\Vert^2 \left\Vert \Hwide%
^{-1} \right\Vert^2 \left\Vert \lambda_{i} \right\Vert^2 \frac{1}{T_0}%
\sum_{t=1}^{T_0} \left\Vert \fhattallt-\Htall^\mathsf{T} f_t \right\Vert^2 \\
& \phantom{=\;\;} +5 \left\Vert \Htall\Hhatmiss\Hwide^{-1}-I_r \right\Vert^2
\left\Vert \lambda_{i} \right\Vert^2 \frac{1}{T_0} \sum_{t=1}^{T_0}
\left\Vert f_t \right\Vert^2 \\
& \phantom{=\;\;} +5 \left\Vert \betahattall-\beta \right\Vert^2 \frac{1}{T_0%
} \sum_{t=1}^{T_0}\left\Vert x_{i,t} \right\Vert^2.
\end{align*}
Assumption \ref{ass:covariates x}(1) and Markov's inequality imply that for
every $m\in\{1,2, \ldots,8\}$,
\begin{align*}
\frac{1}{T}\sum_{t=1}^T \left\Vert x_{i,t} \right\Vert^m=O_\mathbb{P}(1),
\qquad \frac{1}{N}\sum_{i=1}^N \left\Vert x_{i,t} \right\Vert^m=O_\mathbb{P}%
(1), \qquad \frac{1}{TN}\sum_{t=1}^T \sum_{i=1}^N \left\Vert x_{i,t}
\right\Vert^m=O_\mathbb{P}(1).
\end{align*}
Then the conclusions follow from above results, the arguments in the proofs
of Lemmas \ref{lemma:p order sample mean is bounded}(5)--(10), \ref%
{lemma:MSE of ehat with respect to e}, and \ref{lemma:properties of
estimators with covariates}.
\end{pf}

\section{Properties of Bootstrap}

As is described in Step 2 of Algorithm \ref{algo:cov confidence interval}
and explained in Remark \ref{remark:pure factor bootstrap in sec4}, the
bootstrap is conducted within a pure factor framework even for a factor
model with covariates (\textit{i.e.}, the model in Section \ref%
{sec:Estimation and Inference in a Model with Covariates}). By the
construction of resampled observations and Lemma \ref{lemma:properties of
estimators with covariates}, \ref{lemma:properties of residuals with
covariates}, it follows that the properties of bootstrap quantities in
Section \ref{sec:Estimation and Inference in a Model with Covariates} under
Assumptions \ref{ass:factors and loadings}--\ref{ass:error is independent of
factor and x} are identical to those in Section \ref{sec:Estimation and
Inference in a Model without Covariates} under Assumptions \ref{ass:factors
and loadings}--\ref{ass:order conditions}. Therefore, unless clarifying
assumptions, we do not distinguishing between a pure factor model and a
factor model with covariates in this section.

Moreover, since the proof of (ordinary) wild bootstrap can be straightforwardly extended to block wild
bootstrap with extra notations,
we provide the results for (ordinary) wild bootstrap procedure in what
follows to save space.

\subsection{Technical Notes}

This subsection introduces some basic concepts and results of bootstrap
asymptotic analysis. Let $\bS_{N,T}$ be the raw sample, \textit{i.e.}, $%
\left\{ \oy_{i,t} : (i,t)\in\mI \right\}$ for a pure factor model and $%
\left\{ \left( \oy_{i,t}, x_{i,t} \right): (i,t)\in\mI \right\}$ for a
factor model with covariates. For simplicity of notation, we omit the
subscript and just write $\bS$ in stead of $\bS_{N,T}$. Define $\mathbb{P}^*$
and $\mathbb{E}^*$ to be the conditional probability and expectation given $%
\bS$, \textit{i.e.}, $\mathbb{P}^*(\cdot)=\mathbb{P} (\cdot|\bS)$ and $%
\mathbb{E}^*(\cdot)=\mathbb{E}(\cdot|\bS)$. The conditional variance $%
\mathrm{Var}^*(\cdot)$ and conditional covariance $\mathrm{Cov}^*(\cdot)$
are defined analogously.

For a bootstrap statistic $A^*_{N,T}$, we say that $A^*_{N,T}$ is of an
order $o_{\mathbb{P}^*}(1)$ in probability, denoted by $A^*_{N,T}=\ops(1)$,
if and only if for any $\varepsilon>0$ and $\delta>0$,
\begin{align*}
\lim_{N,T\to\infty} \mathbb{P} \left[ \mathbb{P}^* \left( \left\Vert
A^*_{N,T} \right\Vert>\varepsilon \right) >\delta \right] =0.
\end{align*}
We say that $A^*_{N,T}$ is of an order $\Ops (1)$ in probability, denoted by
$A^*_{N,T}=\Ops(1)$, if and only if for any $\delta>0$, there exists a $%
0<M<\infty$ such that
\begin{align*}
\lim_{N,T\to\infty} \mathbb{P} \left[ \mathbb{P}^*\left( \left\Vert
A^*_{N,T} \right\Vert\ge M \right) >\delta \right]=0.
\end{align*}
For a sequence of deterministic numbers $\left\{ c_{N,T} \right\}$,
\begin{align*}
A^*_{N,T}=\ops\left( c_{N,T} \right) \; \Leftrightarrow \; \frac{A^*_{N,T}}{%
c_{N,T}}=\ops(1), \qquad A^*_{N,T}=\Ops\left( c_{N,T} \right) \;
\Leftrightarrow \; \frac{A^*_{N,T}}{c_{N,T}}=\Ops(1).
\end{align*}
One can also see Pages 2891--2892 and Appendix A.1 of %
\citet{cheng2010bootstrap} for measure-theoretic definitions of bootstrap
stochastic orders and relevant measurability issues.

The lemmas below list some properties of bootstrap stochastic orders that
are frequently used in subsequent analyses.

\begin{lemma}
\label{lemma:sufficient conditions for bootstrap stochastic orders} Let $%
A^*_{N,T}$ be a bootstrap statistic. If $\mathbb{E}^* \left( \left\Vert
A^*_{N,T} \right\Vert \right) \xrightarrow{\mathbb{P}} 0$, then $A^*_{N,T}=%
\ops(1)$. If $\mathbb{E}^* \left( \left\Vert A^*_{N,T} \right\Vert \right)
=O_\mathbb{P}(1)$, then $A^*_{N,T}=\Ops(1)$.
\end{lemma}

\begin{pf}{}
See the first paragraph on Page 387 of \citet{chang2003sieve}.
\end{pf}

\begin{lemma}
\label{lemma:operations of bootstrap stochastic orders} Suppose $A_{N,T}=O_%
\mathbb{P}(1)$, $a_{N,T}=o_\mathbb{P}(1)$, $A^*_{1,N,T}=\Ops(1)$, $%
A^*_{2,N,T}=\Ops(1)$, $a^*_{1,N,T}=\ops(1)$, and $a^*_{2,N,T}=\ops(1)$. Then

\begin{enumerate}[label=(\arabic*), labelindent=\parindent, leftmargin=*, nosep]

\item $A^*_{1,N,T}+A^*_{2,N,T}=\Ops(1)$, and $A^*_{1,N,T}+A_{1,N,T}=\Ops(1)$.

\item $A^*_{1,N,T} A^*_{2,N,T}=\Ops(1)$, and $A^*_{1,N,T}A_{1,N,T}=\Ops(1)$.

\item $a^*_{1,N,T}+a^*_{2,N,T}=\ops(1)$, and $a^*_{1,N,T}+a_{1,N,T}=\ops(1)$.

\item $a^*_{1,N,T}a^*_{2,N,T}=\ops(1)$, and $a^*_{1,N,T}a_{1,N,T}=\ops(1)$.

\item $A^*_{1,N,T}+a^*_{1,N,T}=\Ops(1)$, $A^*_{1,N,T}+a_{1,N,T}=\Ops(1)$,
and $A_{1,N,T}+a^*_{1,N,T}=\Ops(1)$.

\item $A^*_{1,N,T}a^*_{1,N,T}=\ops(1)$, $A^*_{1,N,T}a_{1,N,T}=\ops(1)$, and $%
A_{1,N,T}a^*_{1,N,T}=\ops(1)$.
\end{enumerate}
\end{lemma}

\begin{pf}{}
These results follow from the last a few lines on Page 1861 of %
\citet{park2003bootstrap}, Lemma 1 of \citet{chang2003sieve}, and Lemma 3 of %
\citet{cheng2010bootstrap}.
\end{pf}

Now we introduce the definition of conditional convergence in distribution.
For a bootstrap statistic $Z^*_{N,T}$ and a random element $Z$ taking values
in a metric space $\mathbb{D}$, we say that $Z^*_{N,T}$ conditionally
converges in distribution to $Z$ (or equivalently, the conditional
distribution of $Z^*_{N,T}$ weakly converges to that of $Z$) in probability,
denoted by $Z^*_{N,T}\xrightarrow{\mathrm{d}^*} Z$, if and only if for every
bounded Lipschitz continuous function $h:\mathbb{D}\to \mathbb{R}$, we have $%
\mathbb{E}^*\left[ h \left( Z^*_{N,T} \right) \right]\xrightarrow{\mathbb{P}}
\mathbb{E}\left[ h(Z) \right]$ as $N,T\to\infty$. The lemmas below list some
properties of conditional convergence in distribution that are frequently
used in subsequent analyses.

\begin{lemma}
\label{lemma:relationship between bootstrap convergence in P and D} If $%
Z^*_{N,T}\xrightarrow{\mathrm{d}^*} Z$ and $a^*_{N,T}=\ops(1)$, then $%
Z^*_{N,T}=\Ops(1)$ and $Z^*_{N,T}+a^*_{N,T}\xrightarrow{\mathrm{d}^*} Z$.
\end{lemma}

\begin{pf}{}
See Remark 2 of \citet{chang2003sieve}.
\end{pf}

\begin{lemma}
\label{lemma:cdf characterisation of conditional convergence in D} Suppose
that the bootstrap statistic $Z^*_{N,T}$ and random variable $Z$ take their
values in $\mathbb{R}$, and $Z^*_{N,T}\xrightarrow{\mathrm{d}^*} Z$ as $%
N,T\to\infty$. Let $G^*_{N,T}$ denote the conditional cumulative
distribution function of $Z^*_{N,T}$, and $G$ denote the cumulative
distribution function of $Z$, \textit{i.e.}, $G^*_{N,T}(z)=\mathbb{P}%
^*\left( Z^*_{N,T}\le z \right)$ and $G(z)=\mathbb{P}(Z\le z)$ for every $%
z\in\mathbb{R}$. Then $G^*_{N,T}(z)\xrightarrow{\mathbb{P}} G(z)$ as $%
N,T\to\infty$ for every $z\in \mathcal{C}_G$, where $\mathcal{C}_G\subset
\mathbb{R}$ is the set of all continuous points of $G$. Moreover, if $G$ is
everywhere continuous on $\mathbb{R}$, \textit{i.e.}, $\mathcal{C}_G=\mathbb{%
R}$, then
\begin{align*}
\sup_{z\in\mathbb{R}} \left\vert G^*_{N,T}(z)-G(z) \right\vert %
\xrightarrow{\mathbb{P}} 0 \quad \text{as } N,T\to\infty.
\end{align*}
\end{lemma}

\begin{pf}{}
For any $z_0\in \mathbb{R}$ and any $C_0\subset \mathbb{R}$, the distance
between $z_0$ and $C_0$ is measured by
\begin{align*}
d_e \left( z_0, C_0 \right)= \inf_{z\in C_0} \left\vert z_0-z \right\vert.
\end{align*}
For any open set $C_1\subset \mathbb{R}$, construct a sequence of functions $%
\left\{ h_m \right\}_{m=1}^\infty$, so that $h_m(z)=\left[ m d_e \left( z,
C_1^c \right) \right]\wedge 1$ for every $z\in\mathbb{R}$. It is easy to see
that $h_m$ is non-negative, bounded, and Lipschitz continuous with Lipschitz
constant $m$ for every $m\in\bZ_+$, and $h_m \uparrow \mathbbm{1}_{C_1}$ as $%
m\to\infty$. By the monotone convergence theorem,
\begin{align*}
\lim_{m\to\infty} \mathbb{E}\left[ h_m (Z) \right] = \mathbb{E} \left[ %
\mathbbm{1}_{C_1} (Z) \right] =\mathbb{P}\left( Z\in C_1 \right).
\end{align*}
Hence for any $\varepsilon>0$, there exists an $m_0\in\bZ_+$, such that $%
\mathbb{E}\left[ h_{m_0}(Z) \right] > \mathbb{P} \left( Z\in C_1 \right)
-\varepsilon/2$. Since $h_{m_0}\le \mathbbm{1}_{C_1}$, it follows that $%
\mathbb{P}^*\left( Z^*_{N,T} \in C_1 \right) \ge \mathbb{E}^*\left[
h_{m_0}\left( Z^*_{N,T} \right) \right]$ almost surely. By the boundedness
and Lipschitz continuity of $h_{m_0}$ and the definition of conditional
convergence in distribution, $\mathbb{E}^*\left[ h_{m_0}\left( Z^*_{N,T}
\right) \right] \xrightarrow{\mathbb{P}} \mathbb{E}\left[ h_{m_0} \left( Z
\right) \right]$, which implies that
\begin{align*}
\lim_{N,T\to\infty} \mathbb{P} \left[ \mathbb{P}^*\left( Z^*_{N,T} \in C_1
\right) > \mathbb{P}\left( Z\in C_1 \right) -\varepsilon\right] &\ge
\lim_{N,T\to\infty} \mathbb{P} \left( \mathbb{E}^* \left[ h_{m_0}\left(
Z^*_{N,T} \right) \right]> \mathbb{P} \left( Z\in C_1 \right) -\varepsilon
\right) \\
& \ge \lim_{N,T\to\infty} \mathbb{P} \left( \mathbb{E}^* \left[
h_{m_0}\left( Z^*_{N,T} \right) \right] > \mathbb{E}\left[ h_{m_0}(Z) \right]%
-\frac{\varepsilon}{2} \right) =1.
\end{align*}

For any closed set $C_2\subset\mathbb{R}$, because $C_2^c$ is an open set,
the above result implies that
\begin{align*}
\lim_{N,T\to\infty} \mathbb{P} \left[ \mathbb{P}^* \left( Z^*_{N,T}\in C_2
\right) < \mathbb{P} \left( Z\in C_2 \right) +\varepsilon \right] &=
\lim_{N,T\to\infty} \mathbb{P} \left[ 1-\mathbb{P}^* \left( Z^*_{N,T}\in C_2
\right) > 1-\mathbb{P} \left( Z\in C_2 \right) -\varepsilon \right] \\
&= \lim_{N,T\to\infty} \mathbb{P} \left[ \mathbb{P}^* \left( Z^*_{N,T}\in
C_2^c \right) > \mathbb{P} \left( Z\in C_2^c \right) -\varepsilon \right]=1.
\end{align*}

Now consider any Borel set $C_3\subset\mathbb{R}$ with $\mathbb{P}\left[
Z\in \mathrm{cl}\left( C_3 \right) \backslash \mathrm{int} \left( C_3
\right) \right] =0$, where $\mathrm{cl}(\cdot)$ and $\mathrm{int}(\cdot)$
denote the closure and interior of a set, respectively. Then it follows that
\begin{align*}
&\phantom{=\;\;} \lim_{N,T\to\infty} \mathbb{P} \left[ \left\vert \mathbb{P}%
^* \left( Z^*_{N,T}\in C_3 \right) -\mathbb{P} \left( Z\in C_3 \right)
\right\vert < \varepsilon \right] \\
&= \lim_{N,T\to\infty} \mathbb{P} \left[ \mathbb{P} \left( Z\in \mathrm{int}%
(C_3) \right)-\varepsilon < \mathbb{P}^* \left( Z^*_{N,T}\in C_3 \right) <
\mathbb{P} \left( Z\in \mathrm{cl}(C_3) \right)+\varepsilon \right] \\
& \ge 1-\lim_{N,T\to\infty} \mathbb{P} \left[ \mathbb{P}^* \left( Z^*_{N,T}
\in C_3 \right) \le \mathbb{P} \left( Z\in\mathrm{int}(C_3)
\right)-\varepsilon \right] \\
&\phantom{\ge 1 \;} -\lim_{N,T\to\infty} \mathbb{P} \left[ \mathbb{P}^*
\left( Z^*_{N,T} \in C_3 \right) \ge \mathbb{P} \left( Z\in\mathrm{cl}(C_3)
\right)+\varepsilon \right] \\
& \ge 1-\lim_{N,T\to\infty} \mathbb{P} \left[ \mathbb{P}^* \left( Z^*_{N,T}
\in \mathrm{int} (C_3) \right) \le \mathbb{P} \left( Z\in\mathrm{int}(C_3)
\right)-\varepsilon \right] \\
&\phantom{\ge 1 \;} -\lim_{N,T\to\infty} \mathbb{P} \left[ \mathbb{P}^*
\left( Z^*_{N,T} \in \mathrm{cl} (C_3) \right) \ge \mathbb{P} \left( Z\in%
\mathrm{cl}(C_3) \right)+\varepsilon \right]=1,
\end{align*}
which implies that $\mathbb{P}^* \left( Z^*_{N,T}\in C_3 \right)%
\xrightarrow{\mathbb{P}} \mathbb{P} \left( Z\in C_3 \right)$ as $N,T\to\infty
$.

For any $z\in \mathcal{C}_G$, let $C_4=(-\infty, z]$, then $C_4$ is Borel
and satisfies $\mathbb{P}\left[ Z\in \mathrm{cl}\left( C_4 \right)
\backslash \mathrm{int} \left( C_4 \right) \right] =0$. Therefore,
\begin{align*}
G^*_{N,T}(z)=\mathbb{P}^*\left( Z^*_{N,T}\in C_4 \right) \xrightarrow{%
\mathbb{P}} \mathbb{P} \left( Z\in C_4 \right) = G(z)
\end{align*}
as $N,T\to\infty$.

If $G$ is everywhere continuous, \textit{i.e.}, $\mathcal{C}_G=\mathbb{R}$,
then for any $k\in\bZ_+$, there exist $-\infty=z_0<z_1 <\cdots <z_k=\infty$,
so that $G\left( z_i \right)=i/k$ for every $i\in{1,\ldots, k}$. Then for
any $z\in \left[ z_{i-1}, z_i \right]$,
\begin{align*}
G^*_{N,T}(z)-G(z) & \le G^*_{N,T} \left( z_i \right) -G \left( z_{i-1}
\right) =G^*_{N,T} \left( z_i \right) -G\left( z_i \right)+\frac{1}{k}, \\
G^*_{N,T}(z)-G(z) & \ge G^*_{N,T} \left( z_{i-1} \right) -G \left( z_{i}
\right) =G^*_{N,T} \left( z_{i-1} \right) -G\left( z_{i-1} \right)-\frac{1}{k%
},
\end{align*}
which implies that
\begin{align*}
\sup_{z\in\mathbb{R}} \left\vert G^*_{N,T}(z)-G(z) \right\vert \le \frac{1}{k%
}+ \max_{1\le i \le k} \left\vert G^*_{N,T}\left( z_i \right) -G\left( z_i
\right) \right\vert.
\end{align*}
For any $\varepsilon>0$, pick a $k$ such that $1/k <\varepsilon/2$. Since $%
\displaystyle\max_{1\le i \le k} \left\vert G^*_{N,T}\left( z_i \right)
-G\left( z_i \right) \right\vert\xrightarrow{\mathbb{P}} 0$, we have
\begin{align*}
\lim_{N,T\to\infty} \mathbb{P} \left( \sup_{z\in\mathbb{R}} \left\vert
G^*_{N,T}(z)-G(z) \right\vert>\varepsilon \right) \le \lim_{N,T\to\infty}
\mathbb{P} \left( \max_{1\le i \le k} \left\vert G^*_{N,T}\left( z_i \right)
-G\left( z_i \right) \right\vert >\frac{\varepsilon}{2} \right)=0,
\end{align*}
and the proof is complete.
\end{pf}

\subsection{Ancillary Results about Bootstrap Samples}

Let $\Ftid=\left( \ftid_1, \ldots , \ftid_T \right)^\mathsf{T}=\Fhattall$,
and $\Ltid=\left( \ltid_1, \ldots, \ltid_N \right)^\mathsf{T}=\Lhatwide %
\Hhatmiss^\mathsf{T}$. Analogously to the population version, define $%
\Ftidtall=\Ftid$, $\Ltidwide=\Ltid$, $\Ftidwide$ to be a sub-matrix formed
by the first $T_0$ rows of $\Ftid$, and $\Ltidtall$ to be a sub-matrix
formed by the first $N_0$ rows of $\Ltid$. Then let $e_{i,t}^*$ be defined
in Step 2(1)(2) of Algorithms \ref{algo:confidence interval} and \ref%
{algo:cov confidence interval}.

Lemmas \ref{lemma:BHLC sumN ee}--\ref{lemma:BHLC sum lambda e} below verify
(some of) the bootstrap high level conditions in %
\citet{gonccalves2014bootstrapping} and \citet{gonccalves2017bootstrap}.
Lemma \ref{lemma:BHLC sumN ee} corresponds to Condition A*(b) of %
\citet{gonccalves2014bootstrapping} and Condition A.1 of %
\citet{gonccalves2017bootstrap}.

\begin{lemma}
\label{lemma:BHLC sumN ee} If Assumptions \ref{ass:factors and loadings}--%
\ref{ass:order conditions} hold for a pure factor model or Assumptions \ref%
{ass:factors and loadings}--\ref{ass:error is independent of factor and x}
hold for a factor model with covariates, then as $N_0, T_0\to\infty$,

\begin{enumerate}[label=(\arabic*), labelindent=\parindent, leftmargin=*, nosep]

\item For every $t\in \{1,\ldots, T\}$,
\begin{align*}
\sum_{s=1}^T \left[ \mathbb{E}^* \left( \frac{1}{N_0}\sum_{i=1}^{N_0}
e_{i,t}^* e_{i,s}^* \right) \right]^2 =O_\mathbb{P}(1).
\end{align*}

\item
\begin{align*}
\frac{1}{T}\sum_{t=1}^T \sum_{s=1}^T \left[ \mathbb{E}^* \left( \frac{1}{N_0}%
\sum_{i=1}^{N_0} e_{i,t}^* e_{i,s}^* \right) \right]^2 =O_\mathbb{P}(1).
\end{align*}
\end{enumerate}
\end{lemma}

\begin{pf}{}
Since $\left\{ u_{i,t} \right\}$ are i.i.d.\ from $\bN(0,1)$, it follows that
$\mathbb{E} \left( u_{i,t}u_{i,s} \right)=\mathbbm{1}_{\{s=t\}}$. By the
independence between $\left\{ u_{i,t} \right\}$ and the raw sample, and
Lemma \ref{lemma:p order sample mean is bounded}(8),
\begin{align*}
\sum_{s=1}^T \left[ \mathbb{E}^* \left( \frac{1}{N_0}\sum_{i=1}^{N_0}
e_{i,t}^* e_{i,s}^* \right) \right]^2 =\sum_{s=1}^T \left[ \frac{1}{N_0}
\sum_{i=1}^{N_0} \ehat_{i,t} \ehat_{i,s} \mathbb{E} \left( u_{i,t} u_{i,s}
\right) \right]^2 = \left( \frac{1}{N_0}\sum_{i=1}^{N_0} \ehat_{i,t}^2
\right)^2 =O_\mathbb{P} (1).
\end{align*}
Moreover, by the $C_p$ inequality and Lemma \ref{lemma:p order sample mean
is bounded}(9),
\begin{align*}
\frac{1}{T}\sum_{t=1}^T \sum_{s=1}^T \left[ \mathbb{E}^* \left( \frac{1}{N_0}%
\sum_{i=1}^{N_0} e_{i,t}^* e_{i,s}^* \right) \right]^2 =\frac{1}{T N_0^2}
\sum_{t=1}^T \left( \sum_{i=1}^{N_0} \ehat_{i,t}^2 \right)^2 \le \frac{1}{T
N_0}\sum_{t=1}^T \sum_{i=1}^{N_0} \ehat_{i,t}^4 =O_\mathbb{P}(1),
\end{align*}
and the proof is complete.
\end{pf}

Lemma \ref{lemma:BHLC square sumN ee} corresponds to Condition A*(c) of %
\citet{gonccalves2014bootstrapping} and Condition A.2 of %
\citet{gonccalves2017bootstrap}.

\begin{lemma}
\label{lemma:BHLC square sumN ee} If Assumptions \ref{ass:factors and
loadings}--\ref{ass:order conditions} hold for a pure factor model or
Assumptions \ref{ass:factors and loadings}--\ref{ass:error is independent of
factor and x} hold for a factor model with covariates, then as $N_0,
T_0\to\infty$,

\begin{enumerate}[label=(\arabic*), labelindent=\parindent, leftmargin=*, nosep]

\item For every $t\in\{1,\ldots,T\}$,
\begin{align*}
\frac{1}{T}\sum_{s=1}^T \mathbb{E}^* \left[ \left( \frac{1}{\sqrt{N_0}}%
\sum_{i=1}^{N_0} e_{i,t}^* e_{i,s}^* -\mathbb{E}^* \left( e_{i,t}^*
e_{i,s}^* \right) \right)^2 \right] =O_\mathbb{P}(1).
\end{align*}

\item
\begin{align*}
\frac{1}{T^2}\sum_{t=1}^T \sum_{s=1}^T \mathbb{E}^* \left[ \left( \frac{1}{%
\sqrt{N_0}}\sum_{i=1}^{N_0} e_{i,t}^* e_{i,s}^* -\mathbb{E}^* \left(
e_{i,t}^* e_{i,s}^* \right) \right)^2 \right] =O_\mathbb{P}(1).
\end{align*}
\end{enumerate}
\end{lemma}

\begin{pf}{}
Because $\left\{ u_{i,t} \right\}$ are i.i.d.\ as $\bN(0,1)$ and independent
of the raw sample, we have that $e_{i,t}^*$ and $e_{j,s}^*$ are
conditionally independent given the sample whenever $i\ne j$ or $t\ne s$,
and $\mathrm{Var}\left( u_{i,t} u_{i,s} \right)\le \mathbb{E}\left(
u_{i,t}^4 \right)=3$. Then by Cauchy-Schwarz inequality, the $C_p$
inequality, and Lemma \ref{lemma:p order sample mean is bounded}(8)(9),
\begin{align*}
& \frac{1}{T}\sum_{s=1}^T \mathbb{E}^* \left[ \left( \frac{1}{\sqrt{N_0}}%
\sum_{i=1}^{N_0} e_{i,t}^* e_{i,s}^* -\mathbb{E}^* \left( e_{i,t}^*
e_{i,s}^* \right) \right)^2 \right] =\frac{1}{T N_0}\sum_{s=1}^T \mathrm{Var}%
^* \left( \sum_{i=1}^{N_0} e_{i,t}^* e_{i,s}^* \right) \\
& =\frac{1}{T N_0}\sum_{s=1}^T \sum_{i=1}^{N_0} \ehat_{i,t}^2 \ehat_{i,s}^2
\mathrm{Var}\left( u_{i,t} u_{i,s} \right) \le \frac{3}{N_0}\sum_{i=1}^{N_0} %
\left[ \ehat_{i,t}^2 \left( \frac{1}{T}\sum_{s=1}^T \ehat_{i,s}^2 \right) %
\right] \\
& \le 3 \sqrt{\frac{1}{N_0}\sum_{i=1}^{N_0} \ehat_{i,t}^4} \sqrt{\frac{1}{N_0%
}\sum_{i=1}^{N_0} \left( \frac{1}{T}\sum_{s=1}^T \ehat_{i,s}^2 \right)^2}
\le 3 \sqrt{\frac{1}{N_0}\sum_{i=1}^{N_0} \ehat_{i,t}^4} \sqrt{\frac{1}{T N_0%
}\sum_{s=1}^T\sum_{i=1}^{N_0} \ehat_{i,s}^4}=O_\mathbb{P}(1).
\end{align*}
To prove the second claim, we use the $C_p$ inequality and Lemma \ref%
{lemma:p order sample mean is bounded}(9) to conclude that
\begin{align*}
& \frac{1}{T^2}\sum_{t=1}^T \sum_{s=1}^T \mathbb{E}^* \left[ \left( \frac{1}{%
\sqrt{N_0}}\sum_{i=1}^{N_0} e_{i,t}^* e_{i,s}^* -\mathbb{E}^* \left(
e_{i,t}^* e_{i,s}^* \right) \right)^2 \right] \le \frac{3}{T^2 N_0}%
\sum_{i=1}^{N_0} \sum_{t=1}^T\sum_{s=1}^T \ehat_{i,t}^2 \ehat_{i,s}^2 \\
& =\frac{3}{N_0}\sum_{i=1}^{N_0} \left( \frac{1}{T}\sum_{s=1}^T \ehat%
_{i,s}^2 \right)^2 \le \frac{3}{ T N_0} \sum_{s=1}^T\sum_{i=1}^{N_0}\ehat%
_{i,s}^4=O_\mathbb{P}(1),
\end{align*}
and the proof is complete.
\end{pf}

Lemma \ref{lemma:BHLC sum fee} corresponds to Condition A.3 of %
\citet{gonccalves2017bootstrap}.

\begin{lemma}
\label{lemma:BHLC sum fee} If Assumptions \ref{ass:factors and loadings}--%
\ref{ass:order conditions} hold for a pure factor model or Assumptions \ref%
{ass:factors and loadings}--\ref{ass:error is independent of factor and x}
hold for a factor model with covariates, then as $N_0, T_0\to\infty$,
\begin{align*}
\mathbb{E}^* \left( \left\Vert \frac{1}{\sqrt{T N_0}} \sum_{s=1}^T
\sum_{i=1}^{N_0} \ftid_s \left[ e_{i,t}^* e_{i,s}^*-\mathbb{E}^* \left(
e_{i,t}^* e_{i,s}^* \right) \right] \right\Vert^2 \right) =O_\mathbb{P}(1).
\end{align*}
\end{lemma}

\begin{pf}{}
We still start from the i.i.d.\ nature of $\left\{ u_{i,t} \right\}$ and
consider $\mathrm{Cov} \left( u_{i,t}u_{i,s}, u_{j,t}u_{j,q} \right)$. If $%
i\ne j$, then $u_{i,t}u_{i,s}$ and $u_{j,t}u_{j,q}$ are independent of each
other, and hence $\mathrm{Cov} \left( u_{i,t}u_{i,s}, u_{j,t}u_{j,q}
\right)=0$. If $s\ne q$, then (1) $u_{i,s}$ and $u_{j,q}$ are independent of
each other; (2) either $u_{i,t}$ is independent of $u_{i,s}$, or $u_{j,t} $
is independent of $u_{j,q}$. This implies that $\mathbb{E}\left(
u_{i,t}u_{i,s}u_{j,t}u_{j,q} \right)=0$, $\mathbb{E}\left( u_{i,t}u_{i,s}
\right) \mathbb{E}\left( u_{j,t}u_{j,q} \right)=0$, and hence $\mathrm{Cov}
\left( u_{i,t}u_{i,s}, u_{j,t}u_{j,q} \right)=0$. Now we have established
that $\mathrm{Cov} \left( u_{i,t}u_{i,s}, u_{j,t}u_{j,q} \right)\ne 0$ only
if $i=j$ and $s=q$. By this fact and Cauchy-Schwarz inequality,
\begin{align*}
& \mathbb{E}^* \left( \left\Vert \frac{1}{\sqrt{T N_0}} \sum_{s=1}^T
\sum_{i=1}^{N_0} \ftid_s \left[ e_{i,t}^* e_{i,s}^*-\mathbb{E}^* \left(
e_{i,t}^* e_{i,s}^* \right) \right] \right\Vert^2 \right) \\
& = \mathbb{E}^* \left( \frac{1}{T N_0}\sum_{s=1}^T \sum_{q=1}^T
\sum_{i=1}^{N_0} \sum_{j=1}^{N_0} \ftid_s^\mathsf{T} \ftid_q \left[
e_{i,t}^* e_{i,s}^* -\mathbb{E}^* \left( e_{i,t}^* e_{i,s}^* \right) \right] %
\left[ e_{j,t}^* e_{j,q}^*-\mathbb{E}^* \left( e_{j,t}^* e_{j,q}^* \right) %
\right] \right) \\
&= \frac{1}{T N_0} \sum_{s=1}^T \sum_{q=1}^T \sum_{i=1}^{N_0}
\sum_{j=1}^{N_0} \ftid_s^\mathsf{T} \ftid_q \mathrm{Cov}^* \left( e_{i,t}^*
e_{i,s}^*, e_{j,t}^* e_{j,q}^* \right) \\
&= \frac{1}{T N_0} \sum_{s=1}^T \sum_{q=1}^T \sum_{i=1}^{N_0}
\sum_{j=1}^{N_0} \ftid_s^\mathsf{T} \ftid_q \ehat_{i,t} \ehat_{i,s} \ehat%
_{j,t} \ehat_{j,q} \mathrm{Cov} \left( u_{i,t}u_{i,s}, u_{j,t}u_{j,q} \right)
\\
&= \frac{1}{T N_0}\sum_{s=1}^T \sum_{i=1}^{N_0} \ftid_s^\mathsf{T} \ftid_s %
\ehat_{i,t}^2 \ehat_{i,s}^2 \mathrm{Var} \left( u_{i,t}u_{i,s} \right) \\
& \le \frac{3}{N_0} \sum_{i=1}^{N_0} \left[ \ehat_{i,t}^2 \left( \frac{1}{T}%
\sum_{s=1}^T \ftid_s^\mathsf{T} \ftid_s \ehat_{i,s}^2 \right) \right] \\
& \le 3 \sqrt{\frac{1}{N_0}\sum_{i=1}^{N_0} \ehat_{i,t}^4} \sqrt{\frac{1}{%
T^2 N_0} \sum_{i=1}^{N_0} \left( \sum_{s=1}^T \ftid_s^\mathsf{T} \ftid_s %
\ehat_{i,s}^2 \right)^2}.
\end{align*}
By the $C_p$ inequality and Cauchy-Schwarz inequality,
\begin{align*}
& \frac{1}{T^2 N_0} \sum_{i=1}^{N_0} \left( \sum_{s=1}^T \ftid_s^\mathsf{T} %
\ftid_s \ehat_{i,s}^2 \right)^2 \le \frac{1}{T N_0} \sum_{s=1}^T
\sum_{i=1}^{N_0} \left\Vert \ftid_s \right\Vert^4 \ehat_{i,s}^4 \\
& \le \sqrt{\frac{1}{T}\sum_{s=1}^T \left\Vert \ftid_s \right\Vert^8} \sqrt{%
\frac{1}{T}\sum_{s=1}^T\left( \frac{1}{N_0}\sum_{i=1}^{N_0}\ehat_{i,s}^4
\right)^2} \le \sqrt{\frac{1}{T}\sum_{s=1}^T \left\Vert \fhattalls %
\right\Vert^8} \sqrt{\frac{1}{T N_0}\sum_{s=1}^T\sum_{i=1}^{N_0} \ehat%
_{i,s}^8}.
\end{align*}
The desired result follows from Lemma \ref{lemma:p order sample mean is
bounded}(3)(8)(9).
\end{pf}

Lemma \ref{lemma:BHLC sum f lambda e} corresponds to Condition A.4 of %
\citet{gonccalves2017bootstrap}.

\begin{lemma}
\label{lemma:BHLC sum f lambda e} If Assumptions \ref{ass:factors and
loadings}--\ref{ass:order conditions} hold for a pure factor model or
Assumptions \ref{ass:factors and loadings}--\ref{ass:error is independent of
factor and x} hold for a factor model with covariates, then as $N_0,
T_0\to\infty$,
\begin{align*}
\mathbb{E}^* \left( \left\Vert \frac{1}{\sqrt{T N_0}} \sum_{t=1}^T
\sum_{i=1}^{N_0} \ftid_t \ltid_i^\mathsf{T} e_{i,t}^* \right\Vert^2 \right)
=O_\mathbb{P}(1).
\end{align*}
\end{lemma}

\begin{pf}{}
From the definition of Frobenius norm and the properties of trace,
\begin{align*}
& \left\Vert \frac{1}{\sqrt{T N_0}} \sum_{t=1}^T \sum_{i=1}^{N_0} \ftid_t %
\ltid_i^\mathsf{T} e_{i,t}^* \right\Vert^2 = \tr \left( \frac{1}{T N_0}%
\sum_{t=1}^T \sum_{s=1}^T \sum_{i=1}^{N_0}\sum_{j=1}^{N_0} \ftid_t \ltid_i^%
\mathsf{T} \ltid_j \ftid_s^\mathsf{T} e_{i,t}^* e_{j,s}^* \right) \\
& =\frac{1}{T N_0}\sum_{t=1}^T \sum_{s=1}^T \sum_{i=1}^{N_0}\sum_{j=1}^{N_0} %
\tr \left( \ftid_t \ltid_i^\mathsf{T} \ltid_j \ftid_s^\mathsf{T} \right)
e_{i,t}^* e_{j,s}^* = \frac{1}{T N_0}\sum_{t=1}^T \sum_{s=1}^T
\sum_{i=1}^{N_0}\sum_{j=1}^{N_0} \ftid_s^\mathsf{T} \ftid_t \ltid_i^\mathsf{T%
} \ltid_j e_{i,t}^* e_{j,s}^*.
\end{align*}
Because $\left\{ u_{i,t} \right\}$ are i.i.d.\ as $\bN (0,1)$, we have $%
\mathbb{E}\left( u_{i,t}u_{j,s} \right) =\mathbbm{1}_{\{i=j\}}\mathbbm{1}%
_{\{t=s\}}$. By the above facts, the properties of matrix norms, and
Cauchy-Schwarz inequality,
\begin{align*}
& \mathbb{E}^* \left( \left\Vert \frac{1}{\sqrt{T N_0}} \sum_{t=1}^T
\sum_{i=1}^{N_0} \ftid_t \ltid_i^\mathsf{T} e_{i,t}^* \right\Vert^2 \right)
= \mathbb{E}^* \left( \frac{1}{T N_0}\sum_{t=1}^T \sum_{s=1}^T
\sum_{i=1}^{N_0}\sum_{j=1}^{N_0} \ftid_s^\mathsf{T} \ftid_t \ltid_i^\mathsf{T%
} \ltid_j e_{i,t}^* e_{j,s}^* \right) \\
& = \frac{1}{T N_0}\sum_{t=1}^T \sum_{s=1}^T
\sum_{i=1}^{N_0}\sum_{j=1}^{N_0} \ftid_s^\mathsf{T} \ftid_t \ltid_i^\mathsf{T%
} \ltid_j \ehat_{i,t} \ehat_{j,s} \mathbb{E}\left( u_{i,t} u_{j,s} \right) =%
\frac{1}{T N_0}\sum_{t=1}^T \sum_{i=1}^{N_0} \left\Vert \ftid_t
\right\Vert^2 \left\Vert \ltid_i \right\Vert^2 \ehat_{i,t}^2 \\
& \le \sqrt{\frac{1}{T}\sum_{t=1}^T \left\Vert \ftid_t \right\Vert^4 } \sqrt{%
\frac{1}{T}\sum_{t=1}^T \left( \frac{1}{N_0}\sum_{i=1}^{N_0} \left\Vert \ltid%
_i \right\Vert^2 \ehat_{i,t}^2 \right)^2 } \\
& \le \left\Vert \Hhatmiss \right\Vert^2 \sqrt{\frac{1}{T}\sum_{t=1}^T
\left\Vert \fhattallt \right\Vert^4 } \sqrt{\frac{1}{T}\sum_{t=1}^T \left(
\frac{1}{N_0}\sum_{i=1}^{N_0} \left\Vert \lhatwidei \right\Vert^2 \ehat%
_{i,t}^2 \right)^2 }.
\end{align*}
By the $C_p$ inequality and Cauchy-Schwarz inequality,
\begin{align*}
& \frac{1}{T}\sum_{t=1}^T \left( \frac{1}{N_0}\sum_{i=1}^{N_0} \left\Vert %
\lhatwidei \right\Vert^2 \ehat_{i,t}^2 \right)^2 \le \frac{1}{T N_0}%
\sum_{t=1}^T \sum_{i=1}^{N_0} \sum_{t=1}^T \sum_{i=1}^{N_0} \left\Vert %
\lhatwidei \right\Vert^4 \ehat_{i,t}^4 \\
& \le \sqrt{\frac{1}{N_0}\sum_{i=1}^{N_0} \left\Vert \lhatwidei \right\Vert^8%
} \sqrt{\frac{1}{N_0}\sum_{i=1}^{N_0}\left( \frac{1}{T}\sum_{t=1}^T \ehat%
_{i,t}^4 \right)^2} \le \sqrt{\frac{1}{N_0}\sum_{i=1}^{N_0} \left\Vert %
\lhatwidei \right\Vert^8} \sqrt{\frac{1}{T N_0}\sum_{i=1}^{N_0}\sum_{t=1}^T %
\ehat_{i,t}^8}.
\end{align*}
The proof is completed by Lemmas \ref{lemma:pointwise convergence of factors
and loadings in sec3}(5), \ref{lemma:p order sample mean is bounded}%
(3)(4)(9), and Assumption \ref{ass:order conditions}(2).
\end{pf}

Lemma \ref{lemma:BHLC sum lambda e} corresponds to Condition B*(d) of %
\citet{gonccalves2014bootstrapping} and Conditions A.5, A.6 of %
\citet{gonccalves2017bootstrap}.

\begin{lemma}
\label{lemma:BHLC sum lambda e} If Assumptions \ref{ass:factors and loadings}%
--\ref{ass:order conditions} hold for a pure factor model or Assumptions \ref%
{ass:factors and loadings}--\ref{ass:error is independent of factor and x}
hold for a factor model with covariates, then as $N_0, T_0\to\infty$,

\begin{enumerate}[label=(\arabic*), labelindent=\parindent, leftmargin=*, nosep]

\item For every $t\in \{ 1, \ldots, T\}$,
\begin{align*}
\mathbb{E}^* \left( \left\Vert \frac{1}{\sqrt{N_0}} \sum_{i=1}^{N_0}\ltid_i
e_{i,t}^* \right\Vert^2 \right) =O_\mathbb{P}(1).
\end{align*}

\item
\begin{align*}
\frac{1}{T}\sum_{t=1}^T \mathbb{E}^* \left( \left\Vert \frac{1}{\sqrt{N_0}}
\sum_{i=1}^{N_0}\ltid_i e_{i,t}^* \right\Vert^2 \right) =O_\mathbb{P}(1).
\end{align*}
\end{enumerate}
\end{lemma}

\begin{pf}{}
Note that $\mathbb{E}\left( u_{i,t} u_{j,t} \right)=\mathbbm{1}_{\{i=j\}}$ by
the i.i.d.\ $\bN(0,1)$ nature of $\left\{ u_{i,t} \right\}$. By the
properties of matrix norms and Cauchy-Schwarz inequality,
\begin{align*}
& \mathbb{E}^* \left( \left\Vert \frac{1}{\sqrt{N_0}} \sum_{i=1}^{N_0} \ltid%
_i e_{i,t}^* \right\Vert^2 \right) =\mathbb{E}^* \left( \frac{1}{N_0}%
\sum_{i=1}^{N_0} \sum_{j=1}^{N_0} \ltid_i^\mathsf{T} \ltid_j e_{i,t}^*
e_{j,t}^* \right) \\
& = \frac{1}{N_0}\sum_{i=1}^{N_0} \sum_{j=1}^{N_0} \ltid_i^\mathsf{T} \ltid%
_j \ehat_{i,t} \ehat_{j,t} \mathbb{E}\left( u_{i,t}u_{j,t} \right) = \frac{1%
}{N_0} \sum_{i=1}^{N_0} \left\Vert \ltid_i \right\Vert^2 \ehat_{i,t}^2 \\
& \le \left\Vert \Hhatmiss \right\Vert^2 \left( \frac{1}{N_0}
\sum_{i=1}^{N_0} \left\Vert \lhatwidei \right\Vert^2 \ehat_{i,t}^2 \right)
\le \left\Vert \Hhatmiss \right\Vert^2 \sqrt{\frac{1}{N_0}\sum_{i=1}^{N_0}
\left\Vert \lhatwidei \right\Vert^4} \sqrt{\frac{1}{N_0}\sum_{i=1}^{N_0} %
\ehat_{i,t}^4}.
\end{align*}
Then the first claim follows from Lemmas \ref{lemma:pointwise convergence of
factors and loadings in sec3}(5), \ref{lemma:p order sample mean is bounded}%
(4)(8) and Assumption \ref{ass:order conditions}(2).

To show the second claim, we use the above results, Cauchy-Schwarz
inequality and the $C_p$ inequality to conclude that
\begin{align*}
& \frac{1}{T}\sum_{t=1}^T \mathbb{E}^* \left( \left\Vert \frac{1}{\sqrt{N_0}}
\sum_{i=1}^{N_0}\ltid_i e_{i,t}^* \right\Vert^2 \right) \le \left\Vert %
\Hhatmiss \right\Vert^2 \left( \frac{1}{T N_0} \sum_{t=1}^T \sum_{i=1}^{N_0}
\left\Vert \lhatwidei \right\Vert^2 \ehat_{i,t}^2 \right) \\
& \le \left\Vert \Hhatmiss \right\Vert^2 \sqrt{\frac{1}{N_0}\sum_{i=1}^{N_0}
\left\Vert \lhatwidei \right\Vert^4} \sqrt{\frac{1}{N_0}\sum_{i=1}^{N_0}%
\left( \frac{1}{T}\sum_{t=1}^T \ehat_{i,t}^2 \right)^2} \\
& \le \left\Vert \Hhatmiss \right\Vert^2 \sqrt{\frac{1}{N_0}\sum_{i=1}^{N_0}
\left\Vert \lhatwidei \right\Vert^4} \sqrt{\frac{1}{T N_0}%
\sum_{t=1}^T\sum_{i=1}^{N_0} \ehat_{i,t}^4}.
\end{align*}
The proof is completed by Lemmas \ref{lemma:pointwise convergence of factors
and loadings in sec3}(5), \ref{lemma:p order sample mean is bounded}(4)(9)
and Assumption \ref{ass:order conditions}(2).
\end{pf}

In this paper, more bootstrap high level conditions are used in subsequent
contents, and we list them as lemmas below.

\begin{lemma}
\label{lemma:BHLC e} If Assumptions \ref{ass:factors and loadings}--\ref%
{ass:order conditions} hold for a pure factor model or Assumptions \ref%
{ass:factors and loadings}--\ref{ass:error is independent of factor and x}
hold for a factor model with covariates, then for every $m\in\{1,2,\ldots,
8\}$, as $N_0, T_0\to\infty$,

\begin{enumerate}[label=(\arabic*), labelindent=\parindent, leftmargin=*, nosep]

\item
\begin{align*}
\frac{1}{T_0}\sum_{t=1}^{T_0} \left\vert e_{i,t}^* \right\vert^m =\Ops(1)
\quad \text{for every } i\in \{1,\ldots, N\}.
\end{align*}

\item
\begin{align*}
\frac{1}{N_0}\sum_{i=1}^{N_0} \left\vert e_{i,t}^* \right\vert^m =\Ops(1)
\quad \text{for every } t\in \{1,\ldots, T\}.
\end{align*}

\item
\begin{align*}
\frac{1}{T_0 N}\sum_{t=1}^{T_0} \sum_{i=1}^{N} \left\vert e_{i,t}^*
\right\vert^m =\Ops(1).
\end{align*}
\end{enumerate}
\end{lemma}

\begin{pf}{}
Since $\left\{ u_{i,t} \right\}$ are i.i.d.\ as $\bN(0,1)$, we have $\mathbb{E%
} \left( \left\vert u_{i,t} \right\vert^m \right)$ is finite for every $%
m\in\{1,2,\ldots, 8\}$. Hence $\mathbb{E}^* \left( \left\vert e_{i,t}^*
\right\vert^m \right) =\left\vert \ehat_{i,t} \right\vert^m \mathbb{E}\left(
\left\vert u_{i,t} \right\vert^m \right) $. The desired results follow from
Lemmas \ref{lemma:p order sample mean is bounded}(6)(8)(10) and \ref%
{lemma:sufficient conditions for bootstrap stochastic orders}.
\end{pf}

\begin{lemma}
\label{lemma:BHLC sum f e} If Assumptions \ref{ass:factors and loadings}--%
\ref{ass:order conditions} hold for a pure factor model or Assumptions \ref%
{ass:factors and loadings}--\ref{ass:error is independent of factor and x}
hold for a factor model with covariates, then as $N_0, T_0\to\infty$,

\begin{enumerate}[label=(\arabic*), labelindent=\parindent, leftmargin=*, nosep]

\item For every $i\in \{ 1, \ldots, N\}$,
\begin{align*}
\mathbb{E}^* \left( \left\Vert \frac{1}{\sqrt{T_0}} \sum_{t=1}^{T_0}\ftid_t
e_{i,t}^* \right\Vert^2 \right) =O_\mathbb{P}(1).
\end{align*}

\item
\begin{align*}
\frac{1}{N}\sum_{i=1}^N \mathbb{E}^* \left( \left\Vert \frac{1}{\sqrt{T_0}}
\sum_{t=1}^{T_0}\ftid_t e_{i,t}^* \right\Vert^2 \right) =O_\mathbb{P}(1).
\end{align*}
\end{enumerate}
\end{lemma}

\begin{pf}{}
Note that $\mathbb{E}\left( u_{i,t} u_{i,s} \right)=\mathbbm{1}_{\{t=s\}}$ by
the i.i.d.\ $\bN(0,1)$ nature of $\left\{ u_{i,t} \right\}$. By the
properties of matrix norms and Cauchy-Schwarz inequality,
\begin{align*}
& \mathbb{E}^* \left( \left\Vert \frac{1}{\sqrt{T_0}} \sum_{t=1}^{T_0} \ftid%
_t e_{i,t}^* \right\Vert^2 \right) =\mathbb{E}^* \left( \frac{1}{T_0}%
\sum_{t=1}^{T_0} \sum_{s=1}^{T_0} \ftid_t^\mathsf{T} \ftid_s e_{i,t}^*
e_{i,s}^* \right) = \frac{1}{T_0}\sum_{t=1}^{T_0} \sum_{s=1}^{T_0} \ftid_t^%
\mathsf{T} \ftid_s \ehat_{i,t} \ehat_{i,s} \mathbb{E}\left( u_{i,t}u_{i,s}
\right) \\
& = \frac{1}{T_0} \sum_{t=1}^{T_0} \left\Vert \ftid_t \right\Vert^2 \ehat%
_{i,t}^2 = \frac{1}{T_0} \sum_{t=1}^{T_0} \left\Vert \fhattallt %
\right\Vert^2 \ehat_{i,t}^2 \le \sqrt{\frac{1}{T_0}\sum_{t=1}^{T_0}
\left\Vert \fhattallt \right\Vert^4} \sqrt{\frac{1}{T_0}\sum_{t=1}^{T_0} %
\ehat_{i,t}^4}.
\end{align*}
Then the first claim follows from Lemma \ref{lemma:p order sample mean is
bounded}(3)(6) and Assumption \ref{ass:order conditions}(2).

To show the second claim, we use the above results, Cauchy-Schwarz
inequality and the $C_p$ inequality to conclude that
\begin{align*}
& \frac{1}{N}\sum_{i=1}^N \mathbb{E}^* \left( \left\Vert \frac{1}{\sqrt{T_0}}
\sum_{t=1}^{T_0}\ftid_t e_{i,t}^* \right\Vert^2 \right) \le \frac{1}{T_0 N}
\sum_{t=1}^{T_0} \sum_{i=1}^{N} \left\Vert \fhattallt \right\Vert^2 \ehat%
_{i,t}^2 \\
& \le \sqrt{\frac{1}{T_0}\sum_{t=1}^{T_0} \left\Vert \fhattallt \right\Vert^4%
} \sqrt{\frac{1}{T_0}\sum_{t=1}^{T_0}\left( \frac{1}{N}\sum_{i=1}^N \ehat%
_{i,t}^2 \right)^2} \le \sqrt{\frac{1}{T_0}\sum_{t=1}^{T_0} \left\Vert %
\fhattallt \right\Vert^4} \sqrt{\frac{1}{T_0 N}\sum_{t=1}^{T_0}%
\sum_{i=1}^{N} \ehat_{i,t}^4}.
\end{align*}
The proof is completed by Lemma \ref{lemma:p order sample mean is bounded}%
(3)(10) and Assumption \ref{ass:order conditions}(2).
\end{pf}

\subsection{Results about Bootstrap Statistics}

Let $\Dtall^*$, $\Fhattall^*$, $\fhattallt^*$, $\Lhattall^*$, $\lhattalli^*$%
, $\Dwide^*$, $\Fhatwide^*$, $\fhatwidet^*$, $\Lhatwide^*$, $\lhatwidei^*$, $%
\Hhatmiss^*$, $\Chat^*$, $\chat_{i,t}^*$, $\ehat_{i,t}^*$, $L_{k,i}^*$, $%
\Gamhat_t^*$, $\Phihat_i^*$, $\bVhat_{i,t}^*$, and $\sighat_i^{*2} $ be the
bootstrap analogues of $\Dtall$, $\Fhattall$, $\fhattallt$, $\Lhattall$, $%
\lhattalli$, $\Dwide$, $\Fhatwide$, $\fhatwidet$, $\Lhatwide$, $\lhatwidei$,
$\Hhatmiss$, $\Chat$, $\chat_{i,t}$, $\ehat_{i,t}$, $L_{k,i}$, $\Gamhat_t$, $%
\Phihat_i$, $\bVhat_{i,t}$, and $\sighat_i^{2}$, respectively. These
bootstrap analogues are obtained in Setp 2(4) of Algorithms \ref%
{algo:confidence interval} and \ref{algo:cov confidence interval}. Let $%
\Vhattall^*= \Dtall^{*2} $ and $\Vhatwide^*= \Dwide^{*2}$. Moreover, define
the bootstrap version of rotation matrices
\begin{align*}
\Htall^*=\left( \frac{ \Ltidtall^\mathsf{T} \Ltidtall}{N_0} \right) \left(
\frac{\Ftidtall^\mathsf{T} \Fhattall^*}{T} \right) \Vhattall^{*-1}, \qquad %
\Hwide^*=\left( \frac{ \Ltidwide^\mathsf{T} \Ltidwide}{N} \right) \left(
\frac{\Ftidwide^\mathsf{T} \Fhatwide^*}{T_0} \right) \Vhatwide^{*-1}.
\end{align*}

\begin{lemma}
\label{lemma:bootstrap rotation matrices} If Assumptions \ref{ass:factors
and loadings}--\ref{ass:order conditions} hold for a pure factor model or
Assumptions \ref{ass:factors and loadings}--\ref{ass:error is independent of
factor and x} for a factor model with covariates, then as $N_0, T_0\to\infty$%
,

\begin{enumerate}[label=(\arabic*), labelindent=\parindent, leftmargin=*, nosep]

\item $\Htall^*=H_1^*+\ops(1)$, and $\Hwide^*=H_2^*+\ops(1)$, where both $%
H_1^*$ and $H_2^*$ are diagonal matrices with $\pm 1$ on diagonals.

\item $\Vhattall^*=V+\ops(1)$, and $\Vhatwide^*=V+\ops(1)$, where $V$ is
defined in Lemma \ref{lemma:pointwise convergence of factors and loadings in
sec3}.
\end{enumerate}
\end{lemma}

\begin{pf}{}
By the definition and properties of Frobenius norm, Lemma \ref{lemma:p order
sample mean is bounded}(4), and Assumption \ref{ass:order conditions}(2),
\begin{align*}
& \left\Vert \frac{1}{N_0}\sum_{i=1}^{N_0}\lhatwidei \lhatwidei^\mathsf{T}
\right\Vert \le \frac{1}{N_0}\sum_{i=1}^{N_0} \left\Vert \lhatwidei %
\lhatwidei^\mathsf{T} \right\Vert =\frac{1}{N_0}\sum_{i=1}^{N_0} \sqrt{ \tr %
\left( \lhatwidei \lhatwidei^\mathsf{T} \lhatwidei \lhatwidei^\mathsf{T}
\right)} \\
& = \frac{1}{N_0}\sum_{i=1}^{N_0} \sqrt{ \tr \left( \lhatwidei^\mathsf{T} %
\lhatwidei \lhatwidei^\mathsf{T} \lhatwidei \right)} = \frac{1}{N_0}%
\sum_{i=1}^{N_0} \left\Vert \lhatwidei \right\Vert^2 = O_\mathbb{P}(1).
\end{align*}
Moreover,
\begin{align*}
\frac{1}{N_0} \sum_{i=1}^{N_0} \left\Vert \lhattalli \right\Vert^2= \tr %
\left( \frac{1}{N_0} \Lhattall^\mathsf{T} \Lhattall \right) = \tr \left( I_r
\right) =r.
\end{align*}
Let $\epsilon_i=\lhattalli-\Htall^{-1}\Hwide\lhatwidei$ for every $%
i\in\{1,\ldots,N_0\}$. Then by the definition and properties of Frobenius
norm, and Lemma \ref{lemma:MSE between tall and wide loadings in sec3},
\begin{align*}
\left\Vert \frac{1}{N_0} \sum_{i=1}^{N_0} \epsilon_i \epsilon_i^\mathsf{T}
\right\Vert \le \frac{1}{N_0} \sum_{i=1}^{N_0} \left\Vert \epsilon_i
\epsilon_i^\mathsf{T} \right\Vert = \frac{1}{N_0} \sum_{i=1}^{N_0}
\left\Vert \epsilon_i \right\Vert^2 =O_\mathbb{P} \left( \frac{1}{T_0}
\right).
\end{align*}
By the triangle inequality, the properties of matrix norms, Cauchy-Schwarz
inequality, and the above facts,
\begin{align*}
\left\Vert \frac{1}{N_0}\sum_{i=1}^{N_0} \lhattalli \epsilon_i^\mathsf{T}
\right\Vert \le \frac{1}{N_0}\sum_{i=1}^{N_0} \left\Vert \lhattalli %
\right\Vert \left\Vert \epsilon_i \right\Vert \le \sqrt{\frac{1}{N_0}%
\sum_{i=1}^{N_0}\left\Vert \lhattalli\right\Vert^2} \sqrt{\frac{1}{N_0}%
\sum_{i=1}^{N_0}\left\Vert \epsilon_i \right\Vert^2} =O_\mathbb{P}\left(
\frac{1}{\sqrt{T_0}} \right).
\end{align*}
Combining these results with Lemma \ref{lemma:pointwise convergence of
factors and loadings in sec3}(2)(5) yields
\begin{align*}
& \frac{1}{N_0}\Ltidtall^\mathsf{T} \Ltidtall= \Hhatmiss \left( \frac{1}{N_0}
\sum_{i=1}^{N_0} \lhatwidei \lhatwidei^\mathsf{T} \right) \Hhatmiss^\mathsf{T%
} \\
& = \left( \Htall^{-1}\Hwide +\Hhatmiss -\Htall^{-1}\Hwide \right) \left(
\frac{1}{N_0} \sum_{i=1}^{N_0} \lhatwidei \lhatwidei^\mathsf{T} \right)
\left( \Htall^{-1}\Hwide +\Hhatmiss -\Htall^{-1}\Hwide \right)^\mathsf{T} \\
&= \Htall^{-1}\Hwide \left( \frac{1}{N_0} \sum_{i=1}^{N_0} \lhatwidei %
\lhatwidei^\mathsf{T} \right) \Hwide^\mathsf{T} \Htall^{\mathsf{T} -1} \\
& + \left( \Hhatmiss -\Htall^{-1}\Hwide \right) \left( \frac{1}{N_0}
\sum_{i=1}^{N_0} \lhatwidei \lhatwidei^\mathsf{T} \right) \Hwide^\mathsf{T} %
\Htall^{\mathsf{T} -1} \\
& + \Htall^{-1}\Hwide \left( \frac{1}{N_0} \sum_{i=1}^{N_0} \lhatwidei %
\lhatwidei^\mathsf{T} \right) \left( \Hhatmiss -\Htall^{-1}\Hwide \right)^%
\mathsf{T} \\
& +\left( \Hhatmiss -\Htall^{-1}\Hwide \right) \left( \frac{1}{N_0}
\sum_{i=1}^{N_0} \lhatwidei \lhatwidei^\mathsf{T} \right) \left( \Hhatmiss -%
\Htall^{-1}\Hwide \right)^\mathsf{T} \\
&=\Htall^{-1}\Hwide \left( \frac{1}{N_0} \sum_{i=1}^{N_0} \lhatwidei %
\lhatwidei^\mathsf{T} \right) \Hwide^\mathsf{T} \Htall^{\mathsf{T} -1} +O_%
\mathbb{P}\left( \frac{1}{T_0} \right) \\
&= \frac{1}{N_0} \sum_{i=1}^{N_0} \left( \lhattalli-\epsilon_i \right)
\left( \lhattalli-\epsilon_i \right)^\mathsf{T} +O_\mathbb{P}\left( \frac{1}{%
T_0} \right) \\
&= \frac{1}{N_0}\Lhattall^\mathsf{T} \Lhattall -\frac{1}{N_0}%
\sum_{i=1}^{N_0} \lhattalli \epsilon_i^\mathsf{T} -\frac{1}{N_0}%
\sum_{i=1}^{N_0} \epsilon_i \lhattalli^\mathsf{T} +\frac{1}{N_0}%
\sum_{i=1}^{N_0}\epsilon_i\epsilon_i^\mathsf{T} +O_\mathbb{P}\left( \frac{1}{%
T_0} \right) \\
&= \Vhattall +O_\mathbb{P} \left( \frac{1}{\sqrt{T_0}} \right)+O_\mathbb{P}%
\left( \frac{1}{T_0} \right) =\Vhattall+o_\mathbb{P}(1).
\end{align*}

For the pre-treatment subsample, it follows immediately from Lemma \ref%
{lemma:pointwise convergence of factors and loadings in sec3}(5) that
\begin{align*}
\frac{1}{N}\Ltidwide^\mathsf{T} \Ltidwide=\Hhatmiss \left( \frac{1}{N} %
\Lhatwide^\mathsf{T} \Lhatwide \right) \Hhatmiss^\mathsf{T} =\Vhatwide+o_%
\mathbb{P}(1).
\end{align*}
And the desired results follow from the above facts and the proof of Lemma
B.1 of \citet{gonccalves2014bootstrapping}.
\end{pf}

\begin{lemma}
\label{lemma:MSE between bootstrap ftall and ftid} If Assumptions \ref%
{ass:factors and loadings}--\ref{ass:order conditions} hold for a pure
factor model or Assumptions \ref{ass:factors and loadings}--\ref{ass:error
is independent of factor and x} for a factor model with covariates, then for
any integer $m\ge 2$, as $N_0, T_0\to\infty$,

\begin{enumerate}[label=(\arabic*), labelindent=\parindent, leftmargin=*, nosep]

\item For the control subsample,
\begin{align*}
\frac{1}{T} \sum_{t=1}^T \left\Vert \fhattallt^*-\Htall^{*\mathsf{T}} \ftid%
_t \right\Vert^m =\Ops \left( \frac{1}{T} \right).
\end{align*}

\item For the pre-treatment subsample,
\begin{align*}
\frac{1}{T_0} \sum_{t=1}^{T_0} \left\Vert \fhatwidet^*-\Hwide^{*\mathsf{T}} %
\ftid_t \right\Vert^m =\Ops \left( \frac{1}{T_0} \right).
\end{align*}
\end{enumerate}
\end{lemma}

\begin{pf}{}
For $m=2$, the first claim follows by applying Lemma 3.1 of %
\citet{gonccalves2014bootstrapping} to the control subsample. Note that
Lemma 3.1 of \citet{gonccalves2014bootstrapping} relies only on their
Conditions A*(b), A*(c), and B*(d), which have been verified by Lemma \ref%
{lemma:BHLC sumN ee}(2), \ref{lemma:BHLC square sumN ee}(2), and \ref%
{lemma:BHLC sum lambda e}(2) of this paper. One can easily see that
conclusions of Lemma \ref{lemma:BHLC sumN ee}(2), \ref{lemma:BHLC square
sumN ee}(2), and \ref{lemma:BHLC sum lambda e}(2) also hold for the
pre-treatment subsample, which implies that we can apply Lemma 3.1 of %
\citet{gonccalves2014bootstrapping} to the pre-treatment subsample to prove
Claim (2).

Now consider $m>2$. The case for $m=2$ and Assumptions \ref{ass:order
conditions}(2) implies that
\begin{align*}
\sum_{t=1}^{T} \left\Vert \fhattallt^*-\Htall^{*\mathsf{T}} \ftid_t
\right\Vert^2=\Ops(1) \quad \text{and} \quad \sum_{t=1}^{T_0} \left\Vert %
\fhatwidet^*-\Hwide^{*\mathsf{T}} \ftid_t \right\Vert^2=\Ops(1),
\end{align*}
which in turn implies that
\begin{align*}
\sum_{t=1}^{T} \left\Vert \fhattallt^*-\Htall^{*\mathsf{T}} \ftid_t
\right\Vert^m =\Ops(1) \quad \text{and} \quad \sum_{t=1}^{T_0} \left\Vert %
\fhatwidet^*-\Hwide^{*\mathsf{T}} \ftid_t \right\Vert^m =\Ops(1)
\end{align*}
holds for any integer $m>2$. Then the proof is complete.
\end{pf}

\begin{lemma}
\label{lemma:point convergence of bootstrap ftall} If Assumptions \ref%
{ass:factors and loadings}--\ref{ass:order conditions} hold for a pure
factor model or Assumptions \ref{ass:factors and loadings}--\ref{ass:error
is independent of factor and x} for a factor model with covariates, then for
every $t\in \{1,\ldots,T \} $, as $N_0, T_0\to\infty$,
\begin{align*}
\fhattallt^*-\Htall^{*\mathsf{T}}\ftid_t = \Ops \left( \frac{1}{T^{1/4}}
\right).
\end{align*}
\end{lemma}

\begin{pf}{}
The proof is completed by applying Lemma 2 of \citet{gonccalves2017bootstrap}
and the identity before that lemma to the control subsample. Note that Lemma
2 of \citet{gonccalves2017bootstrap} requires their Assumptions 1--2 and
Condition A. Assumptions 1--2 of \citet{gonccalves2017bootstrap} are implied
by Assumptions \ref{ass:factors and loadings}--\ref{ass:order conditions} of
this paper, and Condition A of \citet{gonccalves2017bootstrap} has been
verified by Lemmas \ref{lemma:BHLC sumN ee}--\ref{lemma:BHLC sum lambda e}
of this paper.

The convergence rate $T^{-1/4}$ differs from what is implied by Lemma 2 of %
\citet{gonccalves2017bootstrap}, \textit{i.e.}, $N_0^{-1/2}$, because we use
a looser bound for the term $b_{2t}^*$ defined in the proof of Lemma 2 of %
\citet{gonccalves2017bootstrap}. Specifically, by the triangle inequality
and Cauchy-Schwarz inequality,
\begin{align*}
& \left\Vert b_{2t}^* \right\Vert \le \frac{1}{T}\sum_{s=1}^T \left\Vert f_s
\right\Vert \left\vert \gamma^*_{s,t} \right\vert \le \sqrt{\frac{1}{T}%
\sum_{s=1}^T \left\Vert f_s \right\Vert^2} \sqrt{\frac{1}{T}\sum_{s=1}^T
\gamma_{s,t}^{*2}} \\
& \le \left( \max_{1\le s\le T} \left\Vert f_s \right\Vert \right) \sqrt{%
\frac{1}{T} \sum_{s=1}^T \gamma_{s,t}^{*2}}=O_\mathbb{P} \left( {T^{1/4}}
\right) O_\mathbb{P} \left( \frac{1}{\sqrt{T}} \right) =O_\mathbb{P} \left(
\frac{1}{T^{1/4}} \right),
\end{align*}
and this term turns out to be the dominant term of $\left( \fhattallt^*-%
\Htall^{*\mathsf{T}}\ftid_t\right)$.
\end{pf}

For every $i\in\{1,\ldots,N\}$, define $\ewidei^*=\left( e_{i,1}^*, \ldots,
e_{i,T_0}^* \right)^\mathsf{T}$.

\begin{lemma}
\label{lemma:FFHe} If Assumptions \ref{ass:factors and loadings}--\ref%
{ass:order conditions} hold for a pure factor model or Assumptions \ref%
{ass:factors and loadings}--\ref{ass:error is independent of factor and x}
for a factor model with covariates, then as $N_0, T_0\to\infty$,

\begin{enumerate}[label=(\arabic*), labelindent=\parindent, leftmargin=*, nosep]

\item For every $i\in \{1,\ldots,N\}$,
\begin{align*}
\frac{1}{T_0}\left( \Fhatwide^*-\Ftidwide \Hwide^* \right)^\mathsf{T} \ewidei%
^* =\Ops \left( \frac{1}{\sqrt{T_0}} \right).
\end{align*}

\item
\begin{align*}
\frac{1}{N}\sum_{i=1}^N \left\Vert \frac{1}{T_0}\left( \Fhatwide^*-\Ftidwide %
\Hwide^* \right)^\mathsf{T} \ewidei^* \right\Vert^2 =\Ops \left( \frac{1}{T_0%
} \right).
\end{align*}
\end{enumerate}
\end{lemma}

\begin{pf}{}
By the triangle inequality, Cauchy-Schwarz inequality, and Lemmas \ref%
{lemma:BHLC e}(1), \ref{lemma:MSE between bootstrap ftall and ftid}(2),
\begin{align*}
&\left\Vert \frac{1}{T_0}\left( \Fhatwide^*-\Ftidwide \Hwide^* \right)^%
\mathsf{T} \ewidei^* \right\Vert =\left\Vert \frac{1}{T_0} \sum_{s=1}^{T_0}
\left( \fhatwides^*-\Hwide^{*\mathsf{T}}\ftid_s \right) e_{i,s}^* \right\Vert
\\
& \le \frac{1}{T_0} \sum_{s=1}^{T_0} \left\Vert \fhatwides^*-\Hwide^{*%
\mathsf{T}}\ftid_s \right\Vert \left\vert e_{i,s}^* \right\vert \le \sqrt{%
\frac{1}{T_0}\sum_{s=1}^{T_0}\left\Vert \fhatwides^*-\Hwide^{*\mathsf{T}}%
\ftid_s \right\Vert^2} \sqrt{\frac{1}{T_0}\sum_{s=1}^{T_0}e_{i,s}^{*2}} \\
&=\Ops \left( \frac{1}{\sqrt{T_0}} \right),
\end{align*}
which proves the first claim. For the second claim, we use the above facts
and Lemmas \ref{lemma:BHLC e}(3), \ref{lemma:MSE between bootstrap ftall and
ftid}(2) to conclude that,
\begin{align*}
& \frac{1}{N}\sum_{i=1}^N \left\Vert \frac{1}{T_0}\left( \Fhatwide^*-%
\Ftidwide \Hwide^* \right)^\mathsf{T} \ewidei^* \right\Vert^2 \\
& \le \left( \frac{1}{T_0}\sum_{s=1}^{T_0}\left\Vert \fhatwides^*-\Hwide^{*%
\mathsf{T}}\ftid_s \right\Vert^2 \right) \left( \frac{1}{T_0 N}%
\sum_{s=1}^{T_0}\sum_{i=1}^N e_{i,s}^{*2} \right) =\Ops \left( \frac{1}{T_0}
\right),
\end{align*}
and the proof is complete.
\end{pf}

\begin{lemma}
\label{lemma:HFe} If Assumptions \ref{ass:factors and loadings}--\ref%
{ass:order conditions} hold for a pure factor model or Assumptions \ref%
{ass:factors and loadings}--\ref{ass:error is independent of factor and x}
for a factor model with covariates, then as $N_0, T_0\to\infty$,

\begin{enumerate}[label=(\arabic*), labelindent=\parindent, leftmargin=*, nosep]

\item For every $i\in\{1,\ldots, N\}$,
\begin{align*}
\frac{1}{T_0}\Hwide^{*\mathsf{T}} \Ftidwide^\mathsf{T} \ewidei^*=\Ops\left(
\frac{1}{\sqrt{T_0}} \right).
\end{align*}

\item
\begin{align*}
\frac{1}{N}\sum_{i=1}^N \left\Vert \frac{1}{T_0}\Hwide^{*\mathsf{T}} %
\Ftidwide^\mathsf{T} \ewidei^* \right\Vert^2 =\Ops \left( \frac{1}{T_0}
\right).
\end{align*}
\end{enumerate}
\end{lemma}

\begin{pf}{}
Lemma \ref{lemma:BHLC sum f e}(1) implies that
\begin{align*}
\frac{1}{T_0} \Ftidwide^\mathsf{T} \ewidei^* = \frac{1}{T_0}
\sum_{t=1}^{T_0} \ftid_t e_{i,t}^* =\Ops\left( \frac{1}{\sqrt{T_0}} \right).
\end{align*}
and Claim (1) follows from Lemma \ref{lemma:bootstrap rotation matrices}(1).
Furthermore, Lemma \ref{lemma:BHLC sum f e}(2) implies that
\begin{align*}
\frac{1}{N}\sum_{i=1}^N \left\Vert \frac{1}{T_0} \Ftidwide^\mathsf{T} \ewidei%
^* \right\Vert^2 =\frac{1}{N}\sum_{i=1}^N \left\Vert \frac{1}{T_0}
\sum_{t=1}^{T_0} \ftid_t e_{i,t}^* \right\Vert^2 =\Ops \left( \frac{1}{T_0}
\right),
\end{align*}
and Claim (2) follows by the properties of matrix norms and Lemma \ref%
{lemma:bootstrap rotation matrices}(1).
\end{pf}

\begin{lemma}
\label{lemma:FFFHlam} If Assumptions \ref{ass:factors and loadings}--\ref%
{ass:order conditions} hold for a pure factor model or Assumptions \ref%
{ass:factors and loadings}--\ref{ass:error is independent of factor and x}
for a factor model with covariates, then as $N_0, T_0\to\infty$,

\begin{enumerate}[label=(\arabic*), labelindent=\parindent, leftmargin=*, nosep]

\item For every $i\in \{1,\ldots, N\}$,
\begin{align*}
\frac{1}{T_0} \Fhatwide^{*\mathsf{T}}\left( \Ftidwide-\Fhatwide^* \Hwide%
^{*-1} \right) \ltid_i =\Ops \left( \frac{1}{\sqrt{T_0}} \right).
\end{align*}

\item
\begin{align*}
\frac{1}{N} \sum_{i=1}^N \left\Vert \frac{1}{T_0} \Fhatwide^{*\mathsf{T}%
}\left( \Ftidwide-\Fhatwide^* \Hwide^{*-1} \right) \ltid_i \right\Vert^2 =%
\Ops \left( \frac{1}{T_0} \right).
\end{align*}
\end{enumerate}
\end{lemma}

\begin{pf}{}
By the triangle inequality, the properties of matrix norms, Cauchy-Schwarz
inequality, Lemmas \ref{lemma:p order sample mean is bounded}(3), \ref%
{lemma:MSE between bootstrap ftall and ftid}(2), and Assumption \ref%
{ass:order conditions}(2),
\begin{align*}
& \left\Vert \frac{1}{T_0} \left( \Fhatwide^*-\Ftidwide\Hwide^* \right)^%
\mathsf{T} \Ftidwide \right\Vert =\left\Vert \frac{1}{T_0}\sum_{s=1}^{T_0}
\left( \fhatwides^*-\Hwide^{*\mathsf{T}} \ftid_s \right) \ftid_s^\mathsf{T}
\right\Vert \\
& \le \frac{1}{T_0}\sum_{s=1}^{T_0} \left\Vert \fhatwides^*-\Hwide^{*\mathsf{%
T}} \ftid_s \right\Vert \left\Vert \ftid_s \right\Vert \le \sqrt{\frac{1}{T_0%
}\sum_{s=1}^{T_0}\left\Vert \fhatwides^*-\Hwide^{*\mathsf{T}} \ftid_s
\right\Vert^2} \sqrt{\frac{1}{T_0}\sum_{s=1}^{T_0}\left\Vert \fhattalls %
\right\Vert^2} \\
& =\Ops \left( \frac{1}{\sqrt{T_0}} \right).
\end{align*}
Note that
\begin{align*}
\frac{1}{T_0}\left( \Ftidwide-\Fhatwide^* \Hwide^{*-1} \right)^\mathsf{T} %
\Fhatwide^* =-\Hwide^{*\mathsf{T}-1} \left( b_1^* +b_2^* \right),
\end{align*}
where
\begin{align*}
b_1^*&=\frac{1}{T_0}\left( \Fhatwide^*-\Ftidwide\Hwide^* \right)^\mathsf{T} %
\Ftidwide \Hwide^*, \\
b_2^*&= \frac{1}{T_0}\left( \Fhatwide^*-\Ftidwide\Hwide^* \right)^\mathsf{T}
\left( \Fhatwide^*-\Ftidwide\Hwide^* \right).
\end{align*}
By the above result and Lemma \ref{lemma:bootstrap rotation matrices}(1), $%
b_1^*=\Ops \left( 1 \left/ \sqrt{T_0} \right. \right)$. By the triangle
inequality, the properties of matrix norms, and Lemma \ref{lemma:MSE between
bootstrap ftall and ftid}(2),
\begin{align*}
&\left\Vert b_2^* \right\Vert= \left\Vert \frac{1}{T_0}\sum_{s=1}^{T_0}
\left( \fhatwides^*-\Hwide^{*\mathsf{T}}\ftid_s \right) \left( \fhatwides^*-%
\Hwide^{*\mathsf{T}}\ftid_s \right) ^\mathsf{T} \right\Vert \\
&\le \frac{1}{T_0}\sum_{1}^{T_0} \left\Vert \fhatwides^*-\Hwide^{*\mathsf{T}}%
\ftid_s \right\Vert^2 =\Ops\left( \frac{1}{T_0} \right).
\end{align*}
This completes the proof of Claim (1). Moreover, by the properties of matrix
norms and Lemmas \ref{lemma:pointwise convergence of factors and loadings in
sec3}(5), \ref{lemma:p order sample mean is bounded}(4),
\begin{align*}
& \frac{1}{N} \sum_{i=1}^N \left\Vert \frac{1}{T_0} \Fhatwide^{*\mathsf{T}%
}\left( \Ftidwide-\Fhatwide^* \Hwide^{*-1} \right) \ltid_i \right\Vert^2 \le
\left\Vert \frac{1}{T_0} \Fhatwide^{*\mathsf{T}}\left( \Ftidwide-\Fhatwide^* %
\Hwide^{*-1} \right) \right\Vert^2 \left( \frac{1}{N}\sum_{i=1}^N \left\Vert %
\ltid_i \right\Vert^2 \right) \\
& \le \left\Vert \frac{1}{T_0} \Fhatwide^{*\mathsf{T}}\left( \Ftidwide-%
\Fhatwide^* \Hwide^{*-1} \right) \right\Vert^2 \left\Vert \Hhatmiss %
\right\Vert^2 \left( \frac{1}{N}\sum_{i=1}^N \left\Vert \lhatwidei %
\right\Vert^2 \right) =\Ops \left( \frac{1}{T_0} \right),
\end{align*}
which proves Claim (2).
\end{pf}

\begin{lemma}
\label{lemma:point convergence of bootstrap lamwide} If Assumptions \ref%
{ass:factors and loadings}--\ref{ass:order conditions} hold for a pure
factor model or Assumptions \ref{ass:factors and loadings}--\ref{ass:error
is independent of factor and x} for a factor model with covariates, then for
every $i\in\{1,\ldots, N\}$, as $N_0, T_0\to\infty$,
\begin{align*}
\lhatwidei^*- \Hwide^{*-1}\ltid_i = \Ops \left( \frac{1}{\sqrt{T_0}} \right).
\end{align*}
\end{lemma}

\begin{pf}{}
The result follows immediately from the decomposition
\begin{align}
& \lhatwidei^*- \Hwide^{*-1}\ltid_i= \frac{1}{T_0}\left( \Fhatwide^* -%
\Ftidwide \Hwide^* \right)^\mathsf{T} \ewidei^*  \notag \\
& +\frac{1}{T_0}\Hwide^{*\mathsf{T}} \Ftidwide^\mathsf{T} \ewidei^* +\frac{1%
}{T_0}\Fhatwide^{*\mathsf{T}} \left( \Ftidwide -\Fhatwide^* \Hwide^{*-1}
\right) \ltid_i  \label{eq:decomposition of bootstrap lamwide}
\end{align}
and Lemmas \ref{lemma:FFHe}(1), \ref{lemma:HFe}(1), \ref{lemma:FFFHlam}(1).
\end{pf}

\begin{lemma}
\label{lemma:MSE between bootstrap lamwide and lamtid} If Assumptions \ref%
{ass:factors and loadings}--\ref{ass:order conditions} hold for a pure
factor model or Assumptions \ref{ass:factors and loadings}--\ref{ass:error
is independent of factor and x} for a factor model with covariates, then for
every integer $m\ge2$, as $N_0,T_0\to\infty$,

\begin{enumerate}[label=(\arabic*), labelindent=\parindent, leftmargin=*, nosep]

\item For the pre-treatment subsample,
\begin{align*}
\frac{1}{N}\sum_{i=1}^N \left\Vert \lhatwidei^*- \Hwide^{*-1}\ltid_i
\right\Vert^m= \Ops \left( \frac{1}{T_0} \right).
\end{align*}

\item For the control subsample,
\begin{align*}
\frac{1}{N_0}\sum_{i=1}^{N_0} \left\Vert \lhattalli^*- \Htall^{*-1}\ltid_i
\right\Vert^m= \Ops \left( \frac{1}{T} \right).
\end{align*}
\end{enumerate}
\end{lemma}

\begin{pf}{}
Firstly consider $m=2$. Then Claim (1) can be easily established by the
decomposition \eqref{eq:decomposition of bootstrap lamwide}, the $C_p$
inequality, and Lemmas \ref{lemma:FFHe}(2), \ref{lemma:HFe}(2), \ref%
{lemma:FFFHlam}(2). One can show that these conclusions also hold for the
control subsample, and Claim (2) can be proved analogously.

Now we turn to the case of $m>2$. By the results for the case of $m=2$ and
Assumption \ref{ass:order conditions}(2),
\begin{align*}
\sum_{i=1}^N \left\Vert \lhatwidei^*- \Hwide^{*-1}\ltid_i \right\Vert^2 =\Ops%
(1) \quad \text{and} \quad \sum_{i=1}^{N_0} \left\Vert \lhattalli^*- \Htall%
^{*-1}\ltid_i \right\Vert^2 =\Ops(1),
\end{align*}
which in turn implies that
\begin{align*}
\sum_{i=1}^N \left\Vert \lhatwidei^*- \Hwide^{*-1}\ltid_i \right\Vert^m =\Ops%
(1) \quad \text{and} \quad \sum_{i=1}^{N_0} \left\Vert \lhattalli^*- \Htall%
^{*-1}\ltid_i \right\Vert^m =\Ops(1)
\end{align*}
for any integer $m>2$. And the proof is completed by using Assumption \ref%
{ass:order conditions}(2) again.
\end{pf}

\begin{lemma}
\label{lemma:MSE between bootstrap lamtall and lamwide} If Assumptions \ref%
{ass:factors and loadings}--\ref{ass:order conditions} hold for a pure
factor model or Assumptions \ref{ass:factors and loadings}--\ref{ass:error
is independent of factor and x} for a factor model with covariates, then for
every integer $m\ge 2$, as $N_0, T_0\to\infty$,
\begin{align*}
\frac{1}{N_0}\sum_{i=1}^{N_0} \left\Vert \lhattalli^*-\Htall^{*-1}\Hwide^* %
\lhatwidei^* \right\Vert^m =\Ops \left( \frac{1}{T_0} \right).
\end{align*}
\end{lemma}

\begin{pf}{}
The conclusion is established by the properties of matrix norms, the $C_p$
inequality, Lemmas \ref{lemma:bootstrap rotation matrices}(1), \ref%
{lemma:MSE between bootstrap lamwide and lamtid}, and Assumption \ref%
{ass:order conditions}(2).
\end{pf}

\begin{lemma}
\label{lemma:bootstrap Hmiss} If Assumptions \ref{ass:factors and loadings}--%
\ref{ass:order conditions} hold for a pure factor model or Assumptions \ref%
{ass:factors and loadings}--\ref{ass:error is independent of factor and x}
for a factor model with covariates, then as $N_0, T_0\to\infty$,
\begin{align*}
\Hhatmiss^*-\Htall^{*-1}\Hwide^* = \Ops \left( \frac{1}{\sqrt{T_0}} \right).
\end{align*}
\end{lemma}

\begin{pf}{}
Let $\epsilon_i^*=\lhattalli^*-\Htall^{*-1}\Hwide^* \lhatwidei^*$ for every $%
i\in\{1,\ldots, N_0\}$. By construction,
\begin{align*}
& \Hhatmiss^*=\left( \frac{1}{N_0}\sum_{i=1}^{N_0} \lhattalli^* \lhatwidei^{*%
\mathsf{T}} \right) \left( \frac{1}{N_0}\sum_{i=1}^{N_0} \lhatwidei^* %
\lhatwidei^{*\mathsf{T}} \right)^{-1} \\
& = \left[ \frac{1}{N_0}\sum_{i=1}^{N_0} \left( \Htall^{*-1}\Hwide^* %
\lhatwidei^* +\epsilon_i^* \right) \lhatwidei^{*\mathsf{T}} \right] \left(
\frac{1}{N_0}\sum_{i=1}^{N_0} \lhatwidei^* \lhatwidei^{*\mathsf{T}}
\right)^{-1} \\
&= \Htall^{*-1}\Hwide^* +\left( \frac{1}{N_0}\sum_{i=1}^{N_0} \epsilon_i^* %
\lhatwidei^{*\mathsf{T}} \right) \left( \frac{1}{N_0}\sum_{i=1}^{N_0} %
\lhatwidei^* \lhatwidei^{*\mathsf{T}} \right)^{-1}.
\end{align*}
Moreover,
\begin{align*}
\frac{1}{N_0}\sum_{i=1}^{N_0} \lhatwidei^* \lhatwidei^{*\mathsf{T}}= \Hwide%
^{*-1}\Htall^* \left[ \frac{1}{N_0}\sum_{i=1}^{N_0} \left( \lhattalli^* %
\lhattalli^{*\mathsf{T}} -\lhattalli^* \epsilon_i^{*\mathsf{T}}
-\epsilon_i^* \lhattalli^{*\mathsf{T}} +\epsilon_i^* \epsilon_i^{*\mathsf{T}%
} \right) \right] \Htall^{*\mathsf{T}} \Hwide^{*\mathsf{T}-1}.
\end{align*}

By the triangle inequality, the properties of matrix norms, Cauchy-Schwarz
inequality, Lemmas \ref{lemma:p order sample moments of bootstrap}(2), \ref%
{lemma:MSE between bootstrap lamtall and lamwide}, and Assumption \ref%
{ass:order conditions}(2),
\begin{align*}
& \left\Vert \frac{1}{N_0}\sum_{i=1}^{N_0} \lhattalli^* \epsilon_i^{*\mathsf{%
T}} \right\Vert \le \frac{1}{N_0}\sum_{i=1}^{N_0} \left\Vert \lhattalli^*
\right\Vert \left\Vert \epsilon_i^* \right\Vert \le \sqrt{\frac{1}{N_0}%
\sum_{i=1}^{N_0} \left\Vert \lhattalli^* \right\Vert^2} \sqrt{\frac{1}{N_0}%
\sum_{i=1}^{N_0}\left\Vert \epsilon_i^* \right\Vert^2} =\Ops\left( \frac{1}{%
\sqrt{T_0}} \right).
\end{align*}
Similarly,
\begin{align*}
\left\Vert \frac{1}{N_0}\sum_{i=1}^{N_0} \epsilon_i^* \epsilon_i^{*\mathsf{T}%
} \right\Vert \le \frac{1}{N_0}\sum_{i=1}^{N_0}\left\Vert \epsilon_i^*
\right\Vert^2 =\Ops\left( \frac{1}{T_0} \right).
\end{align*}
Above results imply that
\begin{align*}
\frac{1}{N_0}\sum_{i=1}^{N_0} \lhatwidei^* \lhatwidei^{*\mathsf{T}}= \Hwide%
^{*-1}\Htall^* \Vhattall^* \Htall^{*\mathsf{T}}\Hwide^{*\mathsf{T} -1} +\Ops %
\left( \frac{1}{\sqrt{T_0}} \right).
\end{align*}
One can also show that
\begin{align*}
\left\Vert \frac{1}{N_0}\sum_{i=1}^{N_0} \epsilon_i^* \lhatwidei^{*\mathsf{T}%
} \right\Vert =\Ops\left( \frac{1}{\sqrt{T_0}} \right).
\end{align*}
Then the desired result follows from Lemma \ref{lemma:bootstrap rotation
matrices}.
\end{pf}

\begin{lemma}
\label{lemma:p order sample moments of bootstrap} If Assumptions \ref%
{ass:factors and loadings}--\ref{ass:order conditions} hold for a pure
factor model or Assumptions \ref{ass:factors and loadings}--\ref{ass:error
is independent of factor and x} for a factor model with covariates, then for
every $m\in \{2,3,\ldots,8\}$, as $N_0, T_0\to\infty$,

\begin{enumerate}[label=(\arabic*), labelindent=\parindent, leftmargin=*, nosep]

\item
\begin{align*}
\frac{1}{T}\sum_{t=1}^T \left\Vert \fhattallt^* \right\Vert^m =\Ops(1),
\qquad \frac{1}{T_0}\sum_{t=1}^{T_0} \left\Vert \fhatwidet^* \right\Vert^m =%
\Ops(1).
\end{align*}

\item
\begin{align*}
\frac{1}{N_0}\sum_{i=1}^{N_0} \left\Vert \lhattalli^* \right\Vert^m =\Ops%
(1), \qquad \frac{1}{N}\sum_{i=1}^N \left\Vert \lhatwidei^* \right\Vert^m =%
\Ops(1).
\end{align*}

\item
\begin{align*}
\frac{1}{T_0}\sum_{t=1}^{T_0} \left\vert \ehat_{i,t}^* \right\vert^m =\Ops%
(1) \quad \text{for every } i\in \{1,\ldots, N\}.
\end{align*}

\item
\begin{align*}
\frac{1}{N_0}\sum_{i=1}^{N_0} \left\vert \ehat_{i,t}^* \right\vert^m =\Ops%
(1) \quad \text{for every } t\in \{1,\ldots, T\}.
\end{align*}

\item
\begin{align*}
\frac{1}{T_0}\sum_{t=1}^{T_0} \left\vert \chat_{i,t}^*-\chat_{i,t}
\right\vert^m =\Ops\left( \frac{1}{T_0} \right) \quad \text{for every } i\in
\{1,\ldots, N\}.
\end{align*}
\end{enumerate}
\end{lemma}

\begin{pf}{}
By the $C_p$ inequality, the properties of matrix norms, and Lemmas \ref%
{lemma:p order sample mean is bounded}(3), \ref{lemma:bootstrap rotation
matrices}(1), \ref{lemma:MSE between bootstrap ftall and ftid},
\begin{align*}
&\frac{1}{T}\sum_{t=1}^T \left\Vert \fhattallt^* \right\Vert^m= \frac{1}{T}%
\sum_{t=1}^T \left\Vert \Htall^{*\mathsf{T}} \ftid_t+ \left( \fhattallt^*- %
\Htall^{*\mathsf{T}} \ftid_t \right)\right\Vert^m \\
& \le 2^{m-1}\left[ \left\Vert \Htall^* \right\Vert^m \left( \frac{1}{T}%
\sum_{t=1}^T \left\Vert \fhattallt \right\Vert^m \right) + \left( \frac{1}{T}%
\sum_{t=1}^T \left\Vert \fhattallt^*- \Htall^{*\mathsf{T}} \ftid_t
\right\Vert^m \right) \right] =\Ops(1).
\end{align*}
The rest parts of Claims (1) and (2) can be proved analogously.

To verify Claims (3)--(5), we conduct the following decomposition
\begin{align}
& \ehat_{i,t}^*= e_{i,t}^*-\left( \chat_{i,t}^*-\chat_{i,t} \right)  \notag
\\
&=e_{i,t}^*-\left( \fhattallt^*-\Htall^{*\mathsf{T}}\ftid_t \right)^\mathsf{T%
} \Hhatmiss^* \left( \lhatwidei^*-\Hwide^{*-1} \ltid_i \right) - \ftid_t^%
\mathsf{T} \Htall^* \Hhatmiss^* \left( \lhatwidei^*-\Hwide^{*-1} \ltid_i
\right)  \notag \\
&-\left( \fhattallt^*-\Htall^{*\mathsf{T}}\ftid_t \right)^\mathsf{T} %
\Hhatmiss^* \Hwide^{*-1}\ltid_i -\ftid_t^\mathsf{T} \left( \Htall^* \Hhatmiss%
^* \Hwide^{*-1} -I_r \right) \ltid_i.
\label{eq:decomposition of bootstrap ehat}
\end{align}
By the $C_p$ inequality,
\begin{align*}
& \left\vert \ehat_{i,t}^* \right\vert^m \le 5^{m-1} \left( \left\Vert
e_{i,t}^* \right\Vert^m + \left\Vert \fhattallt^*-\Htall^{*\mathsf{T}}\ftid%
_t \right\Vert^m \left\Vert \Hhatmiss^* \right\Vert^m \left\Vert \lhatwidei%
^*-\Hwide^{*-1} \ltid_i \right\Vert^m \right. \\
& + \left\Vert \fhattallt \right\Vert^m \left\Vert \Htall^* \right\Vert^m
\left\Vert \Hhatmiss^* \right\Vert^m \left\Vert \lhatwidei^*-\Hwide^{*-1} %
\ltid_i \right\Vert^m \\
& + \left\Vert \fhattallt^*-\Htall^{*\mathsf{T}}\ftid_t \right\Vert^m
\left\Vert \Hhatmiss^* \right\Vert^m \left\Vert \Hwide^{*-1} \right\Vert^m
\left\Vert \Hhatmiss \right\Vert^m \left\Vert \lhatwidei \right\Vert^m \\
& \left. + \left\Vert \fhattallt \right\Vert^m \left\Vert \Htall^* \Hhatmiss%
^* \Hwide^{*-1} -I_r \right\Vert^m \left\Vert \Hhatmiss \right\Vert^m
\left\Vert \lhatwidei \right\Vert^m \right),
\end{align*}
and
\begin{align*}
& \left\vert \chat_{i,t}^*-\chat_{i,t} \right\vert^m \le 4^{m-1} \left(
\left\Vert \fhattallt^*-\Htall^{*\mathsf{T}}\ftid_t \right\Vert^m \left\Vert %
\Hhatmiss^* \right\Vert^m \left\Vert \lhatwidei^*-\Hwide^{*-1} \ltid_i
\right\Vert^m \right. \\
& + \left\Vert \fhattallt \right\Vert^m \left\Vert \Htall^* \right\Vert^m
\left\Vert \Hhatmiss^* \right\Vert^m \left\Vert \lhatwidei^*-\Hwide^{*-1} %
\ltid_i \right\Vert^m \\
& + \left\Vert \fhattallt^*-\Htall^{*\mathsf{T}}\ftid_t \right\Vert^m
\left\Vert \Hhatmiss^* \right\Vert^m \left\Vert \Hwide^{*-1} \right\Vert^m
\left\Vert \Hhatmiss \right\Vert^m \left\Vert \lhatwidei \right\Vert^m \\
& \left. + \left\Vert \fhattallt \right\Vert^m \left\Vert \Htall^* \Hhatmiss%
^* \Hwide^{*-1} -I_r \right\Vert^m \left\Vert \Hhatmiss \right\Vert^m
\left\Vert \lhatwidei \right\Vert^m \right),
\end{align*}
Recall that $m\in \{2,3\ldots, 8\}$. Claims (3) and (5) are established by
Lemmas \ref{lemma:pointwise convergence of factors and loadings in sec3}(5), %
\ref{lemma:p order sample mean is bounded}(3), \ref{lemma:BHLC e}(1), \ref%
{lemma:bootstrap rotation matrices}, \ref{lemma:MSE between bootstrap ftall
and ftid}(1), \ref{lemma:point convergence of bootstrap lamwide}, \ref%
{lemma:bootstrap Hmiss} and Assumption \ref{ass:order conditions}(2). Claim
(4) follows from Lemmas \ref{lemma:pointwise convergence of factors and
loadings in sec3}(5), \ref{lemma:p order sample mean is bounded}(4), \ref%
{lemma:BHLC e}(2), \ref{lemma:bootstrap rotation matrices}, \ref{lemma:point
convergence of bootstrap ftall} \ref{lemma:MSE between bootstrap lamwide and
lamtid}(1), \ref{lemma:bootstrap Hmiss} and Assumption \ref{ass:order
conditions}(2).
\end{pf}

\begin{lemma}
\label{lemma:bootstrap Phi and Gamma} If Assumptions \ref{ass:factors and
loadings}--\ref{ass:order conditions} hold for a pure factor model or
Assumptions \ref{ass:factors and loadings}--\ref{ass:error is independent of
factor and x} for a factor model with covariates, then $\Phihat_i^*=\ops %
\left( T^{1/4} \right)$ and $\Gamhat_t^* =\Ops (1)$ as $N_0, T_0\to\infty$.
\end{lemma}

\begin{pf}{}
For every $k\in\{1,\ldots, K\}$, by the triangle inequality, the properties
of matrix norms, Cauchy-Schwarz inequality, Lemma \ref{lemma:p order sample
moments of bootstrap}(1)(3), and Assumption \ref{ass:order conditions}(2),
\begin{align*}
& \left\Vert L_{k,i}^* \right\Vert= \left\Vert \frac{1}{T_0}%
\sum_{s=k+1}^{T_0} \fhattalls^* \fhattallsk^{*\mathsf{T}} \ehat_{i,s}^* \ehat%
_{i,s-k}^* \right\Vert \le \frac{1}{T_0}\sum_{s=k+1}^{T_0} \left\Vert %
\fhattalls^* \ehat_{i,s}^* \right\Vert \left\Vert \fhattallsk^{*\mathsf{T}} %
\ehat_{i,s-k}^* \right\Vert \\
&\le \frac{1}{T_0}\sum_{s=1}^{T_0}\left\Vert \fhattalls^* \right\Vert^2 \ehat%
_{i,s}^{*2} \le \sqrt{\frac{1}{T_0}\sum_{s=1}^{T_0} \left\Vert \fhattalls^*
\right\Vert^4} \sqrt{\frac{1}{T_0}\sum_{s=1}^{T_0} \ehat_{i,s}^{*4} }=\Ops%
(1).
\end{align*}
Then by construction and the triangle inequality,
\begin{align*}
\left\Vert \Phihat_i^* \right\Vert \le \left\Vert L_{0,i}^* \right\Vert+2
\sum_{k=1}^K \left\Vert L_{k,i}^* \right\Vert \le (2K+1) \sqrt{\frac{1}{T_0}%
\sum_{s=1}^{T_0} \left\Vert \fhattalls^* \right\Vert^4} \sqrt{\frac{1}{T_0}%
\sum_{s=1}^{T_0} \ehat_{i,s}^{*4} } =\ops \left( T^{1/4} \right).
\end{align*}

By the triangle inequality, the properties of matrix norms, Cauchy-Schwarz
inequality, Lemma \ref{lemma:p order sample moments of bootstrap}(2)(4), and
Assumption \ref{ass:order conditions}(2),
\begin{align*}
& \left\Vert \Gamhat_t^* \right\Vert=\left\Vert \frac{1}{N_0}%
\sum_{i=1}^{N_0} \lhatwidej^* \lhatwidej^{*\mathsf{T}} \ehat_{j,t}^{*2}
\right\Vert \le \frac{1}{N_0}\sum_{i=1}^{N_0} \left\Vert \lhatwidej^*
\right\Vert^2 \ehat_{j,t}^{*2} \\
& \le \sqrt{\frac{1}{N_0}\sum_{i=1}^{N_0}\left\Vert \lhatwidej^*
\right\Vert^4} \sqrt{\frac{1}{N_0}\sum_{i=1}^{N_0}\ehat_{j,t}^{*4} }=\Ops(1).
\end{align*}
Then the proof is complete.
\end{pf}

\begin{lemma}
\label{lemma:bootstrap sigma2hat} If Assumptions \ref{ass:factors and
loadings}--\ref{ass:order conditions} hold for a pure factor model or
Assumptions \ref{ass:factors and loadings}--\ref{ass:error is independent of
factor and x} for a factor model with covariates, then for every $%
i\in\{1,\ldots, N\}$, we have $\sighat_i^{*2}-\sigma_{i}^2 =\ops(1)$ as $%
N_0, T_0\to\infty$.
\end{lemma}

\begin{pf}{}
By Assumptions \ref{ass:distributions of idiosyncratic errors}(2), \ref%
{ass:moments of idiosyncratic errors}(1) and the fact that $\left\{ u_{i,t}
\right\}$ are i.i.d.\ as $\bN(0,1)$ and independent of the raw sample, $%
\mathbb{E}\left( e_{i,t}^2 u_{i,t}^2 \right)=\sigma_i^2$, $\mathbb{E} \left(
u_{i,t}^4 \right)=3$, and $\mathbb{E}\left( e_{i,t}^2 u_{i,t}^4 \right) \le
3(M+1)$. Then by Markov's inequality, as $N_0, T_0\to\infty$,
\begin{align*}
\frac{1}{T_0}\sum_{t=1}^{T_0} u_{i,t}^4=O_\mathbb{P}(1), \qquad \frac{1}{T_0}%
\sum_{t=1}^{T_0} e_{i,t}^2 u_{i,t}^4=O_\mathbb{P}(1).
\end{align*}
The decomposition \eqref{eq:decomposition of ehat}, the triangle inequality,
Cauchy-Schwarz inequality, and Lemmas \ref{lemma:MSE of ehat with respect to
e}, \ref{lemma:properties of residuals with covariates} imply that
\begin{align*}
& \frac{1}{T_0}\sum_{t=1}^{T_0} \left( c_{i,t}-\chat_{i,t} \right)^2
u_{i,t}^2 \le \sqrt{\frac{1}{T_0}\sum_{t=1}^{T_0} \left( c_{i,t}-\chat_{i,t}
\right)^4} \sqrt{\frac{1}{T_0}\sum_{t=1}^{T_0} u_{i,t}^4}=O_\mathbb{P}
\left( \frac{1}{\sqrt{T_0}} \right) \\
& \left\vert \frac{1}{T_0}\sum_{t=1}^{T_0} \left( c_{i,t}-\chat_{i,t}
\right) e_{i,t} u_{i,t}^2 \right\vert \le \sqrt{\frac{1}{T_0}%
\sum_{t=1}^{T_0}\left( c_{i,t}-\chat_{i,t} \right)^2} \sqrt{\frac{1}{T_0}%
\sum_{t=1}^{T_0} e_{i,t}^2 u_{i,t}^4} =O_\mathbb{P} \left( \frac{1}{\sqrt{T_0%
}} \right).
\end{align*}
By Cauchy-Schwarz inequality and Lemmas \ref{lemma:BHLC e}(1), \ref{lemma:p
order sample moments of bootstrap}(5),
\begin{align*}
\left\Vert \frac{1}{T_0}\sum_{t=1}^{T_0} e_{i,t}^* \left( \chat_{i,t}-\chat%
_{i,t}^* \right) \right\Vert \le \sqrt{\frac{1}{T_0}\sum_{t=1}^{T_0}
e_{i,t}^{*2}} \sqrt{\frac{1}{T_0}\sum_{t=1}^{T_0} \left( \chat_{i,t}-\chat%
_{i,t}^* \right)^2} =\Ops \left( \frac{1}{\sqrt{T_0}} \right).
\end{align*}
Therefore,
\begin{align*}
& \sighat_i^{*2}=\frac{1}{T_0}\sum_{t=1}^{T_0} \ehat_{i,t}^{*2}= \frac{1}{T_0%
}\sum_{t=1}^{T_0} \left[ e_{i,t}^*+\left( \chat_{i,t}-\chat_{i,t}^* \right) %
\right]^2 \\
& =\frac{1}{T_0}\sum_{t=1}^{T_0} e_{i,t}^{*2}+\frac{1}{T_0} \sum_{t=1}^{T_0}
\left( \chat_{i,t}-\chat_{i,t}^* \right)^2 +\frac{2}{T_0}\sum_{t=1}^{T_0}
e_{i,t}^* \left( \chat_{i,t}-\chat_{i,t}^* \right) \\
& =\frac{1}{T_0}\sum_{t=1}^{T_0} e_{i,t}^{*2}+\Ops \left( \frac{1}{\sqrt{T_0}%
} \right) =\frac{1}{T_0}\sum_{t=1}^{T_0} \left[ e_{i,t} +\left( c_{i,t}-\chat%
_{i,t} \right) \right]^2 u_{i,t}^2 +\Ops \left( \frac{1}{\sqrt{T_0}} \right)
\\
& =\frac{1}{T_0}\sum_{t=1}^{T_0} e_{i,t}^2 u_{i,t}^2 +\frac{1}{T_0}%
\sum_{t=1}^{T_0} \left( c_{i,t}-\chat_{i,t} \right)^2 u_{i,t}^2 +\frac{2}{T_0%
}\sum_{t=1}^{T_0} e_{i,t} \left( c_{i,t}-\chat_{i,t} \right) u_{i,t}^2 +\Ops %
\left( \frac{1}{\sqrt{T_0}} \right) \\
& =\frac{1}{T_0}\sum_{t=1}^{T_0} e_{i,t}^2 u_{i,t}^2+ O_\mathbb{P} \left(%
\frac{1}{\sqrt{T_0}} \right) +\Ops \left( \frac{1}{\sqrt{T_0}} \right) =%
\frac{1}{T_0}\sum_{t=1}^{T_0} e_{i,t}^2 u_{i,t}^2+\Ops \left( \frac{1}{\sqrt{%
T_0}} \right).
\end{align*}
And the proof is completed by Assumption \ref{ass:distributions of
idiosyncratic errors}(2) and the ergodic theorem.
\end{pf}

\section{Proof of Main Results}	 \label{sec:Proof of Main Results}
In this section, we provide the proofs of Theorem \ref{thry:coverage} and Theorem \ref{thry:cov coverage}. 

\subsection{Proof of Theorem \ref{thry:coverage}}

For $(i,t)\in\mI_1$, define \begin{align*}
	s_{i,t}=\frac{\chat_{i,t}-y_{i,t}}{\sqrt{\bVhat_{i,t}+\sighat^2_i}}.
\end{align*}
Since $\lbr e_{i,t}:t=1,2,\ldots \rbr$ is strictly stationary,
we have $\V\lp e_{i,t} \rp=\sigma^2_i$ for all $t\in\mathbb{Z}_+$.

\begin{lemma}\label{lemma:convergence of sit}
	If Assumptions \ref{ass:factors and loadings}--\ref{ass:order conditions}
	hold, then $s_{i,t}=-\dfrac{e_{i,t}}{\sigma_{i}} + o_\p(1)$
	as $N_0,T_0\to\infty$ for every
	$(i,t)\in\mI_1$.
\end{lemma}

\begin{pf}{}
	By the decomposition of $\lp \chat_{i,t}-c_{i,t} \rp$ in Equation
	\eqref{eq:decomposition of ehat} and
	Lemma \ref{lemma:pointwise convergence of factors and loadings in sec3}(2)--(5),
	it follows that $\chat_{i,t}-c_{i,t}=o_\p(1)$.
	
	By Cauchy-Schwarz inequality, Assumption \ref{ass:moments of idiosyncratic errors}(1),
	Markov's inequality, and Lemma \ref{lemma:MSE of ehat with respect to e},
	\begin{align*}
		\lv \frac{1}{T_0}\sum_{s=1}^{T_0} e_{i,s}\lp c_{i,s}-\chat_{i,s} \rp \rv
		\le \sqrt{\frac{1}{T_0}\sum_{s=1}^{T_0} e_{i,s}^2}
		\sqrt{\frac{1}{T_0}\sum_{s=1}^{T_0} \lp c_{i,s}-\chat_{i,s} \rp^2}
		=O_\p \lp \frac{1}{\sqrt{T_0}} \rp.
	\end{align*}
	Then $\sighat_i^2$ admits a decomposition
	\begin{align*}
		& \sighat_i^2 =\frac{1}{T_0}\sum_{s=1}^{T_0} \ehat_{i,s}^2=
		\frac{1}{T_0}\sum_{s=1}^{T_0} \lbk e_{i,s} +\lp c_{i,s}-\chat_{i,s} \rp \rbk^2 \\
		& =\frac{1}{T_0}\sum_{s=1}^{T_0} e_{i,s}^2
		+\frac{1}{T_0}\sum_{s=1}^{T_0} \lp c_{i,s}-\chat_{i,s} \rp^2
		+\frac{2}{T_0}\sum_{s=1}^{T_0} e_{i,s} \lp c_{i,s}-\chat_{i,s} \rp
		=\frac{1}{T_0}\sum_{s=1}^{T_0} e_{i,s}^2 +O_\p \lp \frac{1}{\sqrt{T_0}} \rp.
	\end{align*}
	By Assumption \ref{ass:distributions of idiosyncratic errors}(2) and the ergodic
	theorem, $\sighat_i^2 \convp \sigma_i^2$ as $N_0, T_0\to\infty$.
	
	Theorem 6 of \citet{bai2003inferential} implies that
	\begin{align*}
		\fhattallt^\T \lp \frac{\Fhattall^\T \Fhattall}{T}
		\rp^{-1} \Phihat_i \lp \frac{\Fhattall^\T \Fhattall}{T} \rp^{-1} \fhattallt
		&= O_\p(1),	 \\
		\lhatwidei^\T \lp \frac{\Lhatwide^\T	\Lhatwide}{N} \rp^{-1}
		\Gamhat_t \lp \frac{\Lhatwide^\T \Lhatwide}{N} \rp^{-1} \lhatwidei
		&= O_\p(1),
	\end{align*}
	and consequently $\bVhat_{i,t}=o_\p(1)$.
	Therefore, \begin{align*}
		s_{i,t}=\frac{\lp \chat_{i,t}-c_{i,t}\rp +\lp c_{i,t}-y_{i,t}\rp}
		{\sqrt{\sigma_i^2 +\lp \bVhat_{i,t}+ \sighat_i^2-\sigma_i^2 \rp}}
		=\frac{-e_{i,t}+o_\p(1)}{\sqrt{\sigma_i^2+o_\p(1)}}=-\frac{e_{i,t}}{\sigma_i}+o_\p(1)
	\end{align*}
	as $N_0,T_0\to\infty$ for every $(i,t)\in\mI_1$.
\end{pf}

\begin{lemma}\label{lemma:convergence of s*it}
	If Assumptions \ref{ass:factors and loadings}--\ref{ass:order conditions}
	hold, then $s_{i,t}^* =  -\dfrac{e_{i,t}^*}{\sigma_i} +  \ops(1)$
	as $N_0,T_0\to\infty$ for every $(i,t)\in\mI_1$.
\end{lemma}

\begin{pf}{}
	By the decomposition of $\lp \chat_{i,t}^*-\chat_{i,t} \rp$ in Equation
	\eqref{eq:decomposition of bootstrap ehat} and
	Lemmas \ref{lemma:bootstrap rotation matrices},
	\ref{lemma:point convergence of bootstrap ftall},
	\ref{lemma:point convergence of bootstrap lamwide},
	\ref{lemma:bootstrap Hmiss},
	it follows that $\chat_{i,t}^*-\chat_{i,t}=\ops(1)$.
	
	By construction, \begin{align*}
		\bVhat_{i,t}^*&=\frac{1}{T_0} \fhattallt^{*\T} \lp \frac{\Fhattall^{*\T} \Fhattall^*}{T}
		\rp^{-1} \Phihat_i^* \lp \frac{\Fhattall^{*\T} \Fhattall^*}{T} \rp^{-1} \fhattallt^* \\
		&\phantom{=\:\:}	+\frac{1}{N_0} \lhatwidei^{*\T} \lp \frac{\Lhatwide^{*\T}
			\Lhatwide^*}{N} \rp^{-1}
		\Gamhat_t^* \lp \frac{\Lhatwide^{*\T} \Lhatwide^*}{N} \rp^{-1} \lhatwidei^* \\
		&=\frac{1}{T_0}\fhattallt^{*\T} \Phihat_i^* \fhattallt^*
		+\frac{1}{N_0}\lhatwidei^{*\T} \Vhatwide^{*-1} \Gamhat_t^* \Vhatwide^{*-1} \lhatwidei^*.		
	\end{align*}
	Combining this with Lemmas \ref{lemma:bootstrap rotation matrices}(2),
	\ref{lemma:bootstrap Phi and Gamma} yields $\bVhat_{i,t}^*=\ops(1)$.
	
	By Lemma \ref{lemma:bootstrap sigma2hat},  \begin{align*}
		s^*_{i,t}=\frac{\lp \chat_{i,t}^*-\chat_{i,t}\rp +\lp \chat_{i,t}-y^*_{i,t}\rp}
		{\sqrt{\sigma_i^2+ \lbk \bVhat_{i,t}^*+ \lp\sighat_i^*\rp^2-\sigma_i^2 \rbk}}
		=\frac{-e^*_{i,t}+o_{\p^*}(1)}{\sqrt{\sigma_i^2+o_{\p^*}(1)}}
		=-\frac{e^*_{i,t}}{\sigma_i}+o_{\p^*}(1)
	\end{align*}
	as $N_0,T_0\to\infty$ for every $(i,t)\in\mI_1$.
\end{pf}

Let $G_{e,i}$ and $G_{s,i}$ be the cumulative distribution functions
of $e_{i,t}$ and $-\dfrac{e_{i,t}}{\sigma_i}$, respectively, \textit{i.e.},
\begin{align*}
	G_{e,i}(z)=\p\lp e_{i,t}\le z \rp, \qquad
	G_{s,i}(z)=\p\lp -\dfrac{e_{i,t}}{\sigma_i}\le z \rp, \qquad
	z\in\mr.
\end{align*}	
Let $G^*_{e_{i,t}^*}$ and $G^*_{s_{i,t}^*}$ be the conditional cumulative distributions
function of $e_{i,t}^*$ and $s_{i,t}^*$ given $\bS$, respectively, so that
$G^*_{e_{i,t}^*}(z)=\p^*\lp e_{i,t}^* \le z \rp$ and
$G^*_{s_{i,t}^*}(z)=\p^*\lp s_{i,t}^* \le z \rp$ for every $z\in\mr$.

For two (unconditional or conditional) distribution functions $G_1$ and $G_2$,
we measure the distance between
$G_1$ and $G_2$ by the Mallows metric that is defined as
\begin{align*}
	d_2\lp G_1, G_2 \rp=\sqrt{ \inf_{\xi \in \Xi\lp G_1,G_2\rp}
		\int_{\mr^2} \lp z_1-z_2 \rp^2 \ddd \xi \lp z_1, z_2 \rp },
\end{align*}
where $\Xi\lp G_1, G_2 \rp$ is the set of bivariate joint distribution
functions with marginal distribution functions $G_1$ and $G_2$.
Let $\mathscr{Z}\lp G_1, G_2 \rp=\lbr Z\sim \xi: \xi \in \Xi\lp G_1, G_2 \rp \rbr$,
and $\E_Z$ be the expectation with respect to $\xi$ for every $Z\sim \xi$.
Then the Mallows metric can be equivalently expressed as \begin{align*}
	d_2\lp G_1, G_2 \rp=\sqrt{\inf_{Z\in\mathscr{Z}\lp G_1, G_2 \rp}
		\E_Z \lp \lv Z_1- Z_2 \rv^2 \rp}.
\end{align*}

Moreover, let $\rightsquigarrow$ denote weak convergence
of distribution functions. We say $G_{N,T}\rightsquigarrow G$ as $N,T\to\infty$
if and only if for any bounded Lipschitz continuous function $h:\mr\to \mr$, \begin{align*}
	\lim_{N,T\to\infty} \int_{\mr} h(z) \ddd G_{N,T} (z) =\int_{\mr} h(z) \ddd G(z).
\end{align*}
One can see Lemma 2.2 (Portmanteau) of \citet{van1998asymptotic} for equivalent
characterisations of weak convergence.

\begin{lemma}\label{lemma:convergence of distributions of e}
	If Assumptions \ref{ass:factors and loadings}--\ref{ass:order conditions}
	hold, then	$d_2\lp G^*_{e_{i,t}^*}, G_{e,i} \rp\convp 0$ as $T_0\to\infty$
	for every $(i,t)\in\mI_1$.
\end{lemma}

\begin{pf}{}
	Let $\bG_{\ehat,i,T_0}$ and $\bG_{e,i,T_0}$ be empirical distribution
	functions so that \begin{align*}
		\bG_{\ehat,i,T_0}(z)=\frac{1}{T_0}\sum_{s=1}^{T_0}
		\indicator\lbr \ehat_{i,s}-\ehatbar_i \le z \rbr, \qquad
		\bG_{e,i,T_0}=\frac{1}{T_0}\sum_{s=1}^{T_0}
		\indicator\lbr e_{i,s}\le x \rbr, \qquad z\in\mr.
	\end{align*}
	
	We firstly show that given almost every realisation of $\bS$,
	the quantity $z^2$ is uniformly integrable with respect to
	$\lbr \bG_{e,i,T_0} \rbr_{T_0=1}^\infty$,
	\textit{i.e.}, \begin{align*}
		\lim_{\bM\to\infty} \lbk \sup_{T_0\in\bZ_+} \int_{\mr}
		z^2 \indicator_{(\bM,\infty)}\lp z^2 \rp \ddd \bG_{e,i,T_0}(z) \rbk =0.
	\end{align*}
	Consider a fixed $i\in\{N_0+1,\ldots, N\}$.
	By Assumption \ref{ass:distributions of idiosyncratic errors}(2) and the ergodic
	theorem, \begin{align*}
		\bW_{T_0,\bM}\defeq \int_{\mr}
		z^2 \indicator_{(\bM,\infty)}\lp z^2 \rp \ddd \bG_{e,i,T_0}(z)
		=\frac{1}{T_0}\sum_{s=1}^{T_0} e_{i,s}^2 \indicator_{(\bM,\infty)} \lp e_{i,s}^2 \rp
		\convas \E\lbk e_{i,s}^2 \indicator_{(\bM,\infty)} \lp e_{i,s}^2 \rp  \rbk
		\defeq \bW_{\bM}
	\end{align*}
	as $T_0\to\infty$.
	And by Assumption \ref{ass:moments of idiosyncratic errors}(1),
	$\bW_{\bM}\to 0$ as $\bM\to\infty$.
	Therefore, for any $\varepsilon>0$,	there exists $M_1>0$, such that
	$0\le \bW_{M_1} <\varepsilon/2$. Given almost every realisation of $\bS$,
	there exists $M_2>0$, such that $\lv \bW_{T_0,M_1}-\bW_{M_1} \rv<\varepsilon/2$
	for all $T_0>M_2$.
	Note that $\bW_{T_0,\bM}$ is decreasing in $\bM$ for any fixed $T_0$, so $0\le \bW_{T_0,\bM} <\varepsilon$
	holds for all $T_0>M_2$ and $\bM \ge M_1$. Pick $\bM_0=M_1 \vee \lp 1+ \max \lbr e^2_{i,1}, \ldots, e^2_{i,M_2}
	\rbr \rp$.
	Then \begin{align*}
		\sup_{T_0\in\bZ_+} \bW_{T_0,\bM_0}=\lp \max_{1\le T_0 \le M_2} \bW_{T_0,\bM_0} \rp
		\vee \lp \sup_{T_0>M_2} \bW_{T_0,\bM_0} \rp
		=0 \vee \lp \sup_{T_0>M_2} \bW_{T_0,\bM_0} \rp
		<\varepsilon.
	\end{align*}		
	
	By construction, $G^*_{e_{i,t}^*}=\bG_{\ehat,i,T_0}$ for every
	$(i,t)\in\mI_1$. The triangle inequality yields \begin{align*}
		d_2 \lp \bG_{\ehat,i,T_0}, G_{e,i} \rp \le
		d_2 \lp \bG_{\ehat,i,T_0}, \bG_{e,i,T_0} \rp+
		d_2 \lp \bG_{e,i,T_0}, G_{e,i} \rp.
	\end{align*}
	The ergodic theorem implies
	that $\bG_{e,i,T_0}(z)\convas G_{e,i}(z)$ as $T_0\to\infty$
	for every $z\in\mr$. By Lemma 2.11 of \citet{van1998asymptotic}
	and the continuity of $G_{e,i}$ [which follows from Assumption
	\ref{ass:distributions of idiosyncratic errors}(3)],
	\begin{align*}
		\sup_{z\in\mr} \lv \bG_{e,i,T_0}(z)-G_{e,i}(z) \rv \convas 0
	\end{align*}
	as $T_0\to\infty$.
	By Lemma 2.2 of \citet{van1998asymptotic},
	$\p\lp \bG_{e,i,T_0} \rightsquigarrow G_{e,i} \text{ as } T_0\to\infty \rp=1$.
	We have shown that given almost every realisation of $\bS$,
	the quantity $z^2$ is uniformly integrable with respect to
	$\lbr \bG_{e,i,T_0} \rbr_{T_0=1}^\infty$.
	Applying Lemma 8.3 of \citet{bickel1981some} yields that
	$d_2 \lp \bG_{e,i,T_0}, G_{e,i} \rp \convas 0$.
	
	Let $J$ be drawn from a discrete uniform distribution on $\{1,2,\ldots,T_0 \}$.
	Then conditional on $\bS$, we have $\lp \ehat_{i,J}-\ehatbar_i \rp\sim \bG_{\ehat,i,T_0}$
	and $e_{i,J}\sim \bG_{e,i,T_0}$ for every $N_0<i \le N$.
	By the $C_p$ inequality and Lemma \ref{lemma:MSE of ehat with respect to e},
	\begin{align*}
		& \lbk d_2 \lp \bG_{\ehat,i,T_0}, \bG_{e,i,T_0} \rp \rbk^2  \le
		\E^* \lbk \lp \ehat_{i,J}-\ehatbar_i-e_{i,J} \rp^2 \rbk
		= \frac{1}{T_0}\sum_{s=1}^{T_0}\lp \ehat_{i,s}-\ehatbar_i-e_{i,s}\rp^2 \\
		& \le \frac{2}{T_0}\sum_{s=1}^{T_0}\lp \ehat_{i,s}-e_{i,s}\rp^2
		+2 \lp \ehatbar_i \rp^2 = O_\p\lp \frac{1}{T_0} \rp+ 2 \lp \ehatbar_i \rp^2.
	\end{align*}
	Furthermore, by the $C_p$ inequality, Assumptions \ref{ass:distributions of idiosyncratic errors}(2),
	\ref{ass:moments of idiosyncratic errors}(1), the ergodic theorem,
	and Lemma \ref{lemma:MSE of ehat with respect to e},
	\begin{align*}
		&	\lp \ehatbar_i \rp^2=\lbk \frac{1}{T_0}\sum_{s=1}^{T_0}\lp \ehat_{i,s}
		-e_{i,s} \rp +\frac{1}{T_0}\sum_{s=1}^{T_0} e_{i,s} \rbk^2
		\le 2 \lbk \frac{1}{T_0}\sum_{s=1}^{T_0}\lp \ehat_{i,s}
		-e_{i,s} \rp \rbk^2 +2 \lp \frac{1}{T_0}\sum_{s=1}^{T_0} e_{i,s} \rp^2 \\
		& \le \frac{2}{T_0}\sum_{s=1}^{T_0} \lp \ehat_{i,s}-e_{i,s} \rp^2 +o_\p(1)
		=o_\p(1),
	\end{align*}
	and the proof is complete.
\end{pf}

\begin{lemma}\label{lemma:convergence of distributions of s}
	If Assumptions \ref{ass:factors and loadings}--\ref{ass:order conditions}
	hold, then		\begin{align*}
		\sup_{z\in\mr} \lv G^*_{s_{i,t}^*}(z)-G_{s_{i,t}}(z) \rv
		\convp 0
	\end{align*}
	as $N_0,T_0\to\infty$ for every $(i,t)\in\mI_1$.		
\end{lemma}

\begin{pf}{}
	By the proof of Lemma 8.3 of \citet{bickel1981some}, for every
	bounded Lipschitz continuous function $h:\mr\to \mr$ with Lipschitz
	constant $K_L$, \begin{align*}
		\lv \E^* \lbk h \lp e_{i,t}^* \rp \rbk -\E \lbk h \lp e_{i,t} \rp \rbk \rv
		=\lv \int_\mr h(z)\ddd G^*_{e_{i,t}^*} (z) -\int_\mr h(z)
		\ddd G_{e,i} \rv \le K_L d_2 \lp G^*_{e_{i,t}^*}, G_{e,i} \rp.
	\end{align*}
	By Lemma \ref{lemma:convergence of distributions of e}
	and the definition of conditional convergence in distribution,
	we have	$e_{i,t}^* \convds e_{i,t}$ as $T_0\to\infty$
	for every $(i,t)\in\mI_1$.
	By Lemmas \ref{lemma:relationship between bootstrap convergence in P and D}
	and \ref{lemma:convergence of s*it},
	$s_{i,t}^*\convds -\dfrac{e_{i,t}}{\sigma_i}\sim G_{s,i}$
	as $N_0,T_0\to\infty$ for every $(i,t)\in\mI_1$.
	Furthermore, by Theorem 2.7(ii) of \citet{van1998asymptotic}
	and Lemma \ref{lemma:convergence of sit} of this paper,
	$s_{i,t}\convd -\dfrac{e_{i,t}}{\sigma_i}$ as $N_0,T_0\to\infty$ for every
	$(i,t)\in\mI_1$.
	Since $G_{s,i}$ is everywhere continuous by Assumption
	\ref{ass:distributions of idiosyncratic errors}(3),
	we can use Lemma 2.11 of \citet{van1998asymptotic}
	and Lemma \ref{lemma:cdf characterisation of conditional convergence in D} of this paper
	to conclude that \begin{align*}
		\sup_{z\in\mr} \lv G_{s_{i,t}}(z)-G_{s,i}(z)\rv \to 0 \qquad
		\text{and}\qquad
		\sup_{z\in\mr} \lv G^*_{s_{i,t}^*}(z)-G_{s,i}(z)\rv \convp 0
	\end{align*}
	as $N_0,T_0\to\infty$ for every $(i,t)\in\mI_1$. Applying the triangle
	inequality yields the result.
\end{pf}

Now we turn to prove Theorem \ref{thry:coverage}.

\begin{pf}{ of Theorem \ref{thry:coverage}}
	Firstly consider the equal-tailed confidence intervals.
	By Assumption \ref{ass:distributions of idiosyncratic errors}(3)
	and the construction of $s_{i,t}$, the distribution function $G_{s_{i,t}}$
	is everywhere continuous, which implies that $G_{s_{i,t}} \lp s_{i,t} \rp
	\sim \mathrm{Unif}[0,1]$.
	From Lemma \ref{lemma:convergence of distributions of s}, for any $\varepsilon>0$,
	\begin{align*}
		\lim_{N_0,T_0\to\infty} \p \lp \sup_{z\in\mr} \lv G_{s_{i,t}^*}^*(z)
		-G_{s_{i,t}}(z) \rv \ge \frac{\varepsilon}{2} \rp =0.
	\end{align*}
	Note that \begin{align*}
		& \lbr G_{s_{i,t}} \lp s_{i,t} \rp \le \frac{\alpha}{2}-\frac{\varepsilon}{2} \rbr
		\cap \lbr \sup_{z\in\mr} \lv G_{s_{i,t}^*}^*(z)-G_{s_{i,t}}(z) \rv
		<\frac{\varepsilon}{2} \rbr \subset
		\lbr G_{s_{i,t}^*}^* \lp s_{i,t} \rp \le \frac{\alpha}{2} \rbr \\
		&\subset \lbr G_{s_{i,t}} \lp s_{i,t} \rp \le \frac{\alpha}{2}+\frac{\varepsilon}{2} \rbr
		\cup \lbr \sup_{z\in\mr} \lv G_{s_{i,t}^*}^*(z)-G_{s_{i,t}}(z) \rv
		\ge \frac{\varepsilon}{2} \rbr.
	\end{align*}
	Therefore, \begin{align*}
		& \p \lbk G_{s_{i,t}^*}^* \lp s_{i,t} \rp \le \frac{\alpha}{2} \rbk
		\le \p \lbk G_{s_{i,t}} \lp s_{i,t} \rp \le \frac{\alpha}{2}+\frac{\varepsilon}{2} \rbk
		+\p \lbk \sup_{z\in\mr} \lv G_{s_{i,t}^*}^*(z)-G_{s_{i,t}}(z) \rv
		\ge \frac{\varepsilon}{2} \rbk \\
		& = \frac{\alpha}{2}+\frac{\varepsilon}{2}+o(1),
	\end{align*}
	and \begin{align*}
		& \p \lbk G_{s_{i,t}^*}^* \lp s_{i,t} \rp \le \frac{\alpha}{2} \rbk  \ge 1-
		\p \lbk G_{s_{i,t}} \lp s_{i,t} \rp > \frac{\alpha}{2}-\frac{\varepsilon}{2} \rbk
		-\p \lbk \sup_{z\in\mr} \lv G_{s_{i,t}^*}^*(z)-G_{s_{i,t}}(z) \rv
		\ge \frac{\varepsilon}{2} \rbk \\
		& =\frac{\alpha}{2}-\frac{\varepsilon}{2}+o(1).
	\end{align*}
	Similarly, we have \begin{align*}
		1-\frac{\alpha}{2}-\frac{\varepsilon}{2}+o(1) \le
		\p \lbk G_{s_{i,t}^*}^* \lp s_{i,t} \rp \le 1- \frac{\alpha}{2} \rbk
		\le 1-\frac{\alpha}{2}+\frac{\varepsilon}{2}+o(1).
	\end{align*}
	The above facts imply \begin{align*}
		\lv \p \lbk G_{s_{i,t}^*}^* \lp s_{i,t} \rp \le 1- \frac{\alpha}{2} \rbk
		- \p \lbk G_{s_{i,t}^*}^* \lp s_{i,t} \rp \le \frac{\alpha}{2} \rbk
		-(1-\alpha) \rv \le \varepsilon+o(1).
	\end{align*}
	Because $\varepsilon>0$ can be arbitrarily small, the desired result
	follows from the equality \begin{align*}
		& \p\lp \Delta_{i,t}\in \mathrm{EQ}_{1-\alpha,i,t} \rp=
		\p \lp \chat_{i,t}-q_{1-(\alpha/2),i,t} \sqrt{\bVhat_{i,t}+\sighat_i^2}
		\le y_{i,t} \le
		\chat_{i,t}-q_{\alpha/2,i,t} \sqrt{\bVhat_{i,t}+\sighat_i^2} \rp \\
		&=\p\lp s_{i,t}\le q_{1-(\alpha/2),i,t} \rp -\p \lp
		s_{i,t}\le q_{\alpha/2,i,t} \rp
		=\p \lp G^*_{s_{i,t}^*}\lp s_{i,t} \rp \le 1-\frac{\alpha}{2} \rp
		-\p\lp G^*_{s_{i,t}^*}\lp s_{i,t} \rp \le \frac{\alpha}{2} \rp.
	\end{align*}
	
	The result for the symmetric confidence intervals can be proved analogously.
\end{pf}

\subsection{Proof of Theorem \ref{thry:cov coverage}}

For $(i,t)\in\mI_1$, define \begin{align*}
	r_{i,t}=y_{i,t}-x_{i,t}^\T \beta, \qquad
	\rhat_{i,t}=y_{i,t}-x_{i,t}^\T \betahattall, \qquad
	s_{i,t}=\frac{\chat_{i,t}-\rhat_{i,t}}{\sqrt{\bVhat_{i,t}+\sighat^2_i}}.
\end{align*}

\begin{lemma}\label{lemma:sec4 convergence of sit}
	If Assumptions \ref{ass:factors and loadings}--\ref{ass:error is independent of factor and x}
	hold, then $s_{i,t}=-\dfrac{e_{i,t}}{\sigma_{i}} + o_\p(1)$
	as $N_0,T_0\to\infty$ for every	$(i,t)\in\mI_1$.
\end{lemma}

\begin{pf}{}
	By Lemmas \ref{lemma:properties of estimators with covariates},
	\ref{lemma:properties of residuals with covariates} and
	the proof of Lemma \ref{lemma:convergence of sit},
	we can establish that $\chat_{i,t}-c_{i,t} =o_\p(1)$,
	$\bVhat_{i,t}=o_\p(1)$, and $\sighat_i^2-\sigma_i^2=o_\p(1)$
	as $N_0, T_0\to\infty$ for every $(i,t)\in\mI_1$.
	By Theorem 1 of \citet{bai2009panel}, \begin{align*}
		r_{i,t}-\rhat_{i,t}=x_{i,t}^\T \lp \betahattall-\beta \rp
		=O_\p\lp \frac{1}{\sqrt{T N_0}} \rp =o_\p(1).
	\end{align*}
	Therefore, \begin{align*}
		s_{i,t}=\frac{\lp \chat_{i,t}-c_{i,t}\rp+\lp c_{i,t}-r_{i,t}\rp
			+\lp r_{i,t}-\rhat_{i,t} \rp}{\sqrt{\sigma_i^2+\lp \bVhat_{i,t}
				+\sighat_i^2-\sigma_i^2\rp}}
		=\frac{-e_{i,t}+o_\p(1)}{\sqrt{\sigma_{i}^2+o_\p(1)}}
		=-\frac{e_{i,t}}{\sigma_i}+o_\p(1)
	\end{align*}
	as $N_0, T_0\to\infty$ for every $(i,t)\in\mI_1$, and the proof is complete.
\end{pf}

\begin{lemma}\label{lemma:sec4 convergence of s*it}
	If Assumptions \ref{ass:factors and loadings}--\ref{ass:error is independent of factor and x}
	hold, then $s_{i,t}^* =  -\dfrac{e_{i,t}^*}{\sigma_i} +  \ops(1)$
	as $N_0,T_0\to\infty$ for every $(i,t)\in\mI_1$.
\end{lemma}

\begin{pf}{}
	Because the bootstrap is conducted within a pure factor structure regardless
	of the existence of covariates, this lemma follows directly from
	Lemma \ref{lemma:convergence of s*it}.
\end{pf}

\begin{lemma}\label{lemma:sec4 convergence of distributions of e}
	If Assumptions \ref{ass:factors and loadings}--\ref{ass:error is independent of factor and x}
	hold, then	$d_2\lp G^*_{e_{i,t}^*}, G_{e,i} \rp\convp 0$ as $T_0\to\infty$
	for every $(i,t)\in\mI_1$.
\end{lemma}

\begin{pf}{}
	The proof is completed by Lemma \ref{lemma:properties of estimators with covariates},
	\ref{lemma:properties of residuals with covariates} and the proof
	of Lemma \ref{lemma:convergence of distributions of e}.
\end{pf}

\begin{lemma}\label{lemma:sec4 convergence of distributions of s}
	If Assumptions \ref{ass:factors and loadings}--\ref{ass:error is independent of factor and x}
	hold, then		\begin{align*}
		\sup_{z\in\mr} \lv G^*_{s_{i,t}^*}(z)-G_{s_{i,t}}(z) \rv
		\convp 0
	\end{align*}
	as $N_0,T_0\to\infty$ for every $(i,t)\in\mI_1$.		
\end{lemma}

\begin{pf}{}
	The same as the proof of Lemma \ref{lemma:convergence of distributions of s}.
\end{pf}

\begin{pf}{ of Theorem \ref{thry:cov coverage}}
	The proof follows that of Theorem \ref{thry:coverage} and Lemma \ref{lemma:sec4 convergence of sit}-\ref{lemma:sec4 convergence of distributions of s}.
\end{pf}

\end{appendices}

\end{document}